\RequirePackage{ifpdf}
\documentclass[12pt]{JHEP3}

\usepackage{amssymb,amsmath,amsfonts}
\usepackage[final]{showkeys}
\usepackage{amsmath}
\usepackage{graphicx,caption,subcaption}
\usepackage{epsfig}
\usepackage{latexsym}
\def\refeq#1{(\ref{#1})}
\def\hri#1#2{\href{http://arxiv.org/abs/#1}{[ArXiv:#1]#2}}
\def\hre#1#2{\href{http://arxiv.org/abs/#1/#2}{[ArXiv:#1/#2]}}

\def\dt{\partial}
\def\diag{{\rm diag}}
\def\s{\sigma}
\def\cG{{\cal G}}
\def\cD{{\cal D}}
\def\cF{{\cal F}}
\def\be{\begin{equation}}
\def\ee{\end{equation}}
\def\beq{\begin{equation}}
\def\eeq{\end{equation}}
\def\bea{\begin{eqnarray}}
\def\non{\nonumber}
\def\cd{\cdot}

\def\eea{\end{eqnarray}}

\def\l{\lambda}
\def\im{{\rm Im~}}

\def\cC{{\cal C}}
\def\lab{\label}

\def\vf{\varphi}

\def\z{\zeta}
\def\e{\epsilon}
\def\o{\omega}
\def\O{\Omega}
\def\vs{\varsigma}
\def\le{\left}
\def\ri{\right}

\def\th{\theta}

\def\m{\mu}
\def\n{\nu}
\def\O{\Omega}
\def\td{\tilde}
\def\d{\delta}
\def\6{\partial}
\def\de{\partial}

\def\a{\alpha}

\def\t{\tau}
\def\d{\delta}

\def\lab{\label}
\def\parl{\parallel}

\def\<{\langle}
\def\>{\rangle}

\def\g{\gamma}

\title{\vskip -1cm The confining trailing string}
\author{Elias Kiritsis$^{1,3,4}$, Liuba Mazzanti$^2$ and Francesco Nitti$^1$ \\
 ~\\
 $^1$
 \href{http://www.apc.univ-paris7.fr}
{APC, Universit\'e Paris 7}, CNRS/IN2P3, CEA/IRFU, Obs. de Paris, Sorbonne Paris Cit\'e, B\^atiment Condorcet, F-75205, Paris Cedex 13, France (UMR du CNRS 7164).\\
 ~\\
 $^2$
 \href{http://testweb.science.uu.nl/ITF/}{Institute for Theoretical Physics and Spinoza Institute},
Utrecht University, 3508 TD Utrecht, The Netherlands.\\
 ~\\
 $^3$ \href{http://wwwth.cern.ch/}{Theory Group, Physics Department, CERN}, CH-1211, Geneva 23, Switzerland\\
~\\
 $^4$
 \href{http://hep.physics.uoc.gr/}
 {Crete Center for Theoretical Physics}, Department of Physics, University of Crete
 71003 Heraklion, Greece
 }

\abstract{We extend the holographic trailing string picture of a heavy quark to the case of a bulk geometry dual to a confining gauge theory. We compute the classical trailing confining string solution for a
static as well as a uniformly moving quark. The trailing string is infinitely extended and approaches a confining horizon, situated at a critical value of the radial coordinate, along one of the space-time
directions, breaking boundary rotational invariance. We compute the equations for the fluctuations around the classical solutions, which are used to obtain boundary force
correlators controlling the Langevin dynamics of the quark. The imaginary part of the correlators has a non-trivial low-frequency limit, which gives rise to a viscous friction coefficient induced by the confining vacuum. The vacuum correlators are used to define finite-temperature dressed Langevin correlators with an appropriate high-frequency behavior.}
\date{}
\preprint{CCTP-2011-29\\CCQCN-2013-7\\CERN-PH-TH/2013-264}
\keywords{\vspace{-1cm} AdS/CFT, Quark-Gluon Plasma, Langevin diffusion, heavy quarks}

\begin{document}

\maketitle


\section{Introduction, summary and discussion}

The $AdS$/CFT correspondence, or holographic gauge/gravity duality, has offered a new way of looking at quantum large-$N$ gauge theories in the strongly-coupled regime, providing a reformulation in terms of degrees of freedom propagating in a higher-dimensional, curved space-time.

One of the most important practical applications of this framework is the connection between the high-temperature deconfined phase of QCD, explored in heavy-ion collision experiments (the Quark-Gluon Plasma, or QGP), and its dual description in terms of a higher-dimensional black hole. In this context, one can map many of the relevant observables on the gauge theory side (like thermodynamic quantities and hydrodynamic transport coefficients), to corresponding gravitational quantities.

Although this mapping is best understood in a supersymmetric and conformally invariant relative of QCD, namely ${\cal N}= 4$ super-Yang-Mills theory, it is believed that at the qualitative level for some phenomena, such differences may not matter a lot in the deconfined phase. However, deviations from conformal invariance are important close to the deconfinement
transition, and non-conformal models have been developed to improve the holographic description, both in the context of critical string theory (e.g. \cite{witten,mn,ks,sakaisugi}) and following a phenomenological, bottom-up
approach like in \cite{cobi3,hw,ihqcd1,ihqcd2,gubt}.

\subsection{AdS/CFT and heavy quark Langevin diffusion}

One important class of observables that can give insight into the properties
of the QGP are related to the out-of-equilibrium production and subsequent diffusion of a heavy quark (Charm or Bottom). The latter undergoes a process of Brownian-like motion which, in the limit when the relaxation time scale is much longer than the thermal correlation time,
 can be described using a Langevin equation with a viscous friction and a stochastic force $\xi(t)$,
\be\label{intro1}
M_q \ddot{X}(t) + \eta\, \dot X(t) = \xi(t) , \qquad \<\xi(t)\xi(t')\> = \kappa \delta(t-t').
\ee
The resulting particle distribution and energies after hadronization can be interpreted in terms of the original particle energy loss (governed by $\eta$) and of the broadening of its transverse momentum in the plasma (governed by $\kappa$), due to the interactions with the medium.

The AdS/CFT duality, already in its simplest form, offers a way to compute the coefficients $\eta$ and $\kappa$ entering the Langevin equation. This has been carried out in the case of the
supersymmetric and conformally invariant ${\cal N}=4$ Super-Yang-Mills theory, in a series of papers \cite{her,gub1,lrw,tea,gubser,Casal,sonteaney,deboer,iancu} in which a heavy quark moving at constant
velocity $\vec{v}$ is mapped holographically into a classical string extending into the bulk AdS-Schwarzschild black hole geometry, and trailing its uniformly moving endpoint, that represents the boundary quark. In this context, the two-dimensional metric induced on the string world-sheet has a horizon $r_s$, which coincides with the black hole horizon $r_h$ in the static case, whereas it is outside of the latter when the quark has a finite speed. Similarly, in the static case the temperature on the world-sheet is the same as the bulk Hawking temperature, whereas in the moving case the effective world-sheet temperature $T_s$ is lower. Similar results are obtained when this picture is generalized to non-conformal backgrounds, as was done in \cite{hoyos,transport,langevin-1}, and to cases where flavor dynamics is included \cite{Magana:2012kh}.

In this context, the Brownian motion of the boundary quark can be seen as arising
from the effect of the thermal Hawking radiation emanating from the world-sheet horizon, and propagating to the boundary along the string, \cite{deboer,sonteaney}. The viscous parameter $\eta$ and diffusion constant $\kappa$ are computed by the low-frequency limit of the boundary correlators associated to the string fluctuations. The latter are effectively two-dimensional bulk fields living on the string world-sheet, and the retarded correlators are computed by the standard prescription giving in-falling boundary conditions at the (world-sheet) horizon.

\subsection{Beyond the local Langevin equation: the need for UV subtraction}

Many of the qualitative differences between ${\cal N}=4$ SYM theory and real-world QCD (such as supersymmetry and conformal invariance) disappear at finite temperature, possibly making a direct comparison of the results cited above with experiments meaningful. However, the breaking of conformal invariance by temperature is soft, which results in a
conformal equation of state of the ${\cal N}=4$ (conformal) plasma. While this may not be a problem at very high temperatures, where the QCD equation of state also approaches conformality due to asymptotic freedom, things are very different even at the qualitative level around the QCD deconfinement phase transition. There, deviations from a conformal equation of state are significant. Moreover, in the ${\cal N}=4$ plasma the coupling is constant, a fact that has no counterpart in real-world QCD, where the coupling constant runs.

For these reasons, the AdS-Schwarzschild trailing string computation was extended in \cite{langevin-1} to a generic, non-conformal black hole
with a non-constant scalar (dual to the Yang-Mills coupling).
The results
showed that the friction and diffusion coefficients acquire an additional momentum dependence compared to the conformal case.

Moreover, the results of \cite{langevin-1} indicated that for high enough momenta and temperatures -- but still within the reach of LHC heavy-ion experiments -- the simple (Markovian) Langevin equation (\ref{intro1}) is not enough to describe the evolution , because the relaxation process happens on time scales comparable with the thermal correlation time. In this case, the local Langevin equation must be replaced by a {\em generalized} Langevin equation, which includes memory effects, and in which the stochastic force has a finite correlation width,
\be \label{intro2}
M_q \ddot{X}(t) + \int dt G(t-t') X(t') = \xi(t) , \qquad \<\xi(t)\xi(t')\> = G_{sym}(t-t').
\ee
Here, $G(t-t')$ and $G_{sym}(t-t')$ are respectively the retarded and symmetrized correlators of the microscopic force coupling the probe quark to the plasma. The form (\ref{intro1}) is then recovered in the long-time limit, and the coefficients $\eta$ and $\kappa$ are the lowest terms in the Fourier-space low-frequency expansions
\be\label{intro2a}
G(\omega) \sim - i\eta \omega + \ldots, \quad G_{sym}(\omega) \sim \kappa + \ldots, \qquad \mbox{as } \omega \to 0.
\ee

The regimes where the local and the generalized Langevin equations hold are discussed in section \ref{finitemass}, where they are also graphically
represented in figure \ref{validityTp}, for the case of the specific confining model we use for the numerical analysis.

The holographic computation gives access to the full correlators appearing in (\ref{intro2}). One of the results of \cite{langevin-1} was that, under the sole assumption of UV $AdS$ asymptotics of the
background, the imaginary part of the retarded correlator behaves at high-frequency as:
\be\label{intro3}
\im G(\omega) \sim \omega^3, \qquad \mbox{as } \omega \to \infty,
\ee
with a temperature-independent coefficient that has been computed explicitly.
The $\o^3$ behaviour can be understood from dimensional analysis and from the restoration of conformal invariance in the UV.

Equation (\ref{intro3}) has some drastic consequences, since with such a behavior, the standard dispersion relations between the real and imaginary part of the retarded correlator break down: these relations require $G$ to fall off at least as $1/\omega$ at high frequency. This also means that, if one tries to use such a correlator in a numerical simulation of the Langevin process, the short-time effects will immediately dominate the dynamics and it will be impossible to reach a stationary state
not too far from the initial state.

Faced with these problems, in \cite{langevin-2} a subtraction procedure was
adopted, to define {\em dressed} correlators with a softer high-frequency behavior. The physical reason is that that the Langevin evolution of the heavy quark are defined by subtracting from the medium the vacuum contributions, and is the analogue of normal ordering in standard QFT.

 From the observation that the leading divergent term is temperature-independent, it was proposed in \cite{langevin-2} that the physical correlator is obtained by subtracting the corresponding {\em vacuum} contribution:
\be\label{intro4}
G^{(sub)}(\omega) = G(\omega) - G^{(vac)}(\omega).
\ee
The subtracted term is the correlator obtained from the trailing string solution in the vacuum geometry with no black hole. A similar subtraction was
adopted in \cite{Hohler:2011wm} in the case of bulk stress-tensor correlators.

The prescription (\ref{intro4}) can be formally obtained by a change of variables in the heavy-quark path integral, in the limit where the quark mass
is much larger than the frequency. For finite quark mass the UV behavior is softer than in (\ref{intro3}), growing only linearly with frequency.
However, as shown in \cite{langevin-2},
a finite quark mass should be considered as a UV cut-off, above which presumably other kinds of energy loss (e.g. radiative) begin to dominate. In
this work we will consider the case where the quark mass is very large, i.e. we use equation (\ref{intro4}) as the definition of the subtracted
correlator even for finite quark mass $M_q$ in the appropriate frequency range
as discussed in sections \ref{physical} and \ref{finitemass}.

The subtraction corresponds to the requirement that a physical quark in the vacuum does not undergo any dissipation. As was shown in \cite{langevin-2}, the dressed correlator defined in (\ref{intro4}) has indeed the correct high-frequency asymptotics to ensure validity of the appropriate sum rules and dispersion relations.

\subsection{The trailing string in a confining vacuum: summary and discussion of results}

In \cite{langevin-2} the computation of the subtracted (vacuum) term was carried out in the case where the vacuum geometry is dual to a non-confining gauge theory.
In this case the computation is straightforward, since the trailing string profile has a trivial (straight) shape. However, it was found that the same profile, when used in a {\em confining}
geometry\footnote{By this, we mean a geometry that satisfies the general confinement criterion expressed in terms of the holographic Wilson loop \cite{cobi}. In a five-dimensional setup, generated by
a scalar field coupled to the Einstein metric, this criterion translates into a requirement on the scalar field potential in the infrared, as was shown in \cite{ihqcd2}.} leads to unphysical results,
in particular in the {\em low frequency} limit of the correlators. The reason, as we will see, is that the straight string solution used in \cite{langevin-2} is {\em not} the correct saddle-point configuration of the classical trailing string in a confining vacuum.

The motivation and starting point of this paper is to resolve this issue: in this work, we will describe the features of the trailing string configuration in a confining bulk background, study the fluctuations around such a classical configuration, and compute the corresponding vacuum boundary propagators. As we shall see, the physics that emerges has interesting features on its own.

Before we proceed to the detailed discussion of this problem, we outline the main results. We assume a metric in the {\em string frame}\footnote{Note that if the dilaton is non-trivial, as it is in YM, this metric is different from the Einstein metric.} of the form
\be\label{intro5}
ds^2 = b(r)^2\left[dr^2 + \eta_{\mu\nu}dx^\mu dx^\nu\right] , \qquad 0<r<+\infty,
\ee
with $r=0$ corresponding to the UV boundary, and $r=+\infty$ corresponding to the IR (which is generically (mildly) singular in a purely five-dimensional theory).

 In this setting, confining backgrounds are characterized by the existence of a minimum of the {\em string frame} scale factor $b(r)$ at a value
$r=r_m$ of the bulk coordinate. This condition ensures that the Wilson loop, computed holographically, exhibits an area law, with a confining string
tension $\sigma_c$ set by the value of the scale factor at the minimum, i.e. $\sigma_c = b^2(r_m)$ in string units. Here, it is important to make the distinction between the string frame scale factor, which can be non-monotonic, and the {\em Einstein frame} scale factor, which is necessarily monotonically decreasing if the bulk theory does not violate the null energy condition. This distinction is non-trivial when the string dilaton is running, as is the case for the five-dimensional models considered here.
By an abuse of language, we call the surface $r=r_m$ the {\em confining horizon}, since for some aspects of the analysis it plays the same role as a world-sheet horizon.

\subsubsection*{{\em Static quark}}

\begin{figure}[h!]
\begin{center}
\includegraphics[width=12cm]{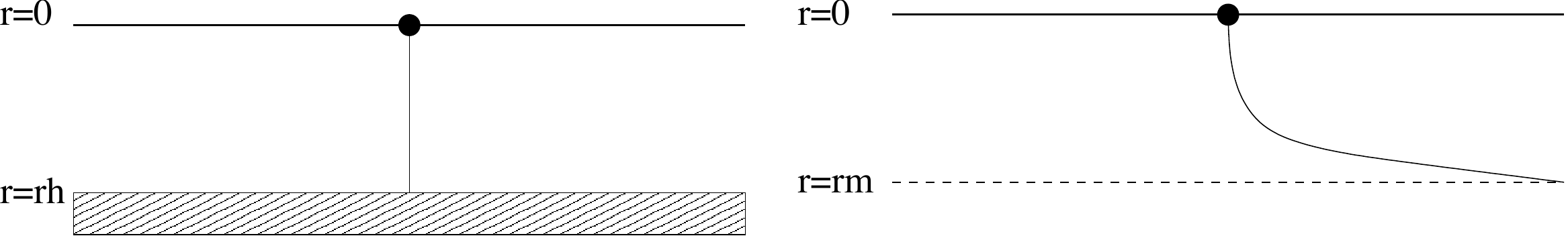}\\
(a)\hspace{5cm}(b)
\caption{The static trailing string configurations in a bulk black hole background (a) and in a bulk confining geometry (b).}\label{introfig}
\end{center}
\end{figure}

We first discuss the case where the boundary probe quark is static. The trailing string in the interior is represented by an embedding of the form
\be
\vec{X}(r,t) = \vec{\xi}(r), \qquad \vec{\xi}(0) = 0,
\ee
and is found by extremizing the Nambu-Goto action. We find there exists a one-parameter family of trailing string solutions, which display a kink in the holographic directions. In this one-parameter set, there is a special solution which does not reach the $r\to \infty$ region, but instead approaches the confining horizon $r_m$ along one of the spatial directions. This is to be contrasted with non-confining geometries, where $b(r)$ is monotonic, and the only allowed string configuration is the trivial one with $\vec{\xi}(r) \equiv 0$ (see figure \ref{introfig}).

\begin{figure}[h!]
\begin{center}
\includegraphics[width=12cm]{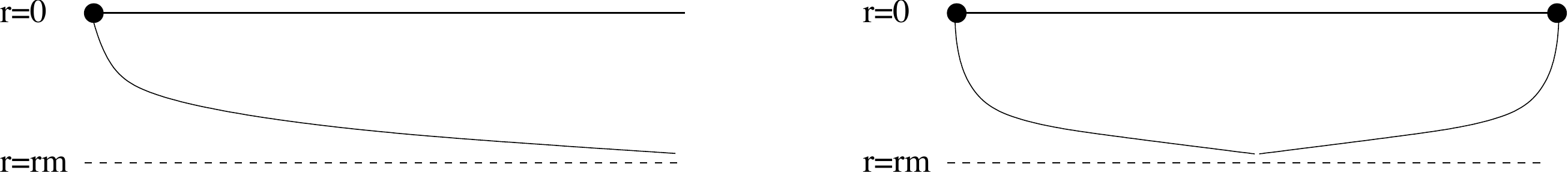}\\
(a)\hspace{5cm}(b)
\caption{The confining trailing string with a single endpoint (a) can be seen as the limit of the Wilson-line configuration connecting a quark and an anti-quark (b)}\label{introfig2}.
\end{center}
\end{figure}

The confining trailing string solution has the following features.
\begin{enumerate}
\item The induced two-dimensional geometry is asymptotically $AdS_2$ in the UV ($r=0$ region) and it becomes asymptotically flat 2-dimensional Minkowski space-time as $r\to r_m$ (See equations (\ref{k6}-\ref{k8})).
The IR asymptotically flat part can be parametrized by time and one of the spatial directions, or equivalently by an exponential change of coordinate $r_m-r \sim e^{-z}$.
In the non-confining case, instead, the string configuration is a straight line that extends from the UV boundary to the IR at $r=\infty$, the worldsheet geometry is still $AdS_2$ in the UV, but there is no confining horizon.
\item From the boundary point of view, the string represents the infinite flux tube which must be attached to a single probe quark in a confining theory. In fact, the energy of the flux tube matches the confining string tension, as it is given by the length of the string measured along the confining horizon $r=r_m$, i.e. $E \sim b^2(r_m) \, L\sim \sigma_c\, L $.
\item The trailing string solution can be understood as {\em half} of the holographic Wilson loop configuration analyzed in \cite{MaldaReyYee,cobi}, where a boundary quark-antiquark pair is connected by a finite length string: in the limit where the anti-quark is sent to infinity along one of the spatial directions, the string approaches indefinitely the confining horizon $r_m$ (see figure \ref{introfig2}). The picture in terms of this unobserved {\em shadow quark} at infinity will play a significant role in what follows.
\item The configuration breaks the boundary $SO(3)$ rotational invariance to the $SO(2)$ around the string direction. This breaking is spontaneous, as the string profile vanishes as we approach the boundary. The way to think about this breaking, and whether one should recover rotational invariance or not (for example by averaging over the possible string directions) will be discussed later.
\end{enumerate}

Boundary correlators governing the dynamics of the quark (and used in the subtraction procedure outlined above) can be found by analysing fluctuations
around the classical string solution,
\be
\vec{X}(t,r) = \vec{\xi}(r) + \delta \vec{X}(t,r).
\ee
In order to compute the correlators we have to chose suitable boundary conditions for the fluctuations in the IR. {\em We argue that the appropriate boundary conditions are infalling at $r_m$, as in
the case of a black hole horizon}. To understand why, consider the two-quark picture of the previous paragraph. We want to study the local retarded response of the medium to a fluctuation in the
position of a single quark at the origin, without any information coming from the ``shadow'' quark at infinity. The condition that the shadow quark does not send any information back along the string
(which is consistent precisely when we consider this quark as being infinitely far, since any signal will take an infinite time to reach it and come back) is encoded in the requirement that the fluctuations travelling along the string are ingoing at spatial infinity (the confining horizon). As for the case of a true black hole horizon, this choice will induce low frequency modes and dissipation effects.

Due to the breaking of isotropy by the background, we have to distinguish between fluctuations that are transverse and longitudinal to the string direction, $\delta X^T$ and $\delta X^L$. Then, the
boundary retarded correlators are computed by the standard holographic prescription,
\be
G^{L,T}(\o) = \lim_{r\to 0} \left[{\cal G}(r)\delta X^{L,R}(-\o,r) \,\de_r \, \delta X^{L,R}(\o,r)\right],
\ee
where ${\cal G}(r)$ is a function of the background metric coefficients only.

Our analysis of the fluctuations can be summarized by following results:
\begin{enumerate}
\item In the {\bf longitudinal sector,} the spectrum is gaped,
\be \label{intro6}
\omega > T_m, \qquad T_m \equiv {1\over 4\pi}\sqrt{2b''(r_m) \over b(r_m)}.
\ee
Correspondingly, the imaginary part of the longitudinal retarded correlator $G^L(\omega)$, identically vanishes at frequency $\omega < T_m$, and there are no long-lived modes in this sector. The explicit form of the correlator is given in equation (\ref{cor8}).
\item In the {\bf transverse sector} on the other hand, there is a continuum of modes starting at $\omega=0$, thus signaling the presence of arbitrarily low frequency excitations\footnote{The transverse
mode spectrum becomes discrete with two normalizable massless modes when the second endpoint is brought to a finite distance, as found in \cite{cobi2}. In the infinitely long string limit the
massless modes become non-normalizable and disappear in a continuum spectrum starting at $\omega=0$.}. Infalling modes at the confining horizon behave exactly as infalling modes in a {\em black-hole} horizon, with the Hawking temperature given by $T_m$ (\ref{intro6}):
\be\label{intro6b}
\delta X^T(t,r) \sim e^{-i\omega t} (r_m-r)^{-i\omega/(4\pi T_m)}.
\ee
The full transverse correlator is given in equation (\ref{cor5}). At low frequencies, one finds a non-vanishing viscous friction coefficient for the boundary quark even at zero temperature, i.e. the quark experiences a drag caused by being immersed in a confining vacuum. At low frequency we have
\be\label{intro7}
 G^T(\omega) \sim - i \sigma_c \omega + \ldots
\ee
therefore the viscous friction coefficient in the transverse sector is equal to the confining string tension (cfr. equation (\ref{intro2a})):
\be\label{intro7a}
\eta^T = \sigma_c.
\ee
\item Despite the similarity of the solution (\ref{intro6b}) with the corresponding black hole infalling modes, the parameter $T_m$ that governs the near-horizon behavior of the transverse fluctuation is {\em not} to be interpreted as a temperature: we show this by an explicit computation of the fluctuation-dissipation relation between the symmetric and the retarded correlators, using the method described in \cite{caron}. We find that the
fluctuation-dissipation relation has the form (equation (\ref{fd23})):
\be\label{intro7b}
G_{sym}(\omega) = - {\rm sign}\,(\omega) \im G(\omega),
\ee
which is the zero-temperature limit of the thermal fluctuation-dissipation relation $G_{sym} = -\coth(\omega/2T)\im G(\omega)$. Therefore, we conclude that the quantum state of the fluctuations is the zero-temperature vacuum. As a consequence, although it displays a non-vanishing friction coefficient $\eta^T$ in the transverse sector given by equation (\ref{intro7a}), the corresponding Langevin diffusion coefficient $\kappa^T$ vanishes, as can be seen by taking the zero-frequency limit of (\ref{intro7b}) and comparing with (\ref{intro2a}).

On the other hand, we can turn on a thermal gas of fluctuations over the confining vacuum, and this will turn on dissipation, as a thermal FD relation at a temperature $T$ now gives:
\be
\kappa^T (T) = 2 T \sigma_c.
\ee
Notice that this is very different from the non-confining case, in which $\im G(\omega) \sim \omega^a$ at low frequency with $a>1$, and even in a thermally excited state $\kappa$ vanishes \cite{langevin-2}.

\item We can recover rotational invariance by averaging over the string direction. This can be done analytically, and the result for the static case is the simple geometric average:
\be\label{intro8}
\< G(\omega)\> = {1\over 3} G^L(\omega) + {2\over 3} G^T(\omega).
\ee

\item The above results were obtained in the case of a five-dimensional vacuum geometry characterized by an infinite range of the conformal coordinate (equation (\ref{intro5})), but can be immediately generalized to different kinds of confining backgrounds where the $r$-coordinate has a finite endpoint $r_{IR}$, and to models in which confinement is driven by some higher-dimensional cycles shrinking to zero size, as in the D4-brane model \cite{witten,sakaisugi} and in the $AdS_6$ soliton geometry \cite{cobi3}.

\end{enumerate}

\subsubsection*{{\em Moving quark}}

In the case where the boundary quark is driven at a constant velocity $\vec{v}$, the salient features of the results described above change little. The trailing string configuration is the same
as in the static case, except that it is dragged along the quark with a constant velocity. One important property of the solutions is that the direction of the string is completely independent of the
direction of the boundary quark velocity. This is unlike the black hole case, where the minimal action solution has the string trailing opposite to the quark velocity. The power dissipated by dragging the quark is velocity- and angle-dependent, and can be computed from the momentum flow along the string (equation (\ref{pi-1})):
\be\label{intro8-i}
{dE\over dt} = \sigma_c {v \cos\theta \over \sqrt{1-v^2\sin^2\theta}}.
\ee
It is straightforward to see that this is in fact the work needed in order to stretch the string in a direction parametrized by $\th$ by an amount $\Delta X= v \Delta t$ in a time $\Delta t$, in the non-relativistic limit where the
right hand side of equation \refeq{intro8-i} is linear in the velocity.

As for the static case, we have computed the string fluctuations and the corresponding boundary correlators. The analysis is complicated by the presence of {\em two} independent spatial directions:
the direction of the quark velocity, and that of the string.
The quark velocity now couples the transverse and longitudinal fluctuations.

Nevertheless, we are able to decouple the equations, and to show that in the diagonal basis the resulting correlators are
essentially a combination of those found for $\vec{v}=0$, with velocity-dependent coefficients and the appropriate Lorentz factors in the correlators. The result for $G(\omega;v,\theta,\varphi)$
depends parametrically on two quantities: the boundary quark velocity $v$ and the relation between the velocity direction and the string direction, parametrized by the angles $(\theta,\varphi)$. The
explicit expression of these correlators can be found in equations (\ref{tdG}) and (\ref{tdGapp}-\ref{Hij}).

\subsubsection*{{\em Consistency and the shadow quark interpretation}}

We have mentioned above that one way to interpret the confining trailing string solution and the fluctuations around it is in terms of an unobserved
``shadow'' quark at infinity. Ingoing boundary conditions at infinity correspond to the requirement that no information is available from the shadow quark.
As it turns out, it is only with this interpretation in mind that our results for finite $v$ can be made consistent with a (generalized) Langevin treatment of the boundary quark. The reason is that the validity of the Langevin treatment imposes a {\em consistency condition} to be obeyed between the classical drag force $\<{\cal F}\>$, computed as the momentum flow along the string, and the linear response encoded in the retarded correlator. As we show, these two quantities must be related by
\be\label{intro9}
{d\< {\cal F}\> \over dv} = -i \left[{d G^\parl(\omega) \over \omega}\right]_{\omega=0},
\ee
where $ G^\parl$ is the correlator between fluctuations parallel to the quark velocity.

It turns out that equation (\ref{intro9}) is violated if we compute both sides with a single string solution at fixed angle, i.e. if we use $G(\omega;v,\theta,\varphi)$ in (\ref{intro9}) (with one
exception in the case of the string aligned with the boundary quark velocity, $\theta=0,\pi$, that we will discuss later). It is also violated if we average uniformly over all possible directions
$(\theta,\varphi)$ of the string. This indicates that the stationary states in which the string direction is fixed, or in which all directions have equal weight, do not provide consistent ensembles
for the response of a single quark, and that we are forgetting to include some other degrees of freedom.

Indeed, we find that the relation (\ref{intro9}) is obeyed if we fully abide to the {\em shadow quark} interpretation of the trailing string: if we assume that the shadow quark has a velocity
$\vec{w}$ independent of $\vec{v}$, and then we average on all the possible $\vec{w}$ with equal weight, the resulting correlator does obey the relation (\ref{intro9}). This corresponds to assuming that we are in complete ignorance about the state of the shadow quark. In contrast, in the computation of the fixed-direction correlators, the hidden assumption is that that both endpoints (the observed quark and the shadow quark) move with the same velocity vector, so that the string angle is fixed. The same holds if we simply average over all directions with equal weight. Computing the propagators in this fashion introduces a hidden correlation between the shadow quark and the observed quark, which makes the interpretation
in terms of the response of a single particle inconsistent.

The resulting correlators obtained by averaging over the shadow quark velocity can be obtained analytically, and take a very simple form in terms of the transverse and longitudinal correlators $G^T(\o)$ and $G^L(\o)$ computed for a static quark. Denoting by $\< G^\parl\>_{w}$ and $\< G^\perp\>_{w}$ the shadow-quark velocity averages of the retarded correlators in the directions parallel and transverse to the {\em observed quark velocity}, we obtain (see equations (\ref{av13a}-\ref{av13b})):
\bea
&& \< G^\parl\>_{w}(\o,v) =
\left[1-{\td{A}(v)\over \g^2}\right]G^L(\o,v) + \td A(v) G^T(\o,v) \label{intro10a}, \\
&& \< G^\perp\>_{w}(\o,v) = \left[1 - {\td{A}(v)\over 2\g^2}\right] G^T(\o,v) + {\td{A}(v)\over 2\g^4} G^L(\o,v) \label{intro10b},
\eea
where $G^{T,L}(\o,v)$ are simple modifications of the longitudinal and transverse correlators for static quarks, \refeq{hatGperp}-\refeq{hatGparl}, and
\be\label{intro11}
\td A(v) = {2\over v^3} \le({\rm arctanh}(v) - v\ri).
\ee
In the static limit $\vec{v}\to 0$, both expressions (\ref{intro10a}-\ref{intro10b}) reduce to the geometric average (\ref{intro8}), as is intuitively expected.

As we have mentioned, one important exception to the above discussion arises in the case where $\theta=0,\pi$ i.e. where the string direction is parallel to the quark velocity. In this case, relation (\ref{intro9}) is verified. This has also a simple interpretation in terms of the shadow quark: it corresponds to a situation where the two quarks are created back to back with opposite velocity and they move apart from each other, becoming infinitely distant in the long-time limit.

Neglecting the backreaction of the interactions on the quark velocities, this corresponds to a situation where
no extra assumption has to be made on the motion of the shadow quark, but rather this is fixed by momentum conservation. This way, no extra correlation (apart from a kinematic constraint) is introduced between the two endpoints, and the single-quark treatment is consistent. Of course, in this case the boundary quark correlators are different from the ones obtained by averaging, and are simply given by $G^{L,T}(v,\theta=\pi)$. Whether to use
these correlators or the averaged ones depends on the physical situation.
For example, the fixed back-to-back direction correlators should presumably be used when one wants to track a heavy quark resulting from a pair-production process in a heavy-ion collision experiment.

Finally, we have numerically computed the finite temperature dressed Langevin correlators in the case of the Improved Holographic QCD model \cite{ihqcdrev}, which was the focus of the analysis of
\cite{langevin-1}. We have defined dressed correlators by subtracting the ($\vec{w}$-averaged) vacuum contribution, and as expected we observe the correct fall-off at large $\omega$. On the other
hand, since the subtracted term has a non-trivial linear behavior at small frequency, this subtraction also renormalizes by a finite amount the transport coefficients computed by the trailing string
in the black hole background, as can be seen in equations (\ref{etaparlrw}-\ref{etaperprw}) and (\ref{etafixbhparl}-\ref{etafixbhperp}).

\subsection{Limits of applicability of the holographic Langevin approach and connection with observation} \label{physical}

To connect this work with heavy-ion collision observations, other issues must be also taken into account. One of the most important is the effect of a finite mass for the quark, which limits the range of applicability of our results,
and in general of the holographic trailing string approach, to a subregion
of parameter space (in terms of temperature, quark momentum and fluctuation frequency). Outside these bounds, which we discuss below, other effects such as flavor backreaction or radiative energy loss start becoming important.

In deriving the results discussed above, we have considered the mass of the quark to be effectively infinite. When a finite quark mass $M_q$ is introduced, the discussion must be modified, and some limitations of the parameter space where our treatment is valid arise. The first two effects described below were discussed in \cite{langevin-1} and \cite{langevin-2}, and concern the
trailing string in the black hole background. The third item in the list is specific to the confining case.
\begin{enumerate}
\item {\bf An upper bound on the quark momentum} \\
We recall that if the quark mass is finite, the endpoint of the string has to be set at a finite, non-zero value of the radial coordinate $r_q \sim 1/M_q$ (which represents the surface where the flavor branes end) rather than $r=0$. Correspondingly, the black hole trailing string solution ceases to exist when the world-sheet horizon $r_s$, which lies outside the black hole horizon for non-zero quark velocity, goes beyond  $r_q$ towards the boundary\footnote{As $r_s$ approaches $r_q$ the string solution develops larger and larger bending at that point. The Nambu-Goto approximation therefore breaks down.}. 
Since $r_s$ decreases with increasing quark velocity, we obtain  an upper bound $p_{max}$ (which depends also on the background temperature) on the quark momentum, above which the trailing string approach breaks down. In the ultra-relativistic limit (which is the regime of momenta of interest for heavy-ion collisions) the bound is schematically
$\sqrt{\g}/T \gtrsim 1/M_q$
where $\gamma$ is the Lorentz contraction factor of the quark. This can then be translated to a bound in momentum
 \be
p< p_{max} \propto M_q^3/T^2.
\ee
Outside this bound, other effects such as radiative energy loss or hard scattering events may dominate over viscous energy loss.

\item {\bf An upper bound on the fluctuation frequency} \\
As shown in \cite{langevin-2}, for finite $M_q$ the holographic computation of the boundary Langevin correlators does not extend to infinite frequency: the boundary on-shell action has
the correct form of a quark kinetic term only under the condition $\omega \g r_q < 1$. This puts a (velocity-dependent) upper bound $\omega_{max}$ on the fluctuation frequency,
\be \label{intro12}
\omega < \omega_{max} \approx M_q/\g
\ee
above which the description of the boundary quark dynamics in terms of a Langevin process breaks down.
\item {\bf A lower bound on the fluctuation frequency} \\
Finally, for the confining trailing string solution a finite quark mass puts an upper bound on the length of the flux tube (thus the proper length of the trailing string),
 $L_{max}\approx 2\gamma M_q / \sigma_c $.
A string longer than $L_{max}$ is unstable to splitting by the production of a new quark-antiquark pair. For a finite quark mass the shadow quark is not infinitely far, but at most a distance $L_{max}$ away.
Therefore, after a time of order $L_{max}$, the signal from the shadow quark will come back, and the infalling boundary condition will be violated. Correspondingly, our treatment holds only for modes of frequency
\be\label{intro13}
\omega > \omega_{min} \approx {\sigma_c \over 2 \g M_q}.
\ee

For a finite but large quark mass, the results we have found indicate dynamics which mimics dissipation on a long but finite intermediate time scale. In order to observe the transition between the dissipative-like behavior and the restoration of the non-dissipative regime, one should compute the response by allowing a signal to start at $t=0$, then propagate to the other end of the string, and then scatter back. In this case we can observe the transition between an ingoing-like solution and a stationary solution of the fluctuation equations setting-in after a time $t\sim L_{max}$.

In the presence light quarks, this bound must be modified because one can form heavy-light mesons by breaking the string. The minimal energy required in this case is not given by the light quark mass, but rather by the chiral condensate $\Lambda_\chi$, which sets the energy for getting a light quark-antiquark pair from the vacuum. Therefore, a more conservative bound is obtained by replacing $2 M_q$ by $\Lambda_\chi$ in equation (\ref{intro13}).
\end{enumerate}
These  bounds depend on the quark velocity as an external parameter; the first bound also depends on the temperature of the plasma.
The detailed form of these bounds will be given in section \ref{finitemass} for the general case, and a numerical analysis is presented in the specific case of the IHQCD model \cite{gkmn3}.
For example, in the specific IHQCD model which we analysed numerically, for the case of an ultra-relativistic charm quark (rest mass $M_q \simeq 1.4 ~ GeV$)
propagating through a plasma at $T\sim 3 T_c$, the frequency and momentum range of applicability of our results are:
\be
6 ~ MeV \lesssim \omega \lesssim 2 ~ GeV, \qquad p \lesssim 270~ GeV,
\ee
The frequency range corresponds to a minimal time interval length $\Delta t \approx 1/\omega_{max} = 0.09 ~ fm$, and to a maximal duration $\Delta t
\approx 1/\omega_{min} = 31 ~ fm$. The latter is much longer than both the relaxation time for the charm at these temperatures, and of the time span
of existence of the QGP fireball.


We leave for future work the computation of the full correlators for finite quark mass, as well as the practical application of these results to realistic situations and comparison with experimental results.

\subsection{Paper structure}

This work is organized as follows. In section \ref{v=0} we describe the general setup, compute the trailing string solution for a static boundary quark, discuss the linearized fluctuation equations
around the classical solution, and obtain the associated boundary retarded correlators.

In section \ref{sec3} we compute the fluctuation-dissipation relation between the
retarded and symmetrized correlator of the trailing string fluctuations, and
show that it corresponds to zero temperature.

In section \ref{BHintro} we discuss the confining trailing string in the case of a uniformly moving quark, and compute the corresponding classical solution and fluctuation equations.

In section \ref{sec5} we compute the boundary retarded correlators
for the uniformly moving quark, introduce different ways of dealing with the angle dependence, and analyze the consistency with the single-quark response theory.

Finally, in section \ref{dressed} we carry out the dressing procedure
of the black hole trailing string correlators in the specific case of Improved Holographic QCD, and we discuss the range of applicability of our results.

Several technical details are presented in the appendices. Appendix \ref{appBH} discusses a generalization to an arbitrary direction of the
trailing string solution in the black hole background.

Appendix \ref{kinkydetails} presents the details of the derivation of the confining trailing string classical solution and fluctuations.

In Appendix \ref{correlators} we give the detailed calculation of the boundary correlators for finite velocity and arbitrary string orientation.

 Appendix \ref{vaverage} contains the details of the averaging procedure over the shadow quark velocity.

In Appendix \ref{antisym} we discuss the effect of a certain antisymmetric term in the quadratic action, and show that it does not contribute to the correlators.

Appendix \ref{Langevin-app} presents a review of the field theory derivation of the Generalized Langevin equation, as well as a proof of the Ward identity that relates the classical drag force and the retarded correlator at zero-frequency.

\section{Static trailing string in confining theories} \label{v=0}

We consider a generic five--dimensional model admitting asymptotically $AdS$ black--hole solutions, where the bulk metric {\em in the string frame} takes the general form
\be\label{metric}
ds^2=b^2(r)\left(-f(r) dt^2+dx_i dx^i+{dr^2 \over f(r)}\right),
\ee
with $b(r) \to \ell/r$ and $f(r)\to 1$ as $r \to 0$ (near boundary region). A concrete realization may be obtained in five dimensions in terms of a class of Einstein-dilaton theories \cite{ihqcd1,ihqcd2,gkmn2}. Among these
theories, the holographic duals of theories that are confining at zero temperature, display a first--order phase transition at a critical temperature $T_c$ between the confining and the non--confining phase. The solutions that minimize the free energy above and below $T_c$ have the following properties :
\begin{itemize}
\item $\mathbf{T<T_c}$\\
In this case $f(r) \equiv 1$. The scale factor $b(r)$ admits an absolute minimum at $r=r_m$, close to which we introduce the following notation:
\be\label{bm}
b(r) = b_m + {1\over 2} b_m'' (r_m - r)^2 + \ldots \qquad r\simeq r_m,
\ee
where $b_m=b(r_m)$ and $b_m''=b''(r_m)$ are two positive constants, \cite{ihqcd1,ihqcd2}. The presence of a minimum in the string frame scale factor is what causes the Wilson loop to exhibit an area law at large separations. At the same time, as it was shown in \cite{gkmn1,gkmn2}, this feature of the vacuum scale factor is
a necessary and sufficient condition to have a first order deconfinement phase transition at $T=T_c$.

To be specific, we take $r$ to be defined in the range $(0,+\infty)$, although this hypothesis can be relaxed: confining backgrounds exist in which (\ref{bm}) is realized and in addition $b(r) \to +\infty$ as a power-law around some finite $r_{IR} > r_m$ \cite{ihqcd2}. All our results immediately generalize to these cases.

\item $\mathbf{T>T_c}$\\
The bulk solution is a black hole: $f(r)$ is a non-trivial function, vanishing at some $r_h>0$, with $b(r_h)$ staying finite:
\be
f(r) \sim 4\pi T (r_h-r) + \ldots, \quad b(r) = b_h + b'_h (r-r_h) + \ldots \qquad r\simeq r_h .
\ee
We assume for simplicity that $b(r)$ is monotonically decreasing in the whole range $0<r<r_h$\footnote{For the ``big''
black hole that dominates the ensemble, this assumption is verified, and although we could not find a general proof this seems always to be the case in all models we have investigated. For the subleading small black hole branch, the scale factor may have a minimum for some $T>T_c$, since in this branch it eventually has to reduce to the zero-temperature scale factor as $r_h \to \infty$. This may have lead to interesting phase transition between
different string kinds of trailing string solutions, but we will not explore this case here.}.
\end{itemize}

\subsection{The static trailing string solution}

Our goal is to analyze the holographic dual description of a probe quark in the boundary theory. We start with the static case, which is technically simpler but already contains all the qualitative and
conceptual features of the general case of a moving quark, which will be discussed in the section \ref{BHintro}. Although technically a single quark in a confining vacuum is forbidden, we can still introduce
it and treat is as a probe, keeping in mind that we will encounter an infinite energy contribution from the flux tube associated to it. This will be reflected nicely in the features of the trailing
string solution in the holographic dual description. In other words, we are considering a probe quark as an infinitely massive object.

The fact that the quark is non dynamical means that its mass is much larger than the temperature and
the confining scale. The gravity dual configuration is a trailing string stretching from the boundary at $r=0$ into the interior \cite{gubser}. The trailing string profile is described in the static gauge by the embedding fields $X^i(r,t)$. It is found by minimizing the Nambu-Goto action, $S_{NG}=-{1\over2\pi\ell_s^2}\int \sqrt{-det (g_{ind})}$, subject to boundary conditions $X^i(r=0,t) =X^i_q(t)$ specified by the motion of the boundary quark. In the static case, this simply means that the endpoint of the string is fixed at the origin, $\vec{X}(r=0,t)=0$.

In the static gauge, with the background metric is specified by (\ref{metric}), the static trailing string profile ansatz is
\be
\label{eq1} X_i = \xi_i(r),
\ee
with the boundary condition $\xi_i(0)=0$.
The induced world-sheet metric and Nambu-Goto action are:
\bea
&&ds^2 = b^2(r)\left[-f(r) dt + \left({1\over f(r)} +|\vec{\xi}'|^2\right)dr^2\right], \label{indmetric} \\
&& \nonumber \\
&& S_{NG} = -{1\over 2\pi \ell_s^2} \int dr dt\, b^2(r)\sqrt{1+ f(r)|\vec{\xi}'|^2},\label{NG}
\eea
where $\ell_s$ is the string length. Since the action only depends on $\xi'_i$, there are three separate integration constants,
\be
C_i \equiv {f(r)b^2(r) \xi_i' \over \sqrt{1+ f(r)|\vec{\xi}'|^2}}.
\ee
The equation for the profile can therefore be reduced to a first order equation:
\be\label{k1}
\xi'_i(r) = {C_i \over \sqrt{f(r)} \sqrt{b^4(r)f(r) -C^2}}, \qquad C \equiv |\vec{C}|.
\ee

In the black hole case, the denominator vanishes at $r_h=0$, and regularity fixes $C = 0$, giving a trivial constant profile $\xi(r) = 0$, showed in
figure \ref{figbhstat} (see Appendix \ref{appBH} for a derivation of this result).

We now concentrate on the $T<T_c$ geometry, and set $f(r) \equiv 1$. Equation (\ref{k1}) becomes:
\be\label{xi'}
\xi_i'(r) = {C_i\over \sqrt{b^4(r) -C^2}}.
\ee
As described above, in the confining case\footnote{In the non-confining case, $b(r)$ is monotonic, and $b(r)\to 0$ in the IR. Then, reality of equation (\ref{xi'}) demands again $C=0$, and the only
solution is the straight string that extends all the way to $r=+\infty$. This case was the one analyzed in \cite{langevin-2}.} $b(r)$ has a global minimum $b_m>0$ at $r=r_m$ and regularity of (\ref{xi'}) allows $C$ to be any constant in the range:
\be\label{k4}
0\leq C \leq b_m^2.
\ee
Moreover, for any non-zero value of $C$ in the allowed range, there is a 2-parameter degeneracy in the choice of the angular direction of the vector $C_i$. Rotational invariance (including boundary conditions) guarantees that the properties of the solution depend only on the modulus $C$ and not on the direction, so we can choose a basis such that the profile takes the form
\be\label{k4-1}
\vec{X}(r) = (\xi(r),0,0) , \qquad \vec{C} = (C,0,0) ,
\ee
and obtain all other solutions by a simple rotation.

We will now discuss the features of the solution as a function of $C$ in the range (\ref{k4}). This will eventually lead us to select one value which minimizes the action. For definiteness, we will assume that, as $r\to \infty$, the scale factor behaves as a power law\footnote{In the confining geometries with a finite range of the $r$-coordinate, the string frame scale factor behaves in the IR $b(r)\sim (r_{IR}-r)^{-\alpha}$ with $\alpha>0$, and a similar discussion to the one presented here applies.}:
 \be\label{k3}
 b(r) \sim r^a , \qquad r\to \infty, \qquad a>0.
 \ee
For all $C$ in (\ref{k4}), strictly less than $b_m^2$, the denominator in (\ref{xi'}) is nowhere vanishing and the string extends to $r \to \infty$, with $\xi' \sim C r^{-2a}$ (this leaves a finite
$\xi(\infty)$ for $a> 1/2$ and note that $a>0$ for confining backgrounds). In the critical case, $C = b_m^2$, on the other hand, using the expansion (\ref{bm}) close to $r_m$ we find,
\be\label{k5}
C= b_m^2 \quad \Rightarrow \quad \xi' \to {1 \over \sqrt{2 b''_m/b_m}(r_m-r)}, \;\; r\to r_m^- .
\ee
Therefore $\xi$ diverges logarithmically at $r_m$: this means that the string becomes
more and more parallel to the $X$ direction and approaches $r_m$ from below, but never goes beyond it.
The solutions corresponding to non-zero $C$ are shown in figure \ref{figkinkystat}. \\

\begin{figure}
\begin{center}
\begin{subfigure}{.5\textwidth}\includegraphics[height=5.3cm]{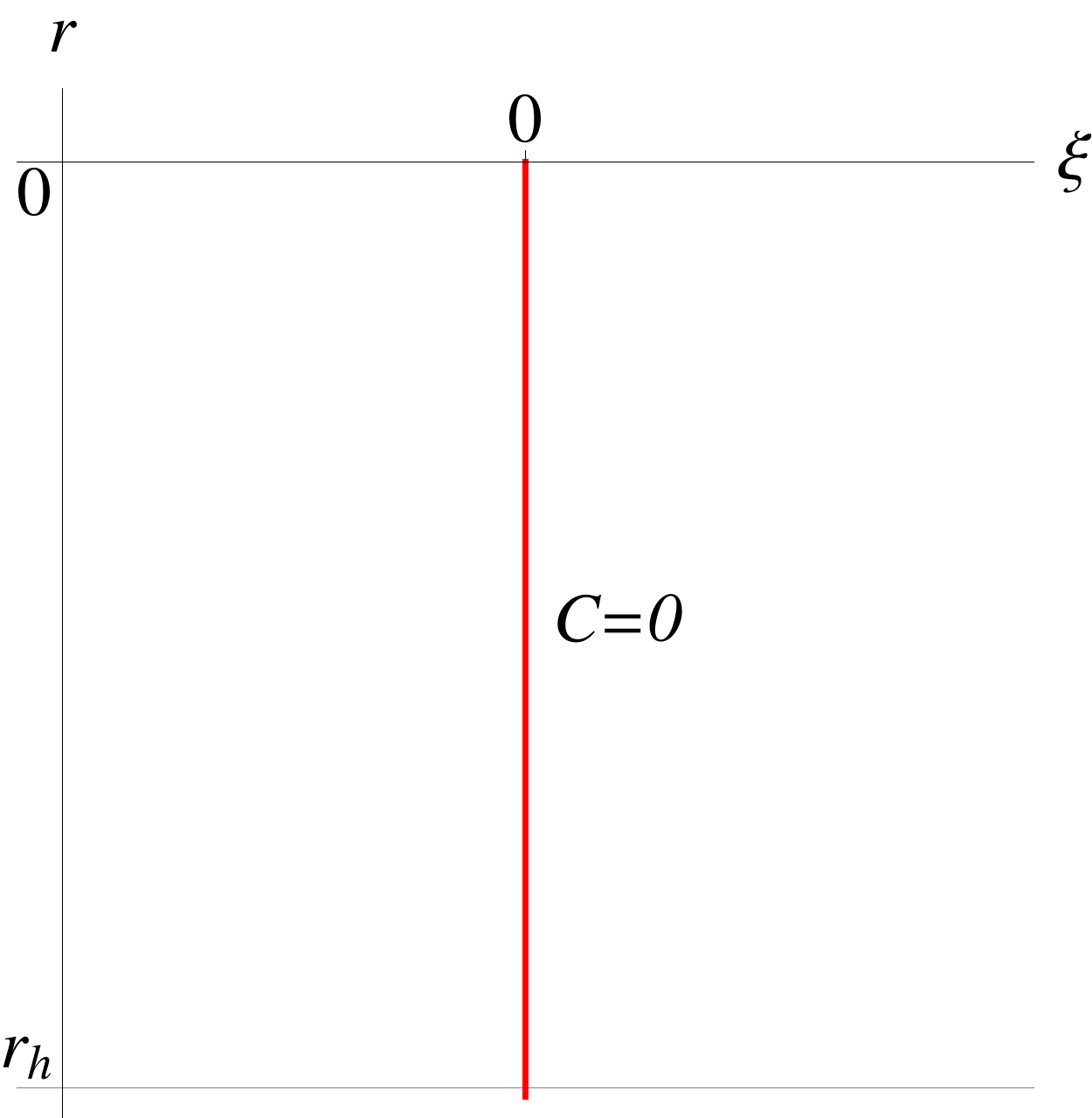}\caption{}\lab{figbhstat}\end{subfigure}~~
\begin{subfigure}{.5\textwidth}\includegraphics[height=5.3cm]{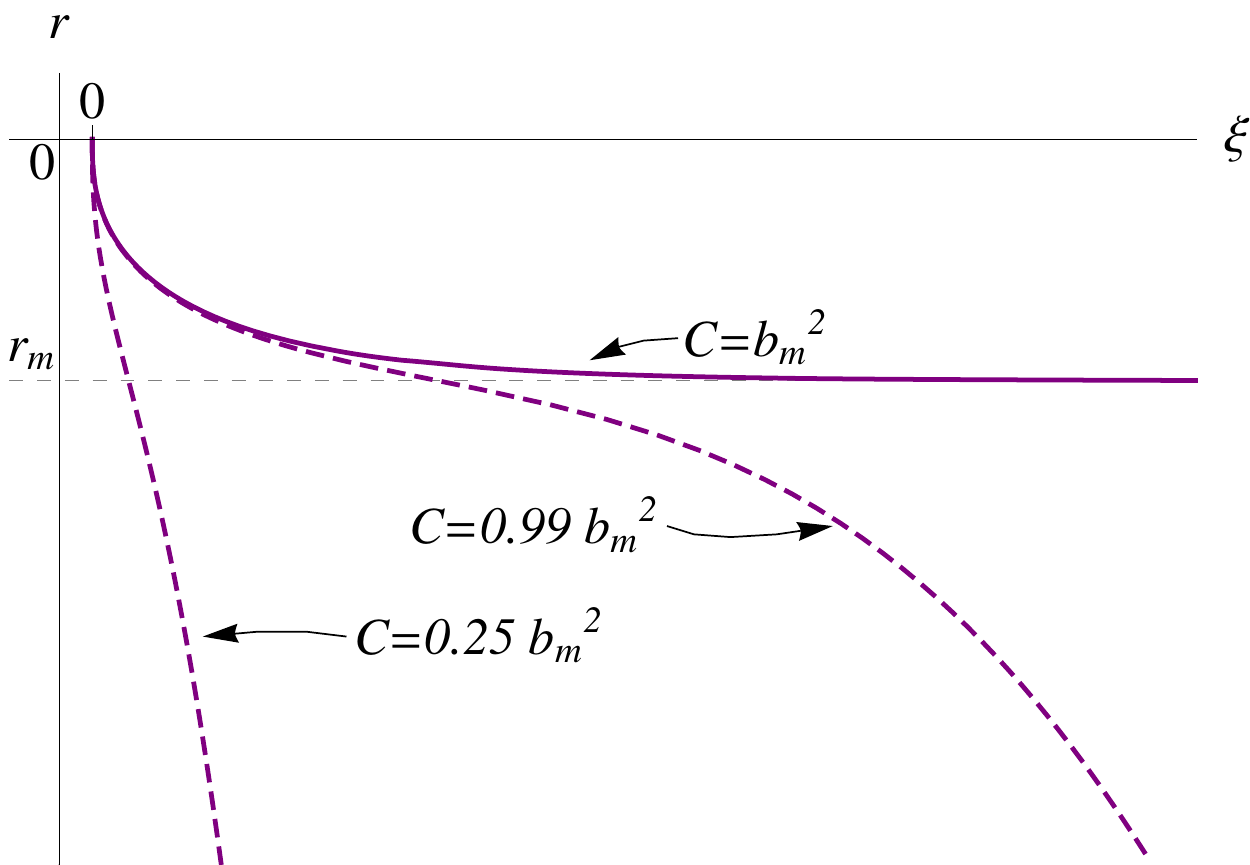}\caption{}\lab{figkinkystat}\end{subfigure}
\caption{\small In figure \protect\ref{figbhstat} we show the profile for a static string $\xi$ in a black hole: a straight line with $C=0$. In figure \protect\ref{figkinkystat} we consider a static string at zero
temperature, where the profile is characterized by the parameter $C$ with $0\leq C \leq b_m^2$. For $0<C<b_m^2$ the string has the shape of a kink and as $C$ approaches zero the string tends to
a straight line. For $C=b_m^2$ the string asymptotes above to $r=r_m$.}
\lab{fig1}
\end{center}
\end{figure}

The induced world-sheet metric (\ref{indmetric}) reads, in the critical case:
\be\label{k6}
ds^2 = b^2(r) \left[- dt^2 + {b^4 dr^2 \over (b^4 - b_m^4)} \right].
\ee
Close to the boundary ($r\to 0$), $b(r)\sim \ell/r$ and (\ref{k6}) is asymptotically $AdS_2$.
 As $r\to r_m$ the metric becomes approximately:
\be\label{k7}
ds^2 \rightarrow b^2_m \left[- dt^2 + {b_m dr^2 \over 2 b''_m (r-r_m)^2} \right].
\ee
Changing variables to
\be
r-r_m = -\exp(-z \sqrt{2b''_m/b_m})\;,
 \ee
 this becomes the flat 2-dimensional metric at $z\to +\infty$
\be\label{k8}
ds^2 \rightarrow b^2_m \left[- dt^2 + dz^2 \right], \quad z \to \infty.
\ee
Indeed, the scalar curvature of this metric vanishes near $r=r_m$:
\be
{\cal R}=-{2\over b^8}\left[(3b_m^4-b^4)b'^2+(b^4-b_m^4)bb''\right]\to -{8(b'')^2\over b_m^4}(r-r_m)^2+\cdots
\ee

The preferred value of $C$ is the one that minimizes the Euclidean Action. Thus, we now evaluate the NG action on these solutions, as a function of $C$. From equation (\ref{NG}), the Euclidean action is:
\be\label{k9}
 S[C] = {1\over 2\pi \ell^2_s} \int d\tau \int_\epsilon^{r_{IR}} dr\, b^2(r) Z(r), \qquad Z(r) \equiv {b^2 \over \sqrt{b^4 -C^2}},
\ee
where $\epsilon$ is the usual UV cutoff and $r_{IR}$ is the endpoint of the trailing string in the $IR$, i.e. $r_{IR} = +\infty$ for $C\neq b_m^2$, and $r_{IR} = r_m$ for $C = b_m^2$.
The action (per unit euclidean  time $\tau$) is always divergent in the IR due to the fact that the string has infinite proper length\footnote{On the other hand, in the black hole case, $r_{IR} = r_h$ and the action per unit time is finite due to the redshift factor along the time direction.}. In fact, we observe immediately from (\ref{k9}) that,
\be\label{k10}
S[C] \sim \left\{\begin{array}{ccc} r^{2a+1} & \quad r\to \infty & \quad C< b_m^2 \\
 z & \quad z \to \infty &\quad C = b_m^2
\end{array}\right.
\ee
In the equation above, we used different coordinates $r$ and $z$ merely for convenience. Shortly everything will be re-expressed in terms of the coordinate-independent proper length $L$.

Both on-shell actions in equation (\ref{k10}) diverge. In order to compare actions at different values of $C$ it is therefore necessary to determine their behavior in terms of the proper length $L$.
We can define $ L = \int \sqrt{g_{rr}} dr $, and again we have two distinct behaviors:
\be\label{k11}
L \sim \left\{\begin{array}{ccc} r^{a+1} & \quad r\to \infty & \quad C< b_m^2 \\
 z & \quad z \to \infty &\quad C = b_m^2
\end{array}\right.
\ee
The proper length is also divergent, but we can now evaluate the string action from (\ref{k10}-\ref{k11}) in terms of $L$ as
\be\label{k12}
S \sim \left\{\begin{array}{cc} L^{2a+1\over a+1} & \quad C< b_m^2 \\
 L &\quad C = b_m^2
\end{array}\right. , \qquad
{S[C=b_m^2] \over S[C]} \sim L^{-{a \over a+1}} \rightarrow 0 \;
\text{as} \; L\to \infty,
\ee
This estimate shows that the solution corresponding to the critical value $C=b_m^2$ is the one that minimizes the Euclidean action, since $a$ is non negative. \\

To summarize, we have found that a single probe quark at rest in a confining vacuum is dual to a infinitely long string, which asymptotically stretches along one of the spatial directions, at a fixed radius $r_m$ in the bulk. We refer to the surface $r=r_m$ as {\em confining horizon}. The energy per unit length, in the asymptotic region, is given by:
\be\label{k12b}
\sigma_c = {b_m^2 \over 2\pi \ell_s^2},
\ee
which is exactly the confining string tension of the boundary theory \cite{ihqcd2}. This has a clear interpretation from the boundary theory perspective: in a confining vacuum, a colored object
such as a quark comes with an infinitely long color flux tube, whose tension is $\sigma_c$, and we have found the holographic realization of this configuration. In fact, this will be confirmed later
by computing the drag force that the string exerts on the quark: we will find in section~\ref{friction} a constant force exactly equal to (\ref{k12b}), i.e. the force is the same as that exerted by a string of constant tension $\sigma_c$ attached to the quark.

\begin{figure}
\begin{center}
\includegraphics[scale=.7]{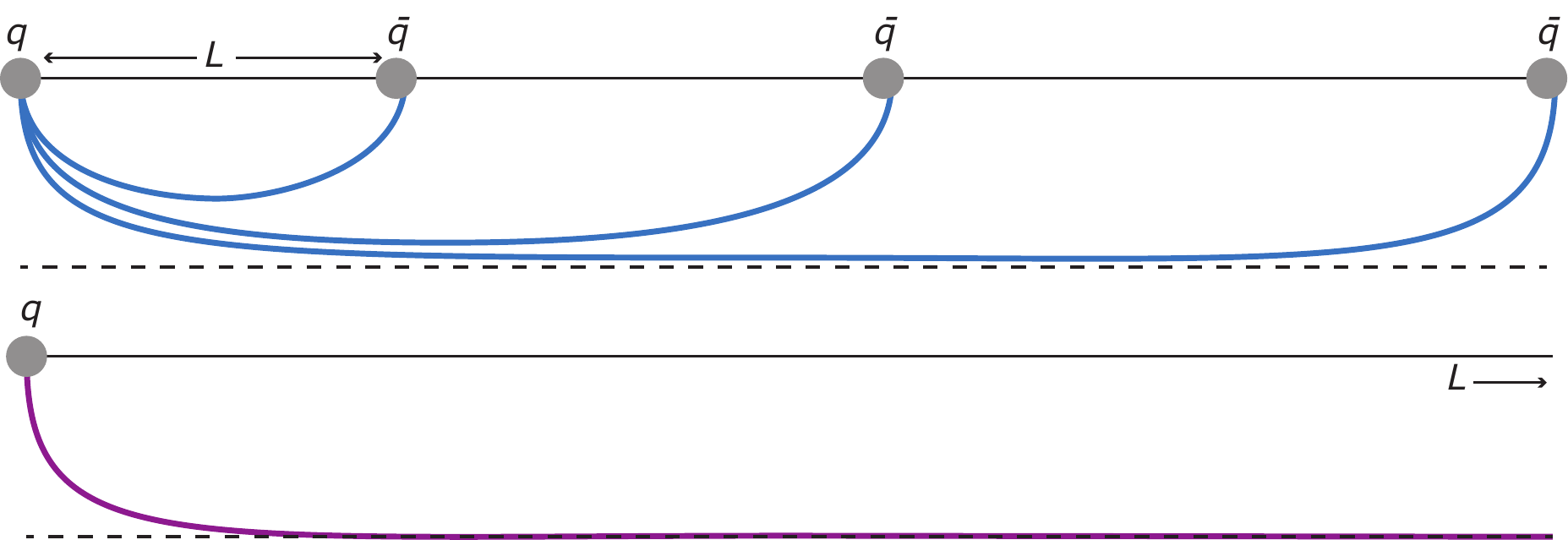}
\end{center}
\caption{\small The string connecting a quark--antiquark pair at zero temperature is illustrated on top, as the distance between the quark and the antiquarks $L$ increases. The bottom figure shows the
$L\to\infty$ limit, where the string has the profile of the confining trailing string with $C=b_m^2$.}
\label{fig2}
\end{figure}

We can think about this configuration in another way, which will be useful later: the stretched string is {\it half} of the flux tube configuration that connects a $q\overline{q}$ pair separated by a
distance $L$ on the boundary, when we send the antiquark at infinite distance away (therefore remaining with a single boundary endpoint), as shown schematically in figure \ref{fig2}. In this sense, the configuration we have found is the $L\to \infty$ limit of the well-known holographic Wilson-line configuration discussed in \cite{MaldaReyYee,cobi}.

The string configuration picks up a direction in the boundary theory (the direction of the flux tube), breaking the full rotational invariance to the $SO(2)$ which leaves the string fixed. This
breaking is spontaneous from the point of view of the boundary theory and is trivially reflected in the bulk theory, since the direction of the confining trailing string is fixed (it does not depend on $r$).

\subsection{Static trailing string fluctuations}

The classical string configuration can be seen as responsible for the classical force acting on the quark. On the other hand, to obtain the correlators governing the boundary Langevin dynamics of the probe quark, as in equation (\ref{intro2}), one must look at fluctuations around the trailing string. These are the holographic dual bulk fields to the boundary operator whose correlations functions enter equation (\ref{intro2}).

 We now look at the fluctuations equations around the confining trailing string found in the previous subsection. We perturb the classical string embedding by writing
\be
X^i = \xi^i(r) + \d X^i(t,r),
\ee
The fluctuation equations can be derived from the expansion of the worldsheet action up to second order in $\d X^i$. The action and the corresponding field equations are given by
\bea
&&S^{(2)} = -{1\over 2\pi \ell_s}\int d^2\sigma \,\,{\cal G}^{ab}_{ij} \de_a \d X^i \de_b \d X^j, \label{S2} \\
&&\de_a\left({\cal G}^{ab}_{ij} \,\de_b\, \d X^j\right)=0, \label{S2-eq}
\eea
where $\sigma^a = (t,r)$.
In the static case we are considering, the kinetic tensor is diagonal, with
\be\label{calG}
{\cal G}_{11} = \left(\begin{array}{cc} - R & 0 \\ 0 & R^3/b^4 \end{array}\right) , \qquad {\cal G}_{22}= {\cal G}_{33} = \left(\begin{array}{cc}- b^4/R & 0\\ 0 & R \end{array}\right),
\ee
where we have defined the function
\be
R(r) = \sqrt{b^4(r) - C^2}.
\ee
From equation (\ref{calG}) we observe that we have different kinetic terms for the longitudinal and transverse fluctuations (with respect to the string direction) and that the field equations are decoupled. We will denote $\d X^1$
and $\d X^{2,3}$ by $\d X^L$ and $d X^T$, respectively. 

Fourier-transforming with respect to time, we have:
\bea
\dt_r\le[ R \, \dt_r \d X^T \ri] + \o^2 {b^4 \over R} \, \d X^T &=& 0, \lab{0v eqperp}\\
\dt_r\le[ {R^{3} \over b^4} \, \dt_r \d X^L \ri] + \o^2 R \, \d X^L &=& 0. \lab{0v eqparl}
\eea
Close to the critical radius $r_m$, the coefficients have the form:
\be\label{nh}
b(r) \simeq b_m, \qquad R(r) \simeq b_m^2 (4\pi T_m) (r_m-r) ,
\ee
where we have introduced for convenience the quantity
\be
\lab{Tm}
T_m \equiv {1\over 4\pi} \sqrt{2b''_m\over b_m}.
\ee
The near-horizon fluctuation equations become
\bea
(\d X^T)'' - {1 \over |r-r_m|}(\d X^T)' + {1 \over |r-r_m|^2} {\o^2 \over (4 \pi T_m )^2 }\d X^T &=& 0, \label{perpeq} \\
(\d X^L)'' - {3\over |r-r_m|}(\d X^L)' + {1 \over |r-r_m|^2} {\o^2 \over (4 \pi T_m )^2 } \d X^L &=& 0. \label{parleq}
\eea

 For the transverse modes
the solutions are ingoing and outgoing waves
\be\lab{kink perp fluc}\d X^T \simeq C_T |r-r_m|^{\pm i\o/(4 \pi T_m)}.\ee

The situation is different for the longitudinal mode: the solution behaves as
\be\lab{kink parl
fluc}\d X^L \simeq C_L |r-r_m|^{-1\pm \sqrt{1-\o^2/(4 \pi T_m)^2}}.\ee Hence, the longitudinal mode is real for $|\o| \leq 4\pi T_m$.

It is instructive to compare equations (\ref{perpeq}-\ref{perpeq}) with those obtained for the trailing string in a black-hole background i.e. when the metric is given by equation (\ref{metric}) with a non-trivial $f(r)$. In that
case, as we have seen, the trailing string has a trivial profile $\xi(r)\equiv 0$, it crosses the bulk horizon at $r=r_h$, and the induced metric is a 2-dimensional black hole with the same temperature $T_h$ as the background:
\be
ds^2_{bh} = b^2(r) \left( - f(r) dt^2 + f(r)^{-1}dr^2\right), \quad f(r_h)=0, \quad f'(r_h) = -4\pi T_h.
\ee
In this case, all fluctuations equations close to the horizon take the form \cite{langevin-1}
\be\label{bheq}
(\d X^i)'' - {1 \over |r-r_h|}(\d X^i)' + {1 \over |r-r_h|^2} {\o^2 \over (4 \pi T_h )^2 }\d X^i.
\ee

Therefore, for the transverse modes close to $r_m$, equation (\ref{perpeq}) looks the same as in the black hole case close to the horizon, with an ``effective''
temperature defined by $T_m$, determined by the features that govern the confinement scale (i.e. $b_m$ and $b_m''$).

Notice however that the world-sheet metric is {\it not} a black hole, since the coefficient of $dt^2$ is nowhere vanishing. Moreover, we have started with a bulk solution at zero temperature. To understand whether $T_m$ can actually be interpreted as a temperature one has to ask what is the ensemble (the density matrix) of the fluctuations. We will see in  Section \ref{sec3} that the fluctuations correlators obey a {\em zero-temperature} fluctuation-dissipation relation, indicating that the modes are actually in the vacuum, and not in a thermal state as is the case for a black hole. This issue cannot be decided
solely by looking at the classical fluctuation equations, since the string induced is conformal to a black hole metric and the field equations are classically conformally invariant. However, at the quantum level the 2D conformal anomaly breaks the degeneracy between conformally equivalent metrics, and as we will see in section \ref{sec3} the fact that the world-sheet induced metric is not that of a black hole will determine the result that the fluctuations are in the $T=0$ vacuum.

Nevertheless, the fact that the near-horizon equation for transverse fluctuations is the same as for a black hole with temperature $T_m$ has some important consequences. Namely, the long wavelength modes show a dissipative behavior in the vacuum: as we will see in subsection (\ref{diffconstapp}), they are subject to a viscous friction which results from the dynamics of confinement.

\subsection{Scrh\"odinger analysis}

Before moving to the computation of the boundary correlators, it is convenient to gain an intuitive picture of the mode solutions and their normalization properties. For this purpose, it is useful to change
variables and coordinates to an analogue Schr\"odinger description, in which normalizability is defined simply by square-summability of the wave-function.

In general, starting from an action for the mode with frequency $\omega$, of the form
\be\label{sch1}
S = -\int dr \, \Big[ A(r) |X'(r)|^2 - B(r) \omega^2 |X(r)|^2 \Big],
\ee
we can change variables as follows:
\be
dz = \sqrt{B(r)\over A(r)}dr, \qquad \Psi = [A(r) B(r)]^{1/4} X.
\ee
Then, the action (\ref{sch1}) takes the form of a canonical Schr\"odinger action with energy $\omega^2$,
\be
S = -\int dz \Big[ |\de_z \Psi|^2 + (V(z) - \omega^2) |\Psi|^2\Big], \qquad V(z) = -(AB)^{1/4} {d^2 \over dz^2} (AB)^{-1/4},
\ee
which leads to the standard equation and normalization condition:
\be
-{d^2\Psi \over dz^2 } + V(z) \Psi = \omega^2 \Psi, \qquad \int dz |\Psi|^2 = 1.
\ee
Once reduced in this form, we can use the quantum mechanical intuition coming from the form of the potential to
have a qualitative understanding of the spectrum.

\begin{figure} \begin{center}
\begin{subfigure}{.5\textwidth}\includegraphics[height=5cm]{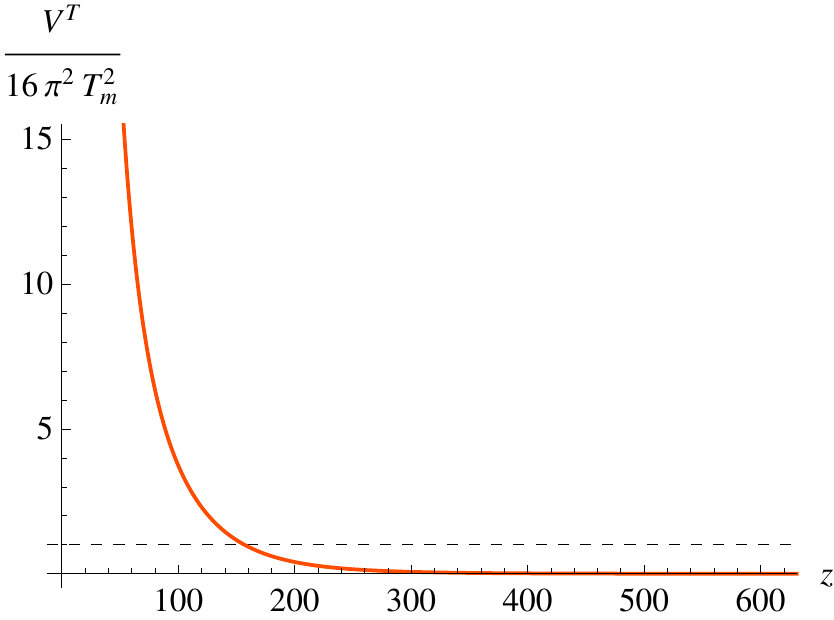}\caption{}\lab{VTfig}\end{subfigure}~~~~
\begin{subfigure}{.5\textwidth}\includegraphics[height=5cm]{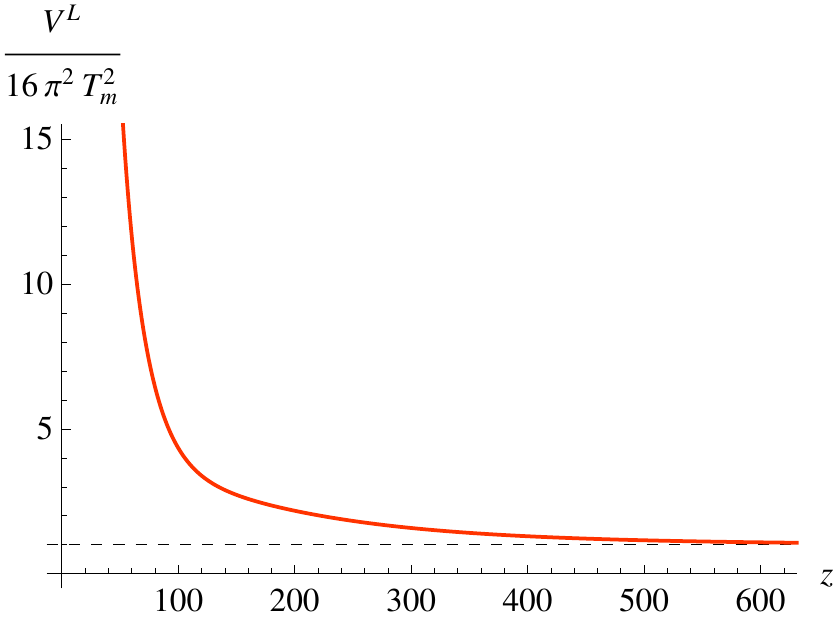}\caption{}\lab{VLfig}\end{subfigure}\\
\caption{The Schr\"odinger potential for the transverse and longitudinal modes, normalized by the factor $(4\pi T_m)^2$ is shown in \protect\ref{VTfig} and \protect\ref{VLfig} respectively, as a function of the
Schr\"odinger coordinate $z$. While for transverse modes the potential exponentially vanishes, it asymptotes to a constant determined by $T_m$, for longitudinal modes.}\lab{potentialfig}
\end{center} \end{figure}

\paragraph{Transverse modes} For the fluctuations corresponding to the $22$ and $33$ components of ${\cal G}$ in (\ref{S2-eq}) we have
$A(r) = R $, $B(r)=b^4/R$. Applying the procedure described above, and the asymptotics (\ref{nh}) we find that the variable $z$ is $z \sim -\log(r_m-r)$ close to $r_m$ and $z\sim r $ in the UV region $r\to
0$. Therefore, the Scrh\"odinger problem is on the half line. The potential is $b^{-1}(d^2 b/dz^2)$. It approximates $1/z^2$ in the UV, and vanishes exponentially as $z\to \infty$.

The spectrum of transverse modes is continuous for $\omega >0$ and at infinity the solutions have oscillatory plane-wave behavior for all frequencies.

\paragraph{Longitudinal modes} For longitudinal fluctuations, we have to look at the $11$ component of ${\cal G}$ in (\ref{S2-eq}). This leads to
$A(r) = R^3/b^4$, $B(r) = R$. The ratio $B/A$ is unchanged, so the coordinate $z$ is the same as for the transverse modes. On the other hand, the potential is now $(R/b)(d^2/dz^2)(b/R)$. It has the
same $1/z^2$ behavior in the UV, but it goes to a constant at infinity,
\be
V(z) \sim (4\pi T_m)^2, \qquad z\to \infty,
\ee
with $T_m$ defined in equation (\ref{Tm}). 
This implies that the spectrum of longitudinal modes is continuous but gaped, and it starts at frequency $\omega_{gap} = 4\pi T_m$. Below this frequency there are no normalizable modes, whereas for
$\omega > \omega_{gap}$ the modes behave as plane waves at infinity.

\vspace{.4cm}
As a concrete example, we show in figure \ref{potentialfig} the Schr\"odinger potentials for transverse and longitudinal fluctuations, computed numerically in the case of the IHQCD model \cite{ihqcd2,gkmn3}. \\

\subsection{Boundary conditions and field theory correlators} \label{diffconstapp}

Having derived the equations governing the fluctuations of the trailing string, we can now compute the corresponding retarded propagators in the boundary field theory in terms of solutions with
appropriate boundary conditions. The boundary propagator is the correlation function of the boundary operator dual to the trailing string fluctuations, i.e. the force operator coupling the probe quark
to the dual field theory \cite{gubser}. This correlator is the same one that enters the Langevin-like equation which can be used to describe the effective dissipative dynamics of the quark after the
field theory is ``integrated out''. More explicitly, the fluctuations of the trailing string compute the Green's functions of the force operators $\cF(t)$ coupling the probe quark to the medium dual
to the bulk theory, (see appendix \ref{Langevin-app} for more details):
\be\label{Gret}
G^{ij}(t) = -i \theta(t) \< 0 |[\cF^i (t), \cF^j (0)]|0\> .
\ee
Here, $G^{ij}(t)$ is the {\em retarded} Green's function, and $\cF$ is the field theory operator coupling to the quark position via the boundary coupling:
\be\label{FdX}
S_{bdr} = \int dt\, \d X^i(t, r=0) \cF^i(t).
\ee

Of particular significance is the imaginary part of the retarded Green's function, which contains no UV divergences (and is therefore scheme-independent) and
from which the diffusion coefficients can be extracted. Explicitly,
the zero-frequency limit of $\im G(\o)$ gives the friction coefficient $\eta$ appearing in the long-time limit of the Langevin equation:
\be\label{cor00}
\im G^{ij}(\omega) \sim - \omega \eta^{ij} + O(\omega^3).
\ee

According to the standard AdS/CFT prescription, given a quadratic bulk action of the form (\ref{S2}),
the boundary retarded correlator in Fourier space is given by
\be\label{cor1}
G^{ij}(\omega) = -{1\over 2\pi \ell_s^2} \left[F^{il}(-\omega,r){\cal G}^{rr}_{lk} \de_r F^{kj}(\omega,r)\right]_{r=0},
\ee
where we have chosen the worldsheet coordinates such that $\cG^{ab}$ is diagonal. In these expressions, $ F^{kj}(\omega,r)$ are the {\em bulk-to-boundary retarded propagators}, i.e. solutions of (\ref{S2-eq}) satisfying unit normalization at $r=0$,
\be\label{cor2}
F^{kj}(\omega,r) = \delta^{kj}
\ee
and, as $r\to r_m$,  {\em infalling} boundary conditions   $F \sim (r_m-r)^{-i(\o/4\pi T_m) t}$, (in the case of complex wave-like solutions),    or {\em normalizable}  boundary condition (in the case of real exponential solutions).

\subsubsection{Infalling modes and the {\em shadow quark}}\lab{shadow}

Before we present the calculation, we comment on the meaning of our choice of infalling boundary conditions at the confining horizon for oscillating modes.
 As we argued before, the configuration we are considering --- a long string stretching from a single boundary quark --- can be thought of as the limit of (the
gravity dual of) a quark-antiquark pair connected by a QCD flux tube, when we send the antiquark an infinite distance away. This is consistent with the case of an external heavy quark of infinite mass,
when the string can be stretched arbitrarily far without breaking. In our case, however, we are interested in the dynamics of a {\em single} quark, and in particular in the response of the medium around the quark to a fluctuation of its position, independently of the antiquark at infinity.

The infalling condition is the natural one in a black hole background, or more generally in the presence of a horizon. However, in our case $r\to r_m$ is not a horizon, but rather the spatial infinity
of an asymptotically-flat region. So what is the physical meaning of this condition in this case? 

As the confining trailing string 
 can be thought, in the boundary theory, as the limit of a color flux  tube 
connecting the quark to an infinitely far  antiquark,
{\em Infalling conditions at $r_m$ correspond to the requirement that no information comes from the antiquark at infinity.} This means that the correlator
we are computing is the one that describes the response of the system to a perturbation of the quark at $X=0$, at times shorter than the time it takes for the perturbation to reach the faraway quark,
and for it to send information back along the string. The choice of infalling boundary condition also means that no information is coming from infinity before the perturbation is turned on, in other
words the correlator is a retarded one.

We are setting up the system so that it is fully specified by a single boundary condition and that we can ignore the response coming from the other quark. This response would manifest itself precisely in a signal that comes back from the ``end'' of the string at infinity. In this situation, choosing ingoing boundary conditions means that we are not allowing the other quark at infinity to react and send a signal back along the string\footnote{For a finite string length this can of course be true only for intermediate time scales, and eventually the fluctuations will become a superposition of incoming and outgoing modes. This will effectively kill all dissipation effects for frequencies smaller than the inverse length of the string, as will be discussed later.}. This interpretation, with the presence of an unobserved {\em shadow quark} at infinity, will turn out to be very important for consistency when we compute the correlators of the fluctuations for a moving quark, in section 5.

 Finally, this argument can also be justified from the fact that the ``retarded'' prescription $\omega \to \omega +i \epsilon$ makes the fluctuation  vanish as $r\to r_m$. In some sense, this boundary condition follows from a regularity requirement.

\subsubsection{Transverse and Longitudinal propagators}

In the static case we are discussing here, the longitudinal and transverse fluctuation equations are decoupled, and the matrix ${\cal G}^{ij}$ (the bulk-to-boundary correlator) is diagonal. This
leads to distinct retarded Green functions in the longitudinal and transverse sectors:
\bea
&&G^T(\o) = -{1\over {2\pi \ell_s^2} }\left[R(r) \de_r {F}^T_0(\omega,r)\right]_{r=0}, \label{GRperp} \\
&&G^L(\o) = -{1\over {2\pi \ell_s^2} }\left[{R^3(r) \over b^4(r)} \de_r {F}^L_0(\omega,r)\right]_{r=0}, \label{GRparl}
\eea
where ${F}^T_0(r,\omega)$ and ${F}^L_0(r,\omega)$ are the solutions of (\ref{0v eqperp}) and (\ref{0v eqparl}) respectively, satisfying the condition (\ref{cor2}) at the boundary and the appropriate boundary conditions at $r_m$.
Below, we analyze separately the fluctuation spectrum for the two kind of modes.

\begin{description}
\item[Transverse modes]

Selecting the infalling mode, the near-horizon form of the solution is:
\be\label{cor3}
{F}^T_0(\omega,r) \sim C_T(\omega) (r_m-r)^{-i\o/(4\pi T_m)}
\ee
where $C_T(\omega)$ is an integration constant which is eventually fixed by the UV normalization.

Since the imaginary part of the quantity in square brakets in (\ref{cor1})  is a conserved flux (see e.g. \cite{langevin-1}), the imaginary part of the retarded Green's function can be obtained given by evaluating equation (\ref{cor3}) at $r_m$, rather than $r=0$, and using
\be \label{cor4}
{1\over 2\pi \ell_s^2} R(r) \,\simeq \, {b_m^2 \over 2\pi \ell_s^2}\,\, 4\pi T_m (r_m-r) , \qquad r\simeq r_m.
\ee
This leads to:
\be\label{cor5}
\im G^T(\omega) = - \o |C_T(\omega)|^2 {b_m^2 \over 2\pi \ell_s^2}.
\ee
\item[Longitudinal modes]

In this case, the solutions in the IR are given by equation (\ref{kink parl fluc}) and one has to separates two regimes:
\begin{itemize}
\item For $\omega > 4\pi T_m$ the solutions are complex, and the infalling boundary conditions select the one of the form
\be\label{cor6}
\hat{F}^L_0(\omega,r) \sim C_L(\omega) (r_m-r)^{-1-i\sqrt{\o^2/(4\pi T_m)^2 -1}}, \qquad r\simeq r_m .
\ee
\item For $0< \o < 4\pi T_m$ on the other hand, the solutions are real and the correct choice is normalizability\footnote{Since we are computing the propagator,  rather than the normal modes, we only impose normalizability at $r_m$, not at $r=0$, so this is not incompatible with the fact that, as stated in the previous section, there are no {\em fully} normalizable modes in this frequency range.} at $r_m$:
 \be\label{cor7}
\hat{F}^L_0(\omega,r) \sim C_L(\omega) (r_m-r)^{-1+\sqrt{1 - \o^2/(4\pi T_m)^2}} , \qquad r\simeq r_m .
\ee

\end{itemize}
Clearly, the retarded Green's function is {\em real} at small frequency,
and the fluctuations along the string direction exhibit a gap equal to $4\pi T_m$:
\be\label{cor8}
\im \, G^L(\omega) = \left\{ \begin{array} {ccc} {\displaystyle - |C_L(\omega)|^2\, \sqrt{{\o^2\over (4\pi T_m)^2}-1}\,  {(4\pi)^{3/2} T_m^3 \over 2\pi \ell_s^2 b_m^{5/2}} },& &\omega > 4\pi T_m, \\
0,
& &\omega < 4\pi T_m. \end{array}\right.
\ee

\end{description}
Below we separately analyze the low-frequency and high-frequency limit of $\im G(\o)$: from the former one can read-off the viscous friction governing the long-time behavior; the latter governs the short-time dynamics and it is relevant for subtracting short-time singularities \cite{langevin-2}.

\subsubsection{Diffusion constants} \lab{diffconst0v}

We define the viscous coefficients from the low-frequency limit of the correlators,
\be\label{cor9a}
G^T \simeq -i\,\eta^T \omega + O(\omega^2), \qquad G^L \simeq -i\,\eta^L \omega + O(\omega^2).
\ee
At small frequencies, $\im G^L $ is identically zero, whereas in the transverse sector the standard procedure of \cite{langevin-1} matches the UV normalization $F^T (\omega,0)=1$ with the
near-horizon behavior (\ref{cor3}) and fixes $C_T(0) =1$. This gives
\be\label{cor9b}
\eta^L = 0 , \qquad \eta^T = {b_m^2 \over {2\pi \ell_s^2}}=\sigma_c.
\ee
While the longitudinal fluctuations are subject to a conservative force, the transverse fluctuations feel a viscous friction force equal to the confining string tension $\sigma_c$.

\subsubsection{High frequency limit}
 At large $\o$, i.e. for $\omega \gg 4\pi T_m$, one can repeat step by step the WKB analysis used in \cite{langevin-1} to solve equations (\ref{0v eqperp}-\ref{0v eqparl}). These equations have the
same form as those studied in \cite{langevin-1} and, as was shown in the aforementioned paper, the high-frequency result only depends on the UV asymptotic of the metric near $r=0$. In particular, if $b(r)\sim \ell/r$ for small $r$, we obtain the universal high-frequency behavior
\be \label{cor10}
\im G^{L,T}(\omega) \simeq - {\ell^2 \over 2\pi \ell_s^2} \o^3 + \ldots ,
\ee
while the IR properties, such as temperature or the confining IR dynamics (in our case, for instance, the terms contributions from $T_m$), give only subleading corrections suppressed by $1/\omega^4$.

\section{Fluctuation-dissipation relations} \label{sec3}

Given the field equations for $\delta X^i$ derived in the previous sections, we would like to understand what kind of density matrix describes the quantum state of the fluctuations. In particular, we would like to understand the role played by the parameter $T_m$ introduced in equation (\ref{Tm}), and whether this
can be truly regarded as an effective temperature. As we will see below, this is not the case: the fluctuations should be considered as in their vacuum state, and not in a thermal ensemble.

One way to clarify this issue is to find the fluctuation-dissipation relation between the retarded and the symmetric correlator in the bulk. This will reveal what is the statistical ensemble that
describes the string fluctuations and as a consequence, the boundary quark fluctuations. Before presenting the explicit calculation, we give an argument based on the
geometric structure of the trailing string solution, which already indicates
that the vacuum state of the fluctuations is not a thermally excited state
from the point of view of a boundary observer.

\subsection{The trailing string geometry}

One way to understand whether the system is in a thermal state, is to pass to Euclidean signature and to check whether there is a regularity condition that forces Euclidean time to be compactified on a circle. For example, this is the case if the induced world-sheet metric is a (Euclidean) 2-dimensional black hole,
\be
ds^2_{bh} = b^2(r)\left[ f(r) d\tau^2 + {dr^2 \over f(r)}\right]
\ee
with $b(r) \to b_h$, $f(r) \to 4\pi T_h (r_h-r)$ as $r \to r_h$. In this metric, the proper distance $d(r_0,r_h)$ from any point $r=r_0$ to the horizon $r=r_h$ is finite,
\be
d(r_0,r_h) = \int_{r_0}^{r_h} dr {b(r) \over \sqrt{f(r)}} < +\infty
\ee
The horizon is, effectively, an {\em end-point} for the string, if we leave the time coordinate running from $-\infty$ to $+\infty$. If we insist on the other hand that the string world-sheet can
only terminate on the UV brane, and has no other boundary in the bulk, we are forced to require that time be compactified on a circle with period $\beta_h \equiv {1 \over T_h}$.

 We repeat this argument in the case of the confining trailing string. Now the 2--dimensional (euclidean) induced metric $g_{ab}$ is (cfr. equation (\ref{k6}))
\be\label{fd1b}
ds^2 = g_{ab}dx^a dx^b = b^2(r)\left[d\tau^2 + {b^4(r) \over R^2(r)}dr^2\right], \qquad R(r) = \sqrt{b^4(r) - b_m^2}.
\ee
Close to the confining horizon $r_m$, $b(r)\simeq b_m$, $R(r)\sim 4\pi T_m b_m^2(r_m-r)$, so that now the proper distance to the confining horizon is
\be
d(r_0,r_m) = \int_{r_0}^{r_m}dr\, {b^2(r)\over R(r)} =\lim_{r\to r_m} \Big(4\pi T_m |\log (r_m-r)| \Big)= +\infty
\ee
The string is infinitely extended, the point $r=r_m$ is not to be considered as an endpoint, and there is no need to compactify time to enforce the condition that the world-sheet has a single boundary. This argument indicates that
the fluctuations can be put in the vacuum, and are not thermally excited\footnote{We thank Jan de Boer for an illuminating discussion about this point.}.

In the next subsection we will see that this argument is confirmed by an explicit Lorentzian calculation of the bulk fluctuations correlators, which shows that the fluctuation-dissipation relation is the one appropriate for
zero temperature.

\subsection{Explicit calculation of the correlators}

In order to find the bulk fluctuation-dissipation relation, we have to separately compute the retarded and symmetrized propagator. The relation between them will reveal which is the density matrix describing the state of the fluctuations. To be precise, we will compute the symmetrized and anti-symmetrized correlators of the operator ${\cal F}$ dual to the string fluctuations,
\be\label{sec3-1}
G_{sym}(t) = {1\over 2}\<\{{\cal F}(t) ,{\cal F}(0)\}\>, \qquad G_{asym}(t) =-i \<[{\cal F}(t) ,{\cal F}(0)]\>,
\ee
the latter being related (in Fourier space) to the retarded correlator $G(\o)$ by:
\be\label{sec3-2}
G_{asym}(\omega) = 2i\im G(\omega).
\ee
It is important that the expectation value of the commutator is independent of the state of the system, and it is determined only by the canonical commutation relations. On the other hand, the anti-commutator is state-dependent. For example, in a thermal ensemble at a temperature $T$, we have the relation
\be\label{sec3-3}
G_{sym}(\o) ={i \over 2} {\rm coth} \le( { \o \over 2T } \ri) G_{asym}(\o).
\ee
One way to determine the state is to separately compute $G_{sym}$ and $G_{asym}$ and read-off the relation between them.

To compute the propagators directly, we use the method described in \cite{caron}, which
does not rely on analytically continuing the fluctuations across horizons. There it is
shown how, {\em once infalling boundary conditions are chosen at the horizon,} the fluctuation-dissipation relation, at late times, is purely determined by the physics in a narrow strip just outside the black-hole horizon.

 The method introduced in \cite{caron} carries over directly to our case: we will impose infalling boundary conditions at $r_m$ (which are the appropriate ones as argued in subsection \ref{shadow}) and, although no horizon is present on the world-sheet, we will see that the fluctuation-dissipation relation is determined by a narrow strip close to the string turning point, $r=r_m$.

We consider the transverse fluctuation and define $X(r,t)\equiv X^T(r,t) $. Its field equation is (see equation~(\ref{0v eqperp}))
\be \label{fd1}
\left[\de_r R \,\de_r - {b^4\over R} \de_t^2\right]X = 0 .
\ee
This equation can be recast in a covariant form with respect to the worldsheet coordinates.
Then, it is easy to check that it is equivalent to
\be \label{transeq}
\left[\de_a b^2 \sqrt{-g}g^{ab} \de_b \right]X = 0,
\ee
where $g_{ab}$ is the induced world-sheet metric (\ref{fd1b}). Equation (\ref{transeq}) can be derived from the following action:
\be\label{transact}
S = -\int d t dr \, b^2 \sqrt{-g} g^{ab}
\de_a X \de_b X.
\ee
This action is covariant with respect to 2-dimensional diffeomorphisms under which the scale factor $b(r)$ transforms as a scalar.

The 2-dimensional induced metric can be put in a conformally flat form by changing coordinates to
\be \label{fd1c}
z = \int dr\, {b^2 \over R}, \qquad ds^2 = b^2(z)\left[-dt^2 + dz^2\right] .
\ee

The conformal factor $b(z)$ interpolates between AdS in the UV and a finite constant in the IR at spatial infinity (see figure \ref{figb0}). It is conformal to half of Minkowski space, as illustrated in figure \ref{fig2d}.

%
\begin{figure}
\begin{center}
\begin{subfigure}{.6\textwidth} \includegraphics[height=5.5cm]{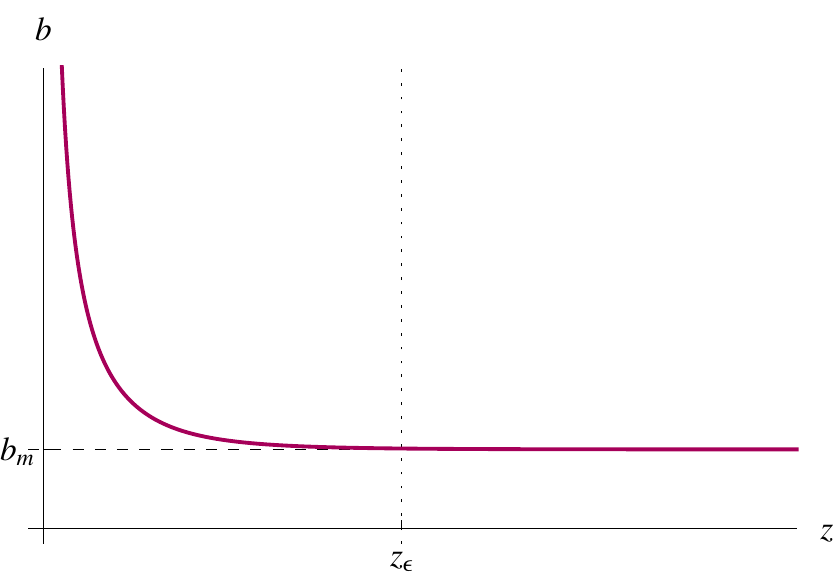} \caption{} \lab{figb0} \end{subfigure}~~
\begin{subfigure}{.4\textwidth} \includegraphics[height=9cm]{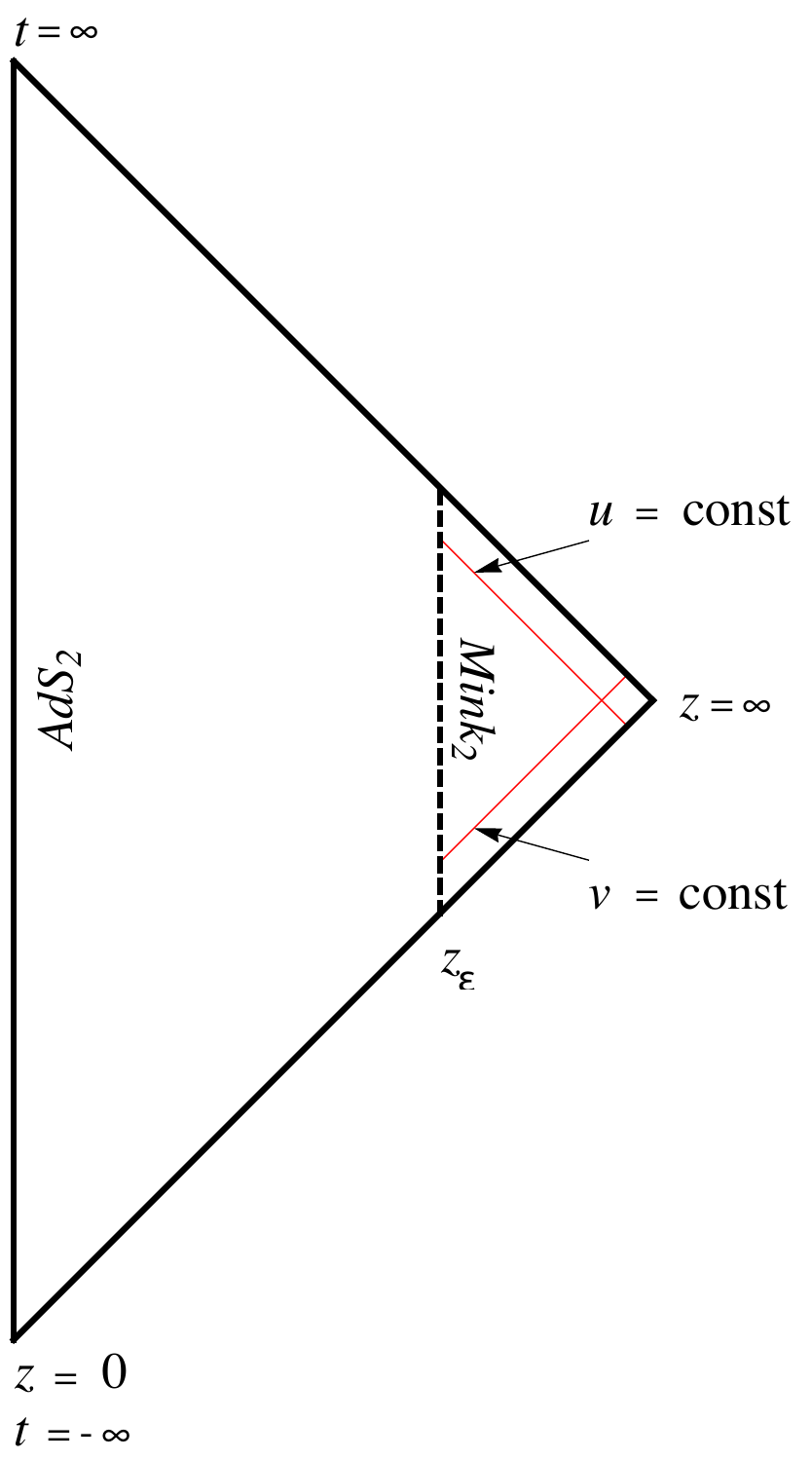} \vspace{-1cm} \caption{} \lab{fig2d} \end{subfigure}
\end{center}
\caption{The 2--dimensional metric (\protect\ref{2dmetricz}) is characterized by a scale factor that asymptotically approaches the constant $b_m$ in the coordinate $z$, as can be seen in figure \protect\ref{figb0}. The Penrose
diagram \protect\ref{fig2d} explicitly shows that the 2--dimensional induced metric in the confining background differs from the $AdS$ black hole that emerges at finite temperature.}\lab{fig2D}\end{figure}

Modes which are incoming from infinity reflect on the boundary and become infalling. For any initial condition made out of a superposition of incoming and outgoing modes, eventually all modes will be
infalling at late times, except those originating at very large $z$. This fact is one of the key points of
the analysis that the authors of \cite{caron} applied to the 2-dimensional black-hole geometry, and most of those considerations apply to our case.

To make this more precise, notice that thanks to the 2-dimensional nature of the problem, equation (\ref{transeq}) and the corresponding action, are invariant under a conformal rescaling of the metric. In particular, they will be the same
for the {\em 2--dimensional black hole} metric
\be \label{fd2}
ds^2_{BH} = (R/b^2) ds^2,
\ee
which has a genuine black hole horizon at $r=r_m$. Explicitly, close to $r_m$ we have
\bea
&& ds^2 \simeq b_m^2\left[-dt^2 + {dr^2 \over (4\pi T_m)^2 (r_m-r)^2}\right],\label{fd3}\\
&& ds^2_{BH} \simeq b_m^2\left[-4\pi T_m (r_m-r) dt^2 + {dr^2 \over 4\pi T_m (r_m-r)}\right] \label{fd4}. 
\eea

This relation allows us to follow the same argument that the authors of \cite{caron} used to determine the bulk fluctuation-dissipation relation
obeyed by the transverse modes. In fact, the null geodesics of the two metrics above are the same, and share the property that {\em only initial data
originating in a space-time region $r_m - \epsilon < r \leq r_m $ are relevant at late times:} the propagation takes a finite time, once it is out of the strip, to arrive at the boundary, reflect back and cross into the strip as an infalling wave. Therefore, {\em if we impose infalling boundary conditions at $r_m$}, so that there are no new outgoing modes originating at finite $t$, the late-time solution will be determined only by the initial data in the strip.

The width of the strip is determined in \cite{caron} by the 2--dimensional black hole temperature $T_h$: for $t\gg 1/T_h $ only initial data in a strip of width $\epsilon \sim e^{-t T_h}$ are important. In our case, the same argument
gives $\epsilon \sim e^{-t T_m}$.

In terms of the asymptotically flat coordinate $z$, where the infalling null geodesics are those of constant $t-z$, the strip becomes the region $z
\gg 1/T_m$ and the 2--dimensional metric reads
\bea
ds^2= b^2(r) \le[ -dt^2 + dz^2 \ri],\lab{2dmetricz}
\eea
with $dz=(b^2/R) dr$.
In the strip, corrections to the Minkowski metric are exponentially small, $\sim e^{-T_m z}$. Inside the strip, the fluctuations behave as free fields in Minkowski space, up to exponentially
small corrections. This is the basic observation that allows to determine the late-time behavior of the correlators, thanks to the equivalence principle.

As shown in \cite{caron}, the bulk fluctuation-dissipation relation is determined by the relation between the {\em stretched horizon correlators}, which are constructed as follows, in terms of the modes in the strip. First, one defines the {\em strip symmetric correlator} $g_{sym}(r_1,t_1|r_2,t_2)$ and {\em strip spectral density} $g_{asym}(r_1,t_1|r_2,t_2)$ . These are solutions of the homogeneous fluctuation equations in the region $r_m-\epsilon<r<r_m$,
\be \label{fd4b}
\left[\de_a b^2(r) \sqrt{-g}g^{ab} \de_b \right]_{1} g(r_1,t_1|r_2,t_2) = 0,
\ee
with reflecting boundary conditions at $r_m-\epsilon$,
\be\label{fd5}
g(r_\epsilon,t_1|r_2,t_2) = g(r_1,t_1|r_\epsilon,t_2) = 0, \qquad r_\epsilon \equiv r_m -\epsilon.
\ee
What will ultimately distinguish between $g_{sym}$ and $g_{asym}$ is their singularity structure on the light-cone.

Next, one constructs the {\em stretched horizon correlators},
\be\label{fd6}
G_h(t_1,t_2) = \left\{\left[b^2(r) \sqrt{-g}g^{rb} \de_b \right]_{1}\left[b^2(r) \sqrt{-g}g^{rb} \de_b \right]_{2} g(r_1,t_1|r_2,t_2)\right\}_{r_1=r_2=r_\epsilon}.
\ee

The symmetrized and anti-symmetrized bulk correlators (the latter coinciding with the imaginary part of the retarded correlator, up to a numerical factor) can be constructed out of these quantities. In particular, the bulk and
boundary fluctuation-dissipation relations are inherited by the corresponding relations among the stretched horizon correlators. Thus,  to determine the fluctuation-dissipation relation on the boundary
theory it is sufficient to compute the horizon correlators (\ref{fd6}).

 We proceed first to the computation of the strip correlators (\ref{fd4b}). It is useful to define light-cone coordinates,
\be \label{fd7}
dv = dt - {b^2 \over R}dr = dt - dz , \qquad du = -dt - {b^2 \over R}dr = -dt - dz,
\ee
in which the metric reads simply
\be \label{fd8}
ds^2 = b^2(r) dudv.
\ee
Inside the strip $r_\epsilon < r < r_m$, these expressions simplify to
\be \label{fd9}
v = t + {1 \over 4\pi T_m} \log (r_m-r), \quad u= - t + {1 \over 4\pi T_m} \log (r_m-r),
\ee
and the flat coordinate is
\be\label{fd10}
z \simeq -{1\over 4\pi T_m}\log(r_m-r).
\ee
The strip extends from $z_\epsilon = (4\pi T_m)^{-1}\log \epsilon$ to infinity (see figure \ref{fig2d} (b)). The metric inside the strip is the flat metric,
\be\label{fd11}
ds^2 \simeq b_m^2 du dv = b_m^2 \left(-dv^2 + 2dv dz\right) , \qquad z_\epsilon \leq z < +\infty,
\ee
up to exponentially small terms, of order $O\left(e^{-4\pi T_m z}\right)$, coming from the expansion of the prefactor around $r_m$, as $b(z) = b_m + (b_m''/2)e^{-4\pi T_m z} + \ldots$ in the large-$z$ region.

Inside the strip, we can consider $b(r)\simeq b_m$ a constant, and equation (\ref{fd4b}) simply becomes:
\be\label{fd12}
\de_{u_1}\de_{v_1} g(v_1,u_1 | v_2,u_2) = 0.
\ee
As $z = -(u+v)/2$, the reflective boundary conditions at $z_\epsilon$ imply that
\be\label{fd13}
g(v_1, -v_1 - 2z_\epsilon|v_2, u_2) = g(v_1, u_1 | v_2, -v_2 - 2z_\epsilon) = 0.
\ee
The solution of (\ref{fd12}) for both $g_{sym}$ and $g_{asym}$ has the same form:
\be\label{fd14}
g(v_1,u_1|v_2,u_2) = f(u_1,u_2) - f(u_1, -v_2-2z_\epsilon) - f(-v_1-2z_\epsilon, u_2) +f(-v_1-2z_\epsilon, -v_2-2z_\epsilon),
\ee
where $f(u_1,u_2)$ is, up to this point, an arbitrary function.

Equation (\ref{fd14}) holds both for the symmetrized and antisymmetrized strip correlators, with appropriate functions $f_{sym}(u_1,u_2)$ and $f_{asym}(u_1,u_2)$. To determine these functions, as in \cite{caron}, we can use free field theory and the equivalence principle in the strip. The symmetrized and anti-symmetrized correlators for a free canonical scalar in Minkowski space are
\bea
&& \< \{\Phi(u_1,v_1),\Phi_2(u_2,v_2)\}\> = -{1\over 4\pi}\log|u_2-u_1||v_2-v_1| \label{fd15},\\
&& \< [\Phi_1(u_1,v_1),\Phi_2(u_2,v_2)]\> = -{1\over 4}\left[{\rm sign}(u_2-u_1) - {\rm sign}(v_2-v_1)\right]. \label{fd16}
\eea
Matching (\ref{fd14}) to the flat space results above fixes the form of the functions $f_{sym}(u_1,u_2)$ and $f_{asym}(u_1,u_2)$:
\be\label{fd17}
f_{sym}(u_1,u_2) = {1 \over 4\pi b_m^2} \log|u_2-u_1|, \quad f_{asym}(u_1,u_2) = {1 \over 4 b_m^2} {\rm sign}(u_1-u_2).
\ee

Next, we can compute the horizon correlators using equations (\ref{fd6}). It is most convenient to do it in terms of the coordinates $(v,z)$. In these variables the flat space-time interval becomes
\be
 g_{ab}\Delta x^a \Delta x^b = \Delta u \Delta v = -(\Delta v)^2 + 2\Delta v\Delta z
\ee
and the strip correlators are given by
\bea
&&g_{sym}(v_1,z_1|v_2,z_2) = {1\over 4\pi b_m^2}\log\left|{ [(v_2 - v_1) - 2(z_2-z_1)](v_2-v_1) \over [(v_2 - v_1) - 2(z_2-z_\epsilon)] [(v_2 - v_1) - (z_\epsilon-z_1)]}\right| , \label{fd18} \\
&& \nonumber \\
&& g_{asym}(v_1,z_1|v_2,z_2) = {1\over 4b_m^2} \left[{\rm sign}(v_2-v_1+ 2(z_2-z_1))- {\rm sign}(v_2-v_1 + 2(z_\epsilon-z_1)) \right. \nonumber \\ && \left.+ {\rm sign}(v_1-v_2 + 2(z_\epsilon-z_2)) - {\rm sign}(v_1-v_2) \right]. \label{fd19}
\eea
In these coordinates, equation (\ref{fd6}) becomes, for both types of horizon correlators,
\be \label{fd20}
G_h(v_1,v_2) = b_m^4\left[(\de_{z_1} + \de_{v_1})(\de_{z_2} + \de_{v_2}) g(v_1,z_1|v_2,z_2) \right]_{z_1=z_2=z_\epsilon} .
\ee
 Inserting expressions (\ref{fd18}) and (\ref{fd19}) in the above equation, we find:
\be\label{fd21}
G_h^{sym}(v_1-v_2) = {b_m^2 \over \pi} \de_{v_1} \de_{v_2} \log|v_2-v_1|, \quad G_h^{asym}(v_1-v_2) = 2 b_m^2 \delta'(v_2-v_1).
\ee
In Fourier space these expression give
\be\label{fd22}
G_h^{sym}(\omega) = b_m^2 |\omega|, \qquad G_h^{asym}(\omega) = - 2 i b_m^2 \omega .
\ee
The Fourier-space relation between the symmetrized and anti-symmetrized propagators is therefore:
\be\label{fd23}
G^{sym}_h(\o) = {i \over 2} \,{\rm sign}(\omega) \, G^{asym}_h(\o),
\ee
{\em i.e. the $T=0$ limit of the thermal fluctuation-dissipation relation (\ref{sec3-3}).}

It is instructive to compare the above calculation to the one in which the world-sheet metric is really a 2-dimensional black hole and to understand how the {\em thermal} fluctuation-dissipation relation appears
in that case. In the strip, the black-hole world-sheet metric and the confining string world-sheet metric are related by a conformal rescaling, (\ref{fd2}), with the conformal factor given by
\be\label{fd24}
{R\over b^2} \simeq 4\pi T_m (r_m-r) = e^{- 4\pi T_m z}.
\ee
Notice that the conformal factor vanishes as we approach the horizon. Since the two-dimensional equation (\ref{fd4}) is unaffected by the rescaling, the computation is unchanged until equation (\ref{fd14}).

The point of departure
is the matching with the free field result: for the black-hole metric this cannot be done in the coordinates $(u,v)$ since these are not locally flat coordinates close to the horizon. In fact the
metric has a non constant scaling factor and takes the form
\be\label{fd25}
ds^2_{BH} = b_m^2 e^{-(u+v) 2\pi T_m} du dv, \qquad u+v \to +\infty.
\ee
To match with equations (\ref{fd15}-\ref{fd16}), in which the space-time interval is flat, we have to go to new coordinates (these are in fact Kruskal coordinates),
\be\label{fd26}
dU = e^{-2\pi T_m \, u }du , \quad dV = e^{-2\pi T_m \, v }dv, \qquad ds^2_{BH} = b_m^2 dU dV.
\ee
The correct matching is done in the $U,V$ coordinates, which are exponentially related to the original $u,v$ coordinates. This only affects $g_{sym}$, for which the function $f_{sym}$ should now be chosen as
\be\label{fd27}
f_{sym} ={1\over 4\pi b_m^2} \log |U_2 - U_1| ={1\over 4\pi b_m^2} \left[\log |1- e^{-2\pi T_m(u_2-u_1)}| - 2\pi T_m u_1 \right].
\ee
On the other hand, $g_{asym}$ is only sensitive to the causal structure, as can be easily seen from equation (\ref{fd16}) and the fact that ${\rm sign}(u_1-u_2) = {\rm sign}(U_1-U_2)$. The
horizon spectral density is unchanged, but the thermal fluctuation-dissipation relation at temperature $T_m$ arises from the exponential in (\ref{fd27}), as shown in the explicit calculation in
\cite{caron} which we will not repeat here:
\bea
G_h^{sym}(\o) ={i \over 2} {\rm coth} \le( { \o \over 2T_m } \ri) G_h^{asym}(\o).
\eea

The argument presented in this section relies on using the induced metric $g_{ab}$ to compute distances inside the strip. On the other hand, as we have pointed out, classical conformal invariance of
the field equation (\ref{transeq}) implies that we could write the action (\ref{transact}) using {\em any} 2-dimensional metric conformally equivalent to the induced metric, but this would
drastically change the final result. Of course, the induced metric is the most natural choice, but this does not resolve the ambiguity. The reason why the induced metric is the {\em correct} metric to use in this computation
can be found in \cite{martinec}, where the authors discuss the quantization of a bosonic string in a black hole background, using the Polyakov action. In that work, it is shown
that one can effectively restrict the path integral to the physical $(d-2)$ {\em transverse} fluctuation modes (which are the only ones we are keeping here) provided that one fixes the gauge on the
world-sheet by identifying the world-sheet metric with the induced metric: in this case, the diffeomorphism ghost determinant and longitudinal modes contributions cancel against each other from the path
integral. If one uses any other conformally related metric, on the other hand, one is forced to keep track of the ghosts and longitudinal modes to obtain the correct results. Since here we are
effectively quantizing the transverse modes only (at the quadratic level the Nambu-Goto and Polyakov form of the action are equivalent, up to a choice of the 2-dimensional metric), consistency requires that we use the {\em induced} metric in equations (\ref{transeq}-\ref{transact}).

Equation (\ref{fd23}) shows that there are no long-time dissipation effects at zero-temperature: in the zero frequency limit we have
\be\label{sec3-4}
G_{asym}(\o) \sim 2i\im G(\omega) \sim -2i \sigma_c\omega \qquad \omega\to 0
\ee
and the Langevin coefficient $\kappa= \lim_{\o\to 0}G_{sym}(\o) = 0 $. However, the fact that $\im G^T(\o)$ vanishes {\em linearly}, means that the Langevin friction coefficient is non-zero. This also means that, as soon as we turn on temperature, i.e. we put by hand the fluctuations in a thermally excited state above the vacuum (the string analog to the thermal $AdS$ gas in the bulk), a non-zero Langevin diffusion constant $\kappa^T(T)$ will arise for the transverse modes, since now the thermal fluctuation-dissipation (\ref{sec3-3}) gives, in the $\omega\to 0$ limit:
\be\label{sec3-5}
\kappa^T(T) = 2T\sigma_c.
\ee

This is very different from what happens in pure $AdS$, and more generally as in the non-confining case: there, as shown in \cite{langevin-2}, at low frequency we always find $\im G(\omega)\sim \omega^{2a+1}$ with $a>0$. Even in a thermal ensemble , inserting this behavior in equation (\ref{fd23}) leads to a vanishing diffusion coefficient $\kappa (T)$.

\section{The trailing confining string for a uniformly moving quark}\lab{BHintro}

We now generalize the discussion to the case where the boundary endpoint of the string moves with a constant velocity $\vec{v}$. The main difference with the previous sections is that isotropy on the boundary is explicitly broken, because the velocity vector picks a preferred direction.

 We first recall briefly what happens in a black-hole bulk geometry. In this case, the trailing string direction is constrained to be parallel to the velocity (see Appendix \ref{appBH} for details), and the string profile  now satisfies the equation:
\be
\xi'(r) = {C\over f(r)}\sqrt{{f(r) - v^2\over b^4(r)f(r) -C^2}}.
\ee
where $C$ is again an integration constant. Now the regularity condition has to be imposed at the point $r_s$, defined by $f(r_s) = v^2$. Since $v<1$, this happens somewhere between the UV boundary ($f(0)=1$) and the horizon ($f(r_h) =0$). Regularity requires $C= v b^2(r_s)$. The induced world-sheet geometry is a
2-dimensional black hole with horizon at $r_s<r_h $ and temperature $T_s$, which in general is smaller than the bulk temperature $T_h$ \cite{langevin-1}.

 We now turn to the case of a confining geometry at $T=0$, and we set again $f(r)=1$.  In this case, as shown in Appendix \ref{Appgen}, the direction of the trailing string does not affect the free energy, and is completely independent of the direction of the velocity. A generic solution will be specified by two  unit vectors, one directed as $\vec{v}$, the other directed as $\vec{\xi}(r)$, plus one integration constant entering the equation for the profile. In the general case, when $\vec{v}$ and $\vec{\xi}$ are not parallel, the solution breaks rotational invariance completely.

We will add one complication at the time, and discuss in the next section the case when $\vec{\xi}(r) \parl \vec{v}$, leaving the general case for section \ref{general}.

\subsection{String parallel to the quark velocity} \label{vparl}

The trailing string solution, specialized to the configuration extending in the direction parallel to the velocity\footnote{This situation is
equivalent to a static trailing string ansatz, where the only preferred direction is the one in which the string bends, namely $X^\parl$.},
is of the form
\be \label{v1} X^\parl = vt + \xi(r), \quad X^\perp = 0,\ee
where $\parl$ and $\perp$ denote directions parallel to the velocity and transverse to the velocity, respectively (while $L$ and $T$ indicate directions longitudinal and transverse with respect to the
direction of the string: in the case of the string parallel to the velocity they coincide).
The Nambu-Goto action is in this case,
\be\label{v2}
S_{NG} = -{1\over 2\pi \ell_s^2} \int dtdr\, b^2 \sqrt{1 - v^2 + \xi^{'2} }.
\ee
From the conservation of worldsheet momentum, the profile $\xi(r)$ satisfies the same equation (\ref{k1}) as in the static case, with $f(r)=1$:
\be\label{v3}
\xi'(r) = {1\over \gamma} {C\over \sqrt{b^4(r) -C^2}}, \qquad \gamma \equiv {1\over \sqrt{1-v^2}}.
\ee
As in the static case, regularity requires $0\leq C\leq b_m^2$, with the action minimized for the critical value $C=b_m^2$.

The force acting on the moving quark can be computed as usual as
the momentum flow along the string. This is given by
\be\label{v4}
\pi_\xi = {\delta S_{NG} \over \delta \xi'} = -{1\over 2\pi \ell_s^2} {b^2 \xi' \over \sqrt{1 - v^2 + \xi^{'2} }} = -{b^2_m \over 2\pi \ell_s^2}.
\ee
We immediately recognize that the right hand side as $\sigma_c$, the confining string tension \cite{ihqcd2}.
The force, i.e. the one-point function of the force operator coupling the quark to the plasma, is then
\be\label{v5}
\< \cF\> = \pi_\xi = -\sigma_c,
\ee
and is directed along the string.
This is {\it not} a frictional force, as it is independent of velocity (in fact, we would have gotten the same result in the $v=0$ case in section \ref{v=0}). It is simply the classical force of a linear confining potential.
This confirms the interpretation of the confining trailing string: a moving quark in a confining vacuum drags along a semi-infinite color flux tube of tension $\sigma_c$ (we can think of it as ending
on an antiquark at infinity, as discussed in subsection \ref{shadow}). The equation of motion is $M\ddot X = \sigma_c$, i.e. the quark feels the constant force due to the string; the energy loss in the time $\Delta t$
is $\Delta E = F v \Delta t = \sigma_c \Delta X$, i.e. the amount of energy needed to extend the flux tube by an amount $\Delta X = v \Delta t$.
There is no dissipative force here, just the coherent effect of color confinement.

The induced world-sheet metric is
\be\label{v6}
ds^2 = b^2(r) \left[- {dt^2 \over \gamma^2} + {b^4 dr^2 \over (b^4 - b_m^4)} \right],
\ee
i.e. the same as (\ref{k6}) except for the time-dilatation factor in $g_{tt}$.

The fluctuation equations look similar to those found in the static case, except for the replacement $\o \to \gamma \o$:

\bea
\dt_r\le[ R \, \dt_r \d X^\perp \ri] + \g^2 \o^2 {b^4 \over R} \, \d X^\perp &=& 0, \lab{0v eqperp-2}\\
\dt_r\le[ {R^{3} \over b^4} \, \dt_r \d X^\parl \ri] + \g^2\o^2 R \, \d X^\parl &=& 0, \lab{0v eqparl-2} \\
 R \equiv \sqrt{b^4(r)-b_m^4}. &&
\eea

\subsection{The general case}\label{general}

We now consider a classical string attached to the boundary and bending in an arbitrary direction, with the end-point moving at velocity $\vec v$ with
components $v^i$:
\be\lab{classic kink}
X^i = v^i t + \xi^i(r), \qquad i=1,2,3.
\ee
Here we present the results, while the details of the computation can be found in Appendix \ref{Appgen}.

\subsubsection{Classical solution}

The induced world-sheet metric and the classical world-sheet action are in this case:

\be
g_{ab} = b^2 \le( \begin{array}{cc} -(1 - |\vec v|^2) & \vec v \cd \vec \xi' \\ \vec v \cd \vec \xi' & 1 + |\vec \xi'|^2 \end{array} \ri), \ee
\be S = - {1\over 2\pi\ell_s^2} \int dtdr\, {b^2} \sqrt{(1 - |\vec v|^2)(1 + |\vec \xi'|^2) + (\vec v \cd \vec \xi')^2}.
\ee
Extremizing this action with respect to the three independent components of $\vec{\xi}(r)$ leads to the same first order equations for each component,
\be\lab{class sol}
\xi'^i = {c^i \over \sqrt{b(r)^4 - C^2}} ,
\ee
where $\vec{c}$ is a {\em constant} vector, whose components are the three integration constants that fix the conserved momenta $\pi^i_\xi$ conjugate to $\xi^i$, and the constant $C$ is given by the equation
\be\label{g1}
C^2 = |\vec c|^2 + \gamma^2 (\vec v \cd \vec c)^2 .
\ee

As usual, regularity imposes $0\leq C \leq b_m^2$. Once $C$ is set by minimizing the action, equation (\ref{g1}) fixes one combination of the three integration constants $c_i$, living two free
parameters that determine the trailing string direction.

Evaluating the action on-shell we obtain:
\be S = - {1\over 2\pi\ell_s^2} \frac1\g \int dtdr\, {b^4 \over \sqrt{b(r)^4 - C^2}}.
\ee
We observe that the action only depends on $C$, and not on the string direction. Following the same reasoning as in section \ref{v=0}, one can show that the Euclidean action is  minimized for $C=b_m^2$ as for the static case.
The two free coefficients out of the three $c^i$ can then be related to the angles that the trailing string forms with the velocity. In particular, if we introduce spherical angular coordinates $(\theta,\varphi)$ with the $x$-axis parallel to $\vec{v}$ such that $\vec{v}=(v,0,0)$, using (\ref{g1}) we easily find:
\be\label{vec-c}
\vec{c} = C \sqrt{1-v^2\over 1-v^2\sin^2\theta} \left(\begin{array}{c}\cos\theta \\ \sin\theta \cos\vf \\ \sin\theta \sin\vf \end{array}\right), \quad C=b_m^2
\ee
The profile of the string in terms of the polar angle is illustrated in figure \ref{fig3}.

\begin{figure}
\begin{center}
\includegraphics[scale=.5]{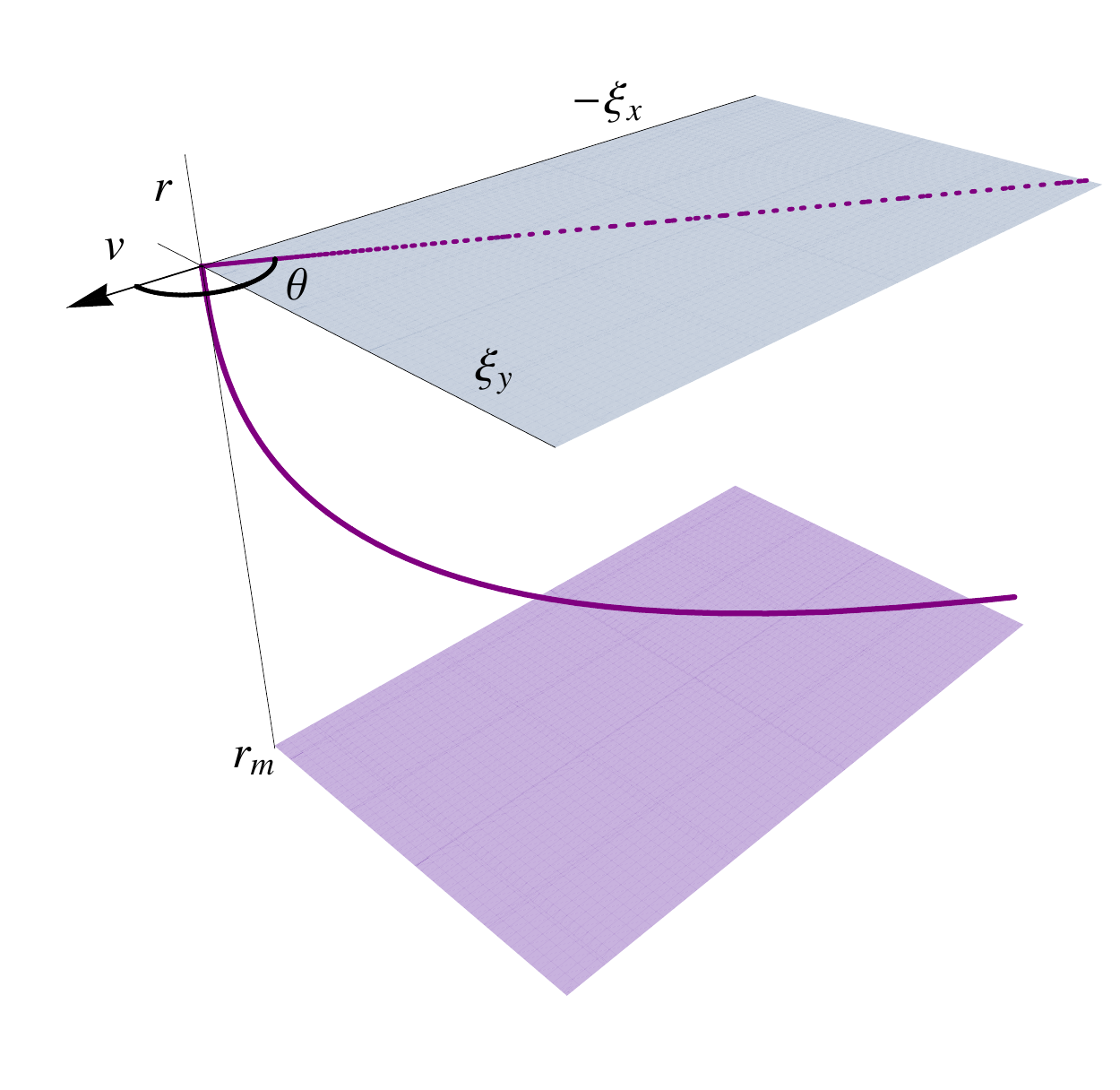}
\end{center}
\caption{\small Here we show the profile of the confining trailing string for $C=b_m^2$. The axes are chosen so that the velocity points in the $x$ direction, and only $x$ and $y$ components of $\xi$ are shown
($\vf=0$).}\lab{fig3}
\end{figure}


As in the previous section we can compute the force acting on the quark, as the momentum flow along the string. A straightforward calculation shows
that the three conserved components of the world-sheet momentum, conjugate to the sting profile, are given by the vector

\be \label{pi}
\vec{\pi}(v,\theta,\vf) = \sigma_c {1\over \sqrt{1-v^2\sin^2\theta}} \left(\begin{array}{c}\cos\theta \\ (1-v^2)\sin\theta \cos\vf \\ (1-v^2)\sin\theta
\sin\vf \end{array}\right).
\ee
Notice that the classical force exerted by the string is velocity and angle-dependent, and in general it is {\em not} directed along the string (whose direction is instead the one of $\vec{c}$, eq
(\ref{vec-c})), except for $\theta=0,\pi$, and/or in the non-relativistic limit $v\ll 1$.
The power dissipated to keep the quark in uniform motion (i.e. the rate of energy loss of the quark to the medium) is
\be\label{pi-1}
\vec{\pi} \cdot \vec{v} = \sigma_c {v\cos\theta \over \sqrt{1-v^2\sin^2\theta}} .
\ee
For small velocities, this coincides with the work against the confining force needed in the time $\Delta t$ to stretch the string by an amount $\Delta X= v \Delta t$.

\subsubsection{Fluctuations}

We can now consider fluctuations in the ansatz \refeq{classic kink}, by writing
\be
X^i = v^i t + \xi^i(r) + \delta X^i(r,t).
\ee
To second order in the fluctuations, the action reads:
\bea\lab{S2b}
S = - {1\over 2\pi\ell_s^2} \int dtdr\, \frac12 \, {\cal G}^{ab}_{ij} \dt_a \d X^i \dt_b \d X^j,
\eea
where the ${\cal G}^{ab}_{ij}$ are obtained by expanding the world-sheet action:
\be \label{calG2}
{\cal G}^{ab}_{ij} = {b^4 \over \sqrt{-\det g_{ab}}} \le[ b^2 g^{ab} \Xi_{ij} - 2 \e^{ab} \a_{ij} \ri].
\ee
Here $\Xi_{ij}$ and $\a_{ij}$ are respectively symmetric and antisymmetric, and their explicit form is given in Appendix (\ref{App-fluctua}).

The fluctuation equations are
\be \lab{generalfluct}
\dt_a \le[ {\cal G}^{ab}_{ij} \dt_b \d X^j \ri] = 0.
\ee
These are coupled equations for the fluctuations $\d X^i$. As shown in Appendix \ref{App-fluctua}, it is possible to introduce new variables that make the system diagonal: first we decompose $\vec{\d X}$ along $\vec{v}$, $\vec{c}$ and the orthogonal direction $\vec{n}$,
\be \d \vec{X} = \d X_v \vec v + \d X_\xi \vec c + \d X_n \vec n,\ee
then we define the linear combinations
\be \label{diagonal}
\d X^L = \d X_\xi + {\vec{v}\cdot \vec{c}\over (1-v^2) C^2 } \, \d X_v, \qquad \d X^T_1 = \d X_n, \qquad d X^T_2 = \d X_v.
\ee
Remarkably, these new variables satisfy the same decoupled equations, (\ref{0v eqperp-2}-\ref{0v eqparl-2}), as those for the longitudinal and transverse modes in the case $\vec{v}$ parallel to $\vec{c}$:
\bea
\dt_r \le[ R \, \dt_r \d X^T \ri] + \g^2 \o^2 {b^4 \over R} \, \d X^T &=& 0, \lab{eqn}\\
\dt_r \le[ {R^{3} \over b^4} \, \dt_r \d X^L \ri] + \g^2 \o^2 R \, \d X^L &=& 0, \lab{eqxi}
\eea
In this case the labels $L,T$ do not refer to the direction of the fluctuations, but rather to the fact that they are governed by the same equations as the fluctuations $L$ and $T$ to the string in
the static case (up to factors of $\g$). Notice however that $\d X_n$ is still transverse to both the velocity and the string direction.

The analysis of (\ref{eqn}) and (\ref{eqxi}) is analogous to the one we carried out in detail in section 2: in particular, the fluctuation $\d X^L$ has real solution for $\g \o < 4\pi T_m$, which leads to a gap in the imaginary part of the retarded Green's function in this direction. On the other hand, the fluctuations $\d X_n$ and $\d X_v$ have a spectrum starting at $\o=0$, and by the same discussion as in the previous section, the corresponding retarded correlator has a non-trivial zero-frequency limit.

\section{Boundary correlators for quark in uniform motion}\label{sec5}

From the solutions of the fluctuation equations with appropriate boundary conditions we can obtain, as was done in the static case, the boundary retarded correlators $G^{ij}$. These will depend on
the direction of the trailing string, as well as on the magnitude and direction of the velocity.
Without loss of generality, we can fix the velocity to be along the $x$-direction, and introduce the polar and azimuthal angles $(\theta,\vf)$ of the trailing string with respect to $\vec{v}$. Our goal is to compute the quantity (see equation (\ref{cor1}))
\be\label{ang1}
G^{ij}(\omega;v;\theta,\vf) = -\left[ F^{ik} (-\omega,r; v,\theta,\vf){\cal G}_{kl}^{rb}(r,v,\theta, \vf)\, {\de_b}\, F^{lj} (\omega,r; v,\theta,\vf)\right]_{r=0},
\ee
where $F^{ij}(r,\omega;v,\theta,\vf)$ are the bulk-to-boundary propagators with
infalling (or normalizable) conditions at $r_m$.

The fact that the fluctuation equations decouple in a suitable basis, and most of all the fact that the corresponding decoupled equations are completely independent of $(\theta,\vf)$, allows us to
write a closed expression for the correlator in terms of two simpler quantities,
\be\label{ang2}
{F}^T(r,\omega;v) \quad \mbox{and} \quad {F}^L(r,\omega;v) ,
\ee
which are the bulk-to-boundary propagators associated to equations (\ref{eqn},\ref{eqxi}). Moreover, they are the same quantities that we used to compute the correlator
in the static case ($v=0$) of section \ref{v=0}, and for the case $\theta=0,\pi$ which we discussed of section \ref{vparl}.
In fact, since $v$ enters in the field equations only through the combination $\gamma \omega$, we have
\be\label{ang3}
{F}^{T,L}(r,\omega;v) = {F}^{T,L}(r,\g \omega;0) \equiv {F}^{T,L}_0(r,\g \omega).
\ee
The functions on the right-hand-side of the last equation are the bulk-to-boundary propagators corresponding to the static quark, obtained in section 2.

 In particular, in the special case of section 3.1, $\theta=0,\pi$ (string parallel to the quark velocity) we recover the results
\be\label{ang4}
 F^{ij} (\omega,r; v,\theta=0,\vf=0) = {v^iv^j \over v^2} {F}^L_0(r,\g\omega) + \left(\delta^{ij} - {v^iv^j \over v^2}\right) {F}^T_0(r,\g\omega).
\ee

In order to obtain the expression for (\ref{ang1}) in terms of the basic propagators ${F}^{T,L}$, we first apply a linear transformation to pass to the coordinates (\ref{diagonal}), which
diagonalize ${\cal G}$, then we write the solutions in terms of the (angle-independent) functions (\ref{ang3}), then we rotate back to the original coordinates. The procedure is explained in detail in
Appendix \ref{correlators}. The crucial observation is that the linear transformation that diagonalizes the field equations is $r$-independent and therefore it commutes with the radial derivative, leading to a result that can be written in closed form:
\be G^{ij}(\omega;v;\theta,\vf) = {1 \over 1-v^2 \sin^2\th} \left( H^{ij}_T(v,\theta,\vf) G^T(\o;v) + H^{ij}_L(v,\theta,\vf) G^L(\o;v) \right) , \lab{tdG} \ee
with $H_{T,L}$ defined as two symmetric $\omega$-independent matrices whose components are given explicitly in Appendix \ref{correlators}, equation (\ref{Hij}). The correlators $G^{T,L}(\omega,v)$ are the
basic boundary Green functions obtained for $\theta=0,\pi$, i.e. from ${F}^{T,L}(\omega,r;v)$, as in equation (\ref{GRperp},\ref{GRparl}) :
\bea G^T(\o;v) &=& - {1\over2\pi\ell_s^2} \left[{R \over \g} F^T(r,-\o;v) \dt_r F^T(r,\o;v) \right]_{r=0},
\lab{hatGperp}\\
G^L (\o;v) &=& - {1\over2\pi\ell_s^2} \left[{\g R^3 \over b^4} F^L(r,-\o;v) \dt_r F^L(r,\o;v) \right]_{r=0}.
\lab{hatGparl} \eea

What is remarkable about equation (\ref{tdG}) is that, although the trailing string solution and fluctuation equations depend non-trivially on the relative direction between the string and the quark velocity, this angular dependence factorizes from the dependence on $\o$ in a quite simple way.

We note that the $\th$- and $\vf$-dependent correlator \refeq{tdG} reduces to $G^{11} = G^L $, $G^{22} = G^{33} = G^T $, (all other components vanish) in the special case $\th=0$ as it is expected.
%

\subsection{Averaging over string orientations}

The Green's function (\ref{tdG}) that we obtained in closed form, although it corresponds to the general solution for the trailing string in the gravity dual, cannot be the end of the story from
the boundary field theory point of view. On the gauge theory side, the only boundary data are the position and velocity of the quark, and there is no place for a dependence on extra angular variables
$\theta,\vf$. If there is no physical principle (other than extremizing the action) which selects a specific direction for the string, then all solutions are physically allowed and, in the spirit
of the $AdS$/CFT correspondence, we should sum their contribution in the path integral. Since all configurations corresponding to different $(\theta,\vf)$ have the same action, this amounts to an
average over all possible string directions, with equal weight.

As we shall see shortly, there is more than one way of performing the average: we will present three distinct possibilities, which give different final answers. Which one is the correct one depends on
the precise definition of the physical problem. However, as we shall see at the end of this section, there is one guiding principle --- the consistency between classical and one-loop result --- that
can eliminate some of the possibilities as unphysical.

No matter what the chosen averaging procedure is, the general way to arrive at angle-independent correlators is given as follows. In AdS/CFT, the boundary theory partition function for the particle
fluctuations is obtained by summing over all possible semi-classical saddle points of the bulk theory (in this case, over all possible string orientations) with some measure $d\mu(\theta, \vf)$ to be
specified: schematically (i.e. ignoring for the moment that we should write the path integrals and effective actions in a double-field formalism),
\be\label{av1}
\exp \{iS_{eff}[\delta \vec{X}_o(\o)]\} = \int d\mu [\theta,\vf] \, \exp \{ i S_{(\theta,\vf)}[\delta \vec{X}] \},
\ee
where $d\mu[\theta,\vf]$ is an appropriate measure, and $S_{(\theta,\vf)}[\delta\vec{X}_o(\o)]$ is the on-shell action for the fluctuation $\delta
\vec X(r,\omega)$ around the string solution specified by the angles $(\theta,\vf)$, subject to the boundary condition
\be\label{av2}
\delta \vec{X}(\omega,r) \to \delta \vec{X}_o(\o) , \quad r\to 0.
\ee
At quadratic order in the fluctuations, the on-shell action is
\be\label{av3}
S_{(\theta,\vf)} = {1\over 2}\int d\omega \, \delta X_o^{i}(-\o) G^{ij}(\o;v,\theta,\vf) \delta X^{j}_o(\o),
\ee
where the propagator is computed holographically by equation (\ref{ang1}).
Since we are working at quadratic order, we may expand the exponential on the right-hand side of (\ref{av1}), perform the average on $G^{ij}(\o,\theta,\vf)$, and re-exponentiate:
\bea Z &\simeq& \int d\mu[\th,\vf] \le\{ 1 + i \int d\o \, G_{ij}(\o;\th,\vf) \, \d X^i_o(-\o) \d X^j_o(\o) \ri\} \\
&=& 1 + i \int d\o \, \<G_{ij}\>_\m(\o) \, \d X^i_o(-\o) \d X^j_o(\o)
\simeq e^{ i \int d\o \, \<G_{ij}\>_\m(\o) \, \d X^i_o(-\o) \d X^j_o(\o) }. \non\eea
obtaining an isotropic boundary action:
\be\label{av4}
S_{eff}[\d \vec{X}(\o)] = {1\over 2} \int d\omega \, \delta X^{i}_o(-\o) \< G^{ij}\>_{\mu} (\o,v) \d X^j_o(\o),
\ee
where
\be\label{av5}
 \< G^{ij}\>_{\mu} (\o,v) = \int d\mu[\theta,\vf] \, G^{ij}(\o;v,\theta,\vf).
\ee
To the order at which we are working, summing over possible string orientations is equivalent to averaging the retarded correlators directly.

Below we present three possible averaging procedures, i.e. distinct choices of the measure $\mu[\th, \vf]$ corresponding to different physical interpretations of the process.

\subsubsection{$\Omega$-averaged correlators}\label{Oav}

The most natural choice is to simply average over all orientations with equal weight,
\be \label{av6}
d \mu[\theta,\vf] = {d\Omega \over 4\pi}.
\ee
This means assuming that all directions of the trailing sting are equally likely, and they are uncorrelated with the direction of the quark velocity. We will denote the average by $\< \quad \>_{\Omega}$.

Thanks to the factorization of the angular variables and frequency displayed in equation (\ref{tdG}), the solid angle integral can be performed analytically. The result is an averaged correlator which does not couple transverse and longitudinal (to the velocity) modes:
\be\label{av7}
\< G^{ij}\>_{\Omega} (\o,v) = {v^iv^j \over v^2} \< G^\parl\>_{\Omega}(\o,v) + \left(\delta^{ij} - {v^iv^j \over v^2}\right) \< G^\perp\>_{\Omega}(\o,v),
\ee
with transverse and longitudinal correlators given by
\bea
&& \< G^\parl\>_{\Omega}(\o,v) = A(v) G^L(\o,v) + \g^2(1-A(v)) G^T(\o,v) \label{av8a} \\
&& \< G^\perp\>_{\Omega}(\o,v) = \frac12 \le[ \le( 1 + A(v) \ri) G^T(\o,v) + {1-A(v)\over\g^2} G^L(\o,v) \ri]. \label{av8b}
\eea
Here $ G^{L,T}(\o,v)$ are the basic boundary correlators for $\theta=\pi$ given in equations (\ref{hatGperp},\ref{hatGparl}), and $A(v)$ is the function
\be\label{av9}
A(v) = {1\over v^3} \le( v - \frac1\g \arcsin v \ri).
\ee

Although it seems the most natural thing to do, we will see in section \ref{consist} that this averaging procedure is {\em inconsistent} with the assumption that the single endpoint description of the string is well-defined (i.e. the effective Langevin description takes into account all other degrees of freedom). Therefore we are led to explore alternative possibilities to eliminate the angle dependence.

\subsubsection{$w$-averaged correlators and the shadow quark picture} \label{wav}

A different possibility for the choice of the integration measure arises if we stick to our physical intuition of the infinite trailing confining string configuration as the limit of a very long string connecting two quarks, one of which is not observed.
If we want to take the shadow quark picture literally, we must average over all states of the endpoint at infinity which are consistent with a stationary state for the string, but otherwise assuming complete ignorance about this state. 
To adhere to this picture, rather than summing over all possible string directions, we can average over all possible {\em uniform velocities} the shadow quark. Under these assumptions, we arrive at a different integration measure, which as we will show in section (\ref{consist}) satisfies the appropriate consistency required by the assumption that we are describing the ensemble of an isolated quark

We start by assuming that the shadow quark has an arbitrary constant velocity $\vec{w}$, and that
the trailing string is along the direction connecting the observed and unobserved quark.
If $\vec{w}$ is not parallel to $\vec{v}$ (the velocity
of the observed quark), at first the trailing string direction will be time-dependent, but in the limit $t\to \infty$ a stationary
state will be reached (see Appendix \ref{vaverage}), with angles $\theta,\vf$ fixed by the relation
\be\label{av10}
\cos \theta = {\vec{v} \over v} \cdot {\vec{w} -\vec{v} \over |\vec{w} -\vec{v}| } = {w \cos\theta_w - v \over (w^2 + v^2 -2 vw\cos\theta_w)^{1/2}}, \qquad \vf=\vf_w.
\ee
Here $w= |\vec{w}|$ and we have introduced $(\theta_w,\vf_w)$ as the polar and azimuthal angle of $\vec{w}$ with respect to $\vec{v}$.

Assuming all possible values of $\vec{w}$ can occur with equal probability means integrating over $dw, d\th_w ,d\vf_w$, with the appropriate
Lorentz-invariant measure, which, as shown in Appendix \ref{vaverage}, is given by
\bea\label{av11}
&&d\mu [w,\th_w,\vf_w] = {1\over \cal N} d\vf_w \, d \cos \th_w \, { w^2 dw \over 2(1-w^2)^2}, \\
&& 0<\theta_w<\pi, \quad 0< \vf_w< 2\pi\quad 0< w< 1, \non
\eea
where ${\cal N}$ is an appropriate normalization factor\footnote{In order to have a finite result, the integrals must be regulated close to $w=1$, but as long as we
regulate both the averages and the normalization in the same way, the result will be finite and will not depend on the details of the regularization.}.

We obtain the desired average by first writing the correlator $G^{ij}$ in equation (\ref{tdG}) as a function of $w,\th_w,\vf_w$ using (\ref{av10}), then integrating
the resulting function with the measure (\ref{av11}),
\be\label{av12}
\< G^{ij}\>_w (\o,v) = \int d\mu [w,\th_w,\vf_w]\, G^{ij}_R(\o;v,\th(v,w,\th_w),\vf_w).
\ee

Using the explicit form of the matrices $H_{T,L}(\th,\vf)$, which can be found in Appendix \ref{correlators}, we find once more that the
integrals can be performed analytically, with the final result
\bea
&& \< G^\parl\>_{w}(\o,v) = \left[1-{\td{A}(v)\over \g^2}\right]G^L(\o,v) + \td A(v) G^T(\o,v) \label{av13a}, \\
&& \< G^\perp\>_{w}(\o,v) = \left[1 - {\td{A}(v)\over 2\g^2}\right] G^T(\o,v) + {\td{A}(v)\over 2\g^4} G^L(\o,v) \label{av13b},
\eea
where again $ G^{T,L}(\o,v)$ are the basic boundary correlators for $\theta=\pi$ and given in equations (\ref{hatGperp},\ref{hatGparl}),
and $\td A(v)$ is the function
\be\label{av14}
\td A(v) = {2\over v^3} \le({\rm arctanh}(v) - v\ri).
\ee

\subsubsection{Fixed-direction correlators} \label{fixed}

As a third example of eliminating the angular dependence from the correlators, we may use some additional input to fix the direction of the trailing string.
For example, in the two-quark picture, we may assume that the quark-antiquark pair is created at rest in a momentum-conserving process,
in which case the two particles have equal and opposite velocities. If these assumptions fit the process under consideration, then
it is natural to pick the particular solution with $\theta=\pi$, rather than averaging over the angles. In this case, the trailing string
solution reduces to the one discussed in section \ref{vparl}, and the
physical correlators are simply given by
\be\label{av15}
G^\parl\Big|_\pi = G^L(\o,v), \qquad G^\perp\Big|_\pi = G^T(\o,v).
\ee

\subsubsection{Static and ultra-relativistic limit}

We will explore what the different average procedures become in the static and high-velocity limit respectively.

As $v\ll 1$, the functions $A(v)$ and $\tilde A(v)$ in (\ref{av9}) and (\ref{av14}) have the following behavior:
\be
A(v) \simeq {1\over 3} + O(v^2), \qquad \tilde{A}(v) \simeq {2\over 3} + O(v^2).
\ee
Inserted in equations (\ref{av8a}-\ref{av8b}) and (\ref{av13a}-\ref{av13b}) this gives the same non-relativistic limit for both angular and $w$-average:
\bea \begin{array}{l} \<G^{\perp}\> \\ \<G^{\parl}\> \end{array} \bigg\} \stackrel{\phantom{sp}v\to0\phantom{sp}}{\longrightarrow} \frac23 G^T +
\frac13 G^L. \lab{0vG}\eea
In this limit we obtain $\<G^{\parl}\>\simeq \<G^{\perp}\>$. In particular the averaged static propagators are completely isotropic. This is a consistency check for our calculations, since all spatial directions should be equivalent for a static quark, once we average over the possible string orientations.

In the ultra-relativistic limit, we obtain
\bea \<G^{\parl}\> \simeq \g^2 \<G^{\perp}\> \simeq G^L \simeq \g^2 G^T, \eea
where we used the fact that $b_m $ is much smaller than $b$ evaluated at the boundary, so that the prefactors in equations \refeq{hatGperp}-\refeq{hatGparl} only differ by a $\g^2$ factor, as
written in the above formula. Moreover, as in the finite-temperature case, the fluctuation equations only
depend on the velocity through the combination $\g\o$, so that the ultra-relativistic limit corresponds to the high-frequency limit. The confining trailing string
correlators then display the same divergence as in the finite-temperature correlators for $v\to1$.

\subsection{High-frequency limits}\lab{highw}

Although different choices of the averaging procedure lead to different results for the boundary correlators, the
high frequency limit is universal.
Equations (\ref{eqn},\ref{eqxi}) and (\ref{hatGperp},\ref{hatGparl}) have the same form as those studied in the black hole backgrounds in \cite{langevin-2}, with the
same asymptotic limit $R\sim b^2$ as $r\to 0$. The results found in \cite{langevin-2} using the WKB method hold and, as $\o \to \infty,$ one finds
\be \label{hf1}
G^L(\o;v) \simeq \g^2 G^T(\o;v), \qquad \o \to \infty, \ee
up to $O(\o^{-4})$ corrections. Using this relation in both equations (\ref{av8a},\ref{av8b}) and (\ref{av13a},\ref{av13b})
one simply obtains:
\be \label{hf2}
\<G^\parl \>(\o;v) \to G^L(\o;v), \quad \<G^\perp\>(\o;v) \to G^T(\o) , \quad \o\to \infty.
\ee
The same is clearly true for the fixed direction correlator of section \ref{fixed}.

Therefore, the zero-temperature confining string correlators share the same high-frequency universal behavior of the finite-temperature correlators.
This result is crucial if we want to use the zero-temperature correlator to define dressed Langevin correlators with an acceptable (i.e. vanishing at least as $1/\omega$)
high-frequency behavior, as was done in \cite{langevin-2} in the non-confining case. This point will be addressed in section \ref{dressed}.

\subsection{Low-frequency limit and friction coefficients}
\lab{friction}

In subsection \ref{diffconst0v} we have derived the friction coefficients in the static limit $v\to 0$. The friction coefficients in the directions perpendicular and parallel to the trailing string were computed by the zero-frequency limit of
the correlators ${G}^{T,L}(\omega)$. Using those results, we want to obtain the
friction coefficients in the general case (before, and after averaging over directions).

Before averaging, for a generic string direction labelled by $(\theta,\vf)$ we have angle-dependent friction coefficients defined by
\be\label{diff1}
\eta^{ij}(v,\theta,\vf) = -\lim_{\o\to 0} {\im G^{ij}(\o;v,\th,\vf) \over \o},
\ee
where $ G^{ij}(\o;v,\th,\vf)$ is given in (\ref{tdG}). On the other hand, since $F(r,\omega; v) = F(r,\g\omega; 0)$, equations (\ref{hatGperp},\ref{hatGparl}) have the same low-frequency limits as in the static case\footnote{The extra $\g$ factor in the denominator of equation~(\ref{hatGperp}) cancels out.}, i.e. those given in (\ref{cor9b}):
\be\label{diff2}
\eta^T \equiv - \lim_{\o\to 0}{\im G^T(\o,v)\over \o} = \sigma_c, \qquad \eta^L= - \lim_{\o\to 0} {\im G^L(\o,v)\over \o} = 0.
\ee
We find
\be\label{diff3}
\eta^{ij}(v,\theta,\vf) = {H^{ij}_T(v,\theta,\vf) \over 1-v^2 \sin^2\th} \, \sigma_c,
\ee
where $H_T$ is given explicitly in Appendix \ref{correlators}. We observe that there are off-diagonal friction coefficients coupling transverse and longitudinal
directions.

Averaging over directions removes off-diagonal components and decouples transverse and longitudinal friction coefficients, and we can use a decomposition analogous to equation \refeq{av7}.
However, unlike the case of the high-frequency limit, which was universal, the friction coefficients will depend on the averaging procedure.
\begin{description}
\item[$\Omega$-average ---]
Integrating equation (\ref{diff3}) over the solid angle (or using directly the low-frequency limit on the averaged correlators (\ref{av8a},\ref{av8b}))
we find
\bea \< \eta^{\parl}\>_\Omega &=& \g^2(1- A(v)) \sigma_c, \label{etaparl-av} \\
 \< \eta^{\perp}\>_\Omega &=& {1+ A(v)\over 2} \sigma_c, \label{etaperp-av} \eea
where $A(v)$ has been defined in \refeq{av9}.

In the non relativistic limit $v\to 0$, expanding $A(v)$ and $\gamma$ one finds that
the diffusion constants become isotropic, and reduce to the geometric average of the two quantities in (\ref{diff2}):
\be
\<\eta^\perp\>_\O \simeq \<\eta^\parl\>_\O \simeq {2 \over 3}\sigma_c, \qquad \mbox{as } v\to 0.
\ee

This is the same result we would find in the static case if we average over all possible string directions.

\item[$w$-average ---]
If we instead integrate using the measure (\ref{av11}), we find (as can be derived from equations (\ref{av13a},\ref{av13b})):
\bea \<\eta^{\parl}\>_w &=& \tilde{A}(v) \, \sigma_c \label{etaparl-w}, \\
 \<\eta^{\perp}\>_w &=& \left[1-{\tilde{A}(v) \over 2\g}\right]\, \sigma_c, \label{etaperp-w} \eea
with $\tilde A(v)$ as defined in \refeq{av14}.

In the non-relativistic limit we again find the isotropic result
\be
\<\eta^\perp\>_w \simeq \<\eta^\parl\>_w \simeq {2 \over 3}\sigma_c, \qquad \mbox{as } v\to 0.
\ee

\item[Fixed angle ---]
Finally, if we fix the trailing string to have the same direction of the quark velocity ($\theta=\pi$) we find, using (\ref{diff3}) and the explicit form
of the matrix $H_T$, equation \refeq{Hij},
\be
\eta^\parl\Big|_\pi = 0, \qquad \eta^\perp\Big|_\pi = \sigma_c, \label{eta-fix}
\ee
which is independent on the magnitude of the velocity\footnote{One may be puzzled by the fact that this result is not isotropic as $v\to0$, but this is not surprising, as it was obtained as the zero-velocity limit of a system with a preferred direction.}.
\end{description}

\subsection{Drag force consistency conditions}\lab{consist}

Throughout the paper we have seen that there are two ways in which the frictional force arises from the holographic computation:
\begin{itemize}
\item From a {\em classical} computation of the momentum flow along the string, which gives the velocity dependent of the force on the moving quark,
\be\label{cc1}
F_q^i = \pi^i(v) = {\delta S \over \delta \xi^{i'}}\Big|_{on-shell};
\ee
\item From a {\em one-loop} computation of the boundary retarded propagator, whose imaginary part gives the velocity dependent matrix of friction coefficients,
\be\label{cc2}
{\im G^{ij} \over \omega}\Big|_{\o\to 0} = - \eta^{ij}(v).
\ee
\end{itemize}
In the standard case of a pure $AdS$ black hole, one finds a trivial velocity dependence and a very simple relation between the right hand sides of (\ref{cc1}) and (\ref{cc2}):
\be\label{cc3}
\vec{\pi} = -\eta \vec{v} , \qquad \eta^{ij}v_i = \eta v^i,
\ee
where $\eta$ is a constant. This means that the classical friction coefficients agrees with the one-loop calculation. This is as it should be, as the two quantities reflect the same physics.

In the general case, the force cannot be described simply as a viscous friction, since the dependence on the velocity is much more complicated than proportionality (\ref{cc3}). However, the agreement between the classical and one-loop result must extend to any situation that has a consistent stationary-state behavior. As shown in Appendix \ref{Langevin-app}, the compatibility between the classical force and the longitudinal component of the one-loop result can be obtained as a Ward identity in the double-field formalism, and reads
\be \label{cc4}
{d \over d v} \pi^i = - \eta^{ij}\, {v_j\over v}.
\ee

Clearly this relation is satisfied for the case of viscous friction with a constant coefficient $\eta$, when $\pi^i = - \eta v^i$ and the propagator behaves at low frequency as $G^{ij} = -i \o \eta (v^i v^j/v^2) + (transverse) + O(\o^2)$, and one recovers (\ref{cc3}).
 However, (\ref{cc4}) is completely general and must hold whenever the ensemble describing the process is well defined, and whenever a stationary limit can be reached (i.e. $\vec{\pi}$ does not depend on time and the limit $\o \to 0$ of the correlators is well-defined), with no other assumption on the velocity dependence of the force and the propagator.

\subsection{Consistency of the holographic correlators}

One may think that consistency of the holographic computation should ensure that (\ref{cc4}) is obeyed, but this in fact is not the case. We analyze
equation (\ref{cc4}) for the various (averaged and non-averaged) correlators defined in section 5.1.

\subsubsection*{Fixed-direction correlators}

We start with the $(\th, \vf)$-dependent quantities, before averaging. The drag force $\pi^i(v, \th, \vf)$ is given by equation (\ref{pi}),
\be \label{cc5}
\vec{\pi}(v,\theta,\vf) =  {\sigma_c\over (1-v^2\sin^2\theta)^{1/2}} \left(\begin{array}{c}\cos\theta \\ (1-v^2)\sin\theta \cos\vf \\ (1-v^2)\sin\theta \sin\vf \end{array}\right).
\ee
Differentiating with respect to $v$ we obtain
\be\label{drag3}
{d \over dv} \pi^j (v) = - {v \, \sigma_c \over (1-v^2\sin^2\th)^{3/2}}\left(\begin{array}{c} \sin^2\th \cos\th \\ \left[(1+v^2)\sin^2\th -2 \right]\sin\th \cos\vf \\ \left[(1+v^2)\sin^2\th -2
\right]\sin\th \sin\vf \end{array}\right).
\ee

On the other hand, the longitudinal\footnote{We remind the reader that we are assuming the velocity vector to be along the $1$-direction.} component of the friction coefficient $\eta^{1i}(v,\th, vf)$
are found via equations \refeq{diff3} and \refeq{Hij}:
\be\label{drag5}
\eta^{j 1}= { \sigma_c \over (1-v^2\sin^2\th)}\left(\begin{array}{c} \sin^2\th \\ \cos\th \sin\th \cos\vf \\ \cos\th \sin\th \sin\vf \end{array}\right).
\ee

Comparing equations (\ref{drag5}) and (\ref{drag3}) we immediately observe that the $(\th, \vf)$-dependent drag force does not obey the consistency relation (\ref{cc4}), {\em except} for the special cases
$\theta=0,\pi$ (string parallel to the velocity), where both sides vanish identically.

One may ask what happens after averaging over the angles, and {\em then} checking whether (\ref{cc4}) if satisfied. We will consider two possible averaging procedures, as already discussed in section
\ref{Oav} and \ref{wav} for the correlators: the naive average over $(\theta,\vf)$ and the average over the unobserved quark velocity.

\subsubsection*{$\O$-average}

Upon averaging over the solid angle, all components of $\pi$ average to zero for any $v$:
\be
\< \pi_i \>_{\Omega} = \int {d\O\over 4\pi} \pi_i(\th, \vf) = 0, \ee
i.e, the average drag force due to strings uniformly distributed around all direction vanishes identically.
Therefore,
\be
{d \over dv} \< \pi^j \>_{\O} = 0.
\ee

On the other hand, while the $12$- and $13$-components of $\<\eta^{ij}\>_\O$ vanish, the $11$-component is positive and has a non-zero angular average. The result is given by equation
(\ref{etaparl-av}),
and the consistency relation (\ref{cc4}) fails.

\subsubsection*{$w$-average}

The second option is to average over the unobserved quark velocity, as explained in Appendix \ref{vaverage}. We first have to express $\cos\theta$ and $\sin\th$ in terms of $(w,\th_w\vf_w)$ in (\ref{cc5}), then take the $w$-average using (\ref{a7}) and (\ref{a8}). This time, we obtain a non-trivial drag force along the velocity direction:
\be
\< \pi^\parl \>_w = \left[-{1\over v} + \left({1\over v^2} -1\right) {\rm arctanh}\, v\right]\sigma_c, \qquad \< \pi^\perp \>_w =0.
\ee
Differentiating with respect to $v$ we obtain
\be
{d \over d v} \< \pi^\parl \>_w = 2{\sigma_c \over v^3 }\left(v- {\rm arctanh}\, v\right)
\ee
which matches exactly with the longitudinal component $\< \eta^{\parl} \>_w$ of the one-loop result, equation (\ref{etaparl-w}).

\subsubsection*{Consistency and the shadow quark picture}

To summarize, we have arrived to the following conclusions:
\begin{enumerate}
\item The consistency relation (\ref{cc4}) {\em fails } for the $(\theta,\vf)$-dependent correlators
coming from a string at an arbitrary angle with the velocity, {\em except}
for the case $\theta=\pi$ considered in section (\ref{fixed}) (or $\th=0$);
\item For direction-averaged correlators, (\ref{cc4}) {\em fails} if we use the simple $\Omega$-average of section (\ref{Oav});
\item On the other hand, the correlators obtained by the $w$-average of section (\ref{wav}) satisfy (\ref{cc4}).
\end{enumerate}

The consistency condition (\ref{cc4}) gives a criterion to select appropriate average procedures. Surprisingly, the most natural option, averaging over all possible string directions with equal weight, does not pass this consistency test. There is a physical explanation for this fact, which comes from the picture of the confining trailing string as a limit of the quark-antiquark system.

Using the non--averaged correlators means that we are choosing an arbitrary fixed direction for the string, But fixing a direction for the string means that we are {\em forcing} the unobserved quark
at infinity to move parallel to the observed quark, with the same velocity: this is the only way the trailing string stays at a fixed arbitrary angle with the observed quark velocity. But this
introduces a {\em correlation} between the two quarks. This does not agree with the point of view that we took for the whole computation, i.e. the fact that we want to observe an {\em isolated} quark
interacting with the confining vacuum. Moreover, this cannot possibly lead to a well-defined ensemble for the observed quark, because by fixing the string direction we are effectively coupling it to a
degree of freedom that is not included in the
medium and is not being ``integrated out.''

This problem survives the average over the angles if we use the naive $\O$-average: for each term in the average, we introduced a correlation between the observed quark and the one at infinity, i.e. we are assuming some specific configuration of a degree of freedom which is not included in the statistical ensemble.

On the other hand, the $w$-average was defined precisely as the averaging procedure for which we make the least assumptions (in fact, none) about the unobserved quark motion (other than the fact that it is uniform, which is required under the hypothesis that the system reaches a stationary state).
Assuming maximal ignorance about the state of the shadow quark is the procedure that agrees with the holographic prescription of infalling boundary conditions at the string turning point, which we
have used to compute the correlators.

We are therefore led to the statement that the confining trailing string description of the single quark system is consistent only under the
{\em shadow quark hypothesis:} it is the limit of a very long quark-antiquark bound state where all correlation is lost since the pair formation, and no information has yet been able to travel from one quark to the other along the string. Under this hypothesis, the gravity dual gives
 a consistent boundary Langevin dynamics for the single quark. This discussion confirms our interpretation of the retarded correlator as the one defined
by infalling conditions.

As a final remark, we have already noticed that the special configuration in which the trailing string is directed as the quark velocity, satisfies the drag force consistency condition, and in
principle defines a consistent ensemble. We believe this is because it is possible to define the physical process in a way that it selects this configuration without introducing extra
correlation between the two quarks: that is, if the quark-antiquark system is produced at rest and assumed to conserve momentum. In this case the string can only stretch along the direction of the velocity. This presumably gives a way of selecting a subset of possible physical states without assuming an arbitrary correlation between the observed and the shadow quark.

It would be interesting to find out whether there are other averaging procedures that
give consistent Langevin dynamics, and to formulate general conditions which
guarantee that (\ref{cc4}) is satisfied.

\section{The dressed Langevin correlators in the confining holographic theory}\label{dressed}

The aim of this section is to give the quantitative results about the Langevin correlators for a heavy quark moving through a gluon plasma. As was observed in \cite{langevin-2}, the correlators at
finite temperature have a universal divergent behavior at high frequencies, that translates into an unphysical behavior in terms of time. It was argued in that paper, from first principles, that
valid subtraction prescription is to consider the dressed correlator $G(\o)-G^{(vac)}(\o)$, where $G(\o)$ is the correlator in a plasma at temperature $T$ and $G^{(vac)}(\o)$ is the correlator in the vacuum.

In section \ref{highw} we indeed pointed out that the vacuum correlators for confining strings share the same high--frequency behavior as the black--hole correlators. The dressed correlator,
defined as the difference of the two, is well behaved in the high--frequency regime. More precisely, it vanishes as $1/\o$.

As opposed to the results in \cite{langevin-1} for the non--confining strings, the vacuum correlators for confining strings display a non trivial low--frequency limit. We derived, in section
\ref{friction}, the friction constants for the different proposed averaging procedures. For generic velocities of the quark, these constants are non zero for all averaged correlators, while in the non--confining string case the friction coefficient vanishes in the vacuum. Therefore, the dressing method in the confining case also modifies the low--frequency limit, by shifting the
coefficient of the linear term in $\o$.

In the remainder of this section, we will first obtain the dressed friction coefficient in generic models, then present the numerical results for the dressed correlators that give a consistent result for the drag force, as explained in section \ref{consist}, namely for
the $w$--averaged and $\th=\pi$ correlators. To perform the numerical computations, we  consider a probe quark with a large and finite mass $M_q$ that implies a cutoff $r_q\sim1/M_q$, close
to $r=0$, in our holographic description. More specifically, $r_q$ is implicitly defined by $M_q={1\over 2\pi\ell_s^2}\int_{r_q}^{r_*} dr b^2$, with $r_*=r_m$ at zero temperature and $r_*=r_h$ at
finite temperature (the dependence of the quark mass on the temperature was analyzed in \cite{transport} for the confining Improved Holographic QCD model).

The results we present here can be derived for any asymptotically $AdS$ background with IR expansion given by equation \refeq{k3}. We are interested in the case $a>0$, that corresponds to
confining theories at low temperatures. The UV expansion (including the case of finite temperature  backgrounds) reads:
\beq\lab{UVasy}
b(r) \simeq \frac{\ell}{r} h(r), \qquad f(r) \simeq 1 - \frac\cC4 r^4, \qquad \mbox{as } r\to0.
\eeq
The function $h(r)$  denotes the subleading contribution to the asymptotic expansion in the UV, as it was defined in \cite{langevin-2}: $rh'/h\to0$, as $r\to0$. Moreover, it was shown in
\cite{gkmn2} that in five dimensions the constant $\cC$ is  essentially  the product of the entropy density $s$ times the temperature $T$,
\beq\lab{calc}
\cC = \le( M_P^3V_3N^2 \ri) sT,
\eeq
where $M_P$ is the 5--dimensional Plank mass, $V_3$ is the volume of the 3--dimensional space, $N$ is the rank of the gauge group of the dual gauge theory.

\subsection{Friction from the dressed correlators}\lab{dressed eta}

As reminded above, the low--frequency behavior of the confining string correlators is non trivial and given by equations \refeq{etaparl-w}-\refeq{etaperp-w} and \refeq{eta-fix}, in the $w$--averaging
and fixed--angle cases, respectively.

We here recall the value for the friction coefficient in the black--hole background analyzed in \cite{langevin-1}, corresponding to the finite--temperature,
deconfined, phase:
\bea
\eta^\parl_{bh} &=& {b_s^2\over2\pi\ell_s^2} \le(1 + 4 v^2 {b_s' \over b_s f_s'} \ri), \\
\eta^\perp_{bh} &=& {b_s^2\over2\pi\ell_s^2},
\eea
with $b_s=b(r_s)$ ($r_s$ is the worldsheet horizon and was defined at the beginning of section \ref{BHintro} by the equation $f(r_s)=v^2$) and similarly for $b_s',f_s'$. For the dressed correlators,
we obtain the friction constants by subtracting the values of the black--hole and vacuum coefficients.

These expressions for the renormalized friction coefficients are valid in the infinite mass approximation. Should we consider a moving quark with finite mass, the vacuum value would be zero, as argued
in section \ref{finitemass}, and the result would just be the finite temperature values obtained in \cite{langevin-1}.

\paragraph{$w$--average ---} Averaging over the velocities of the shadow quark we obtain the renormalized friction constants,
\bea
\eta^\parl_{w} &=& {\g^2\over2\pi\ell_s^2} \le[ b_s^2 \le(1 + 4 v^2 {b_s' \over b_s f_s'} \ri) {1 \over \g^2} - {\td A(v) \over \g^2} b_m^2 \ri], \lab{etaparlrw}\\
\eta^\perp_{w} &=& {1\over2\pi\ell_s^2} \le[ b_s^2 - \le( 1 - {\td A(v) \over 2\g^2 } \ri) b^2_m \ri]. \lab{etaperprw}
\eea
Since $r_m$ is the location of the minimum of the scale factor, $b_m$ and $b_s$ are at most of the same order and, as the temperature or the velocity increases, $b_s$ becomes larger and larger with
respect to $b_m$. Moreover, we note that the factor $1-\td A/2\g$ is never larger than unity, and it equals 1 at $v=1$ (See figure \ref{figeta} (a). Hence $\eta^\perp_{w}>0$ for all velocities, as shown by the plot in figure
\ref{figeta} (b). In these plots, we normalized $\eta^\perp_{w}$ by the factor $\s_s\equiv b_s^2/(2\pi\ell_s^2)$ that corresponds to the value of the friction coefficient at finite temperature, and is related to the
effective string tension by $\s_s=\s_c(b_s/b_m)^2$.

Since $\td A(v)/\g^2$ is non-negative and bounded from above by the value $\td A(0)=2/3$, as shown by the plot in figure \ref{figA}, also $\eta^\parl_{w}$ is always positive.

\begin{figure}
\begin{center}
\includegraphics[height=5cm]{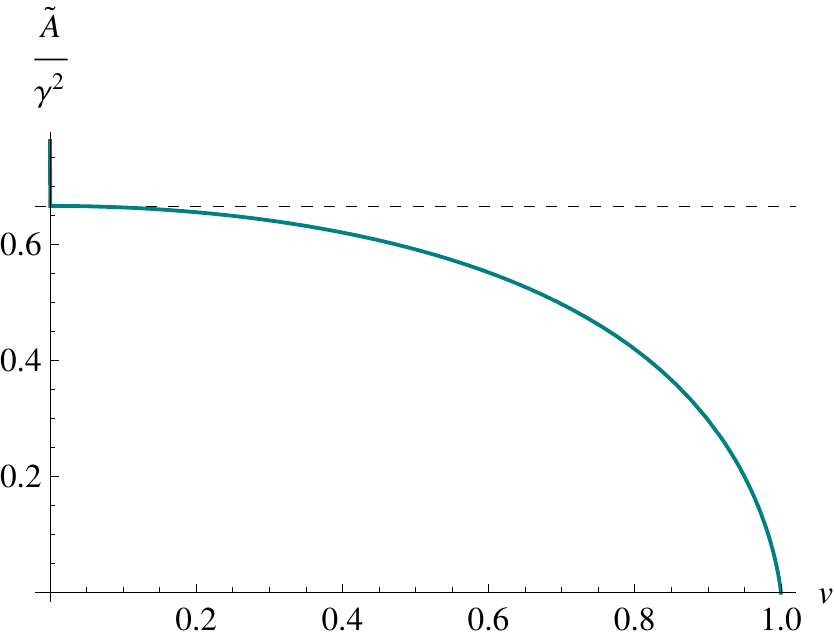}
\caption{In \protect\ref{figA} we observe that the function $\td A(v)$ diverges logarithmically as $v\to1$, so that $\td A(v)/\g^2$ vanishes in this limit and is a monotonically decreasing function bounded from
above by the value at $v\to0$, $\td A(0)=2/3$.}\lab{figA}
\end{center}
\end{figure}

In fact, the factor $\td A(v)$ in $\eta^\perp_{w}$ diverges logarithmically in the ultra--relativistic limit but $\td A(v)/\g^2$ vanishes. This means that the contribution from the vacuum
correlators are subleading for ultra--relativistic quarks ($b_s$ is divergent, $b_s \sim (\ell \pi T)^2 \g$). In this approximation, $r_s\to0$ and we can use the expansion of the
asymptotic $AdS$ region. In particular we know that $\le( 1+ 4 v^2 b'/(b f') \ri)\to \g^2$ close to the boundary. The expressions \refeq{etaparlrw}--\refeq{etaperprw} read
\bea\lab{ultrarel}
\eta^\parl_{w} &\simeq& \g^2 \eta^\perp_{w} \simeq {\ell^2\over4\pi\ell_s^2} h_s^2 \cC^\frac12 \g^3, \mbox{ as } v\to1.
\eea
The function $h(r)$ defined in \refeq{UVasy} appears through the factor $h_s=h(r_s)$ in \refeq{etaparlrw} (it depends on the specific model) and $\cC$ is given by equation \refeq{calc}.


\paragraph{Fixed angle ---} For the configuration in which the quark is dragged in the direction of the velocity, the result is
\bea\lab{etafixbhparl}
\eta^\parl_{\pi} &=& {b_s^2\over2\pi\ell_s^2} = \eta^\parl_{bh}, \\
\eta^\perp_{\pi} &=& {1\over2\pi\ell_s^2} \le[ b_s^2 \le(1 + 4 v^2 {b_s' \over b_s f_s'} \ri) - b_m^2 \ri] = \eta^\perp_{bh} - \s_c \lab{etafixbhperp}.
\eea
For the confining string parallel to the velocity, the friction constant are manifestly positive, since $b_s > b_m$ in the black--hole background above the critical temperature.

Also in this case, the ultra--relativistic asymptotics are given by equation \refeq{ultrarel}, because the subtracted terms are subleading in this limit.

\subsection{Dressed Langevin correlators in Improved Holographic QCD}

The analysis up to this point has been completely general. Here, we give concrete numerical results in the case of the Improved Holographic QCD model, i.e. the 5--dimensional dilaton--gravity model described in \cite{gkmn3}, and already used to compute the bare Langevin correlators in \cite{langevin-1}. This model has the virtue of being confining, conformal at high energy, and
displaying a first order deconfining phase transition at finite temperature. Moreover, it was shown in \cite{gkmn3} to reproduce quite accurately the thermodynamics and low-lying glueball spectrum of pure Yang-Mills theory, and is a good candidate for a phenomenological study of quark diffusion through a plasma in the context of holography.

\begin{figure}
\begin{center}
\includegraphics[height=5cm]{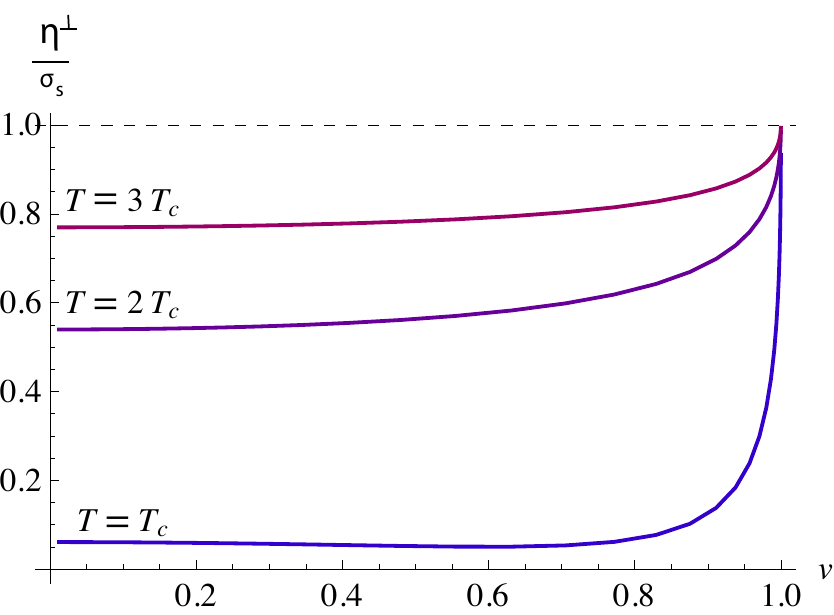}
\caption{The drag coefficient for the directions transverse to the velocity, normalized by the bare drag coefficient $\s_s\equiv b_s^2/(2\pi\ell_s^2)$, is
plotted as a function of the velocity for $T=T_c,2T_c,3T_c$: as the temperature or the velocity grow, $\eta^\perp$ approaches the finite temperature
result $\s_s$. A similar plot holds for
$\eta^\parl/\g^2$.} \label{figeta}
\end{center}
\end{figure}

First, we look at the effect of the contribution of the subtraction to the low-frequency diffusion constant $\eta$. This is displayed in figure \ref{figeta}. As we can see, the effect of the subtraction is quite important at small velocities, but it is almost negligible as the quark velocity approaches unity. As the heavy quark probes encountered in heavy ion collision experiments
are in the ultra-relativistic regime, this result indicates that the subtraction has a negligible effect when comparing the holographic computation of the diffusion coefficient with experimental results.

Next, we discuss the full dressed correlators. The correlators we present here are computed by giving the quarks a finite mass, thus in both the black hole and confining case the
string endpoint is taken to lie at a finite radius $r_q$. In the ultra-relativistic limit $p\gg M_q$, which is the situation most relevant for heavy ion experiments, $r_q$ is related to the quark mass by \cite{langevin-1}
\be\label{rq}
r_q \simeq {\ell^2 \over 2\pi\ell_s^2}{h_q^2 \over M_q},
\ee
where $h_q=h(r_q)$, and $h(r)$ is defined in equation (\ref{UVasy}).

The applicability of the results of the previous sections in the case of a finite mass can be problematic in certain regions
of parameter spaces (namely, very low frequencies, and very high frequencies and momenta). These issues, and the range of validity of the results presented here, will be discussed at greater length in the next subsection.

\subsubsection{Dressed correlators with fixed shadow quark}

The results for the dressed correlators obtained using a fixed angle, $\th = \pi$, defining the confining trailing string to be antiparallel with respect to the velocity are shown in figure \ref{figfix}, where we
compare dressed correlators corresponding to different temperatures of the plasma (same quark mass and velocity), figure \ref{Tfix}, or different momenta for the moving quark (same temperature and
quark mass), figure \ref{vfix}, or else different quark mass, namely Charm and Bottom (same plasma temperature and same momenta), figure \ref{Mfix}.

\begin{figure} \begin{center}
\begin{subfigure}{.5\textwidth}\includegraphics[height=6cm]{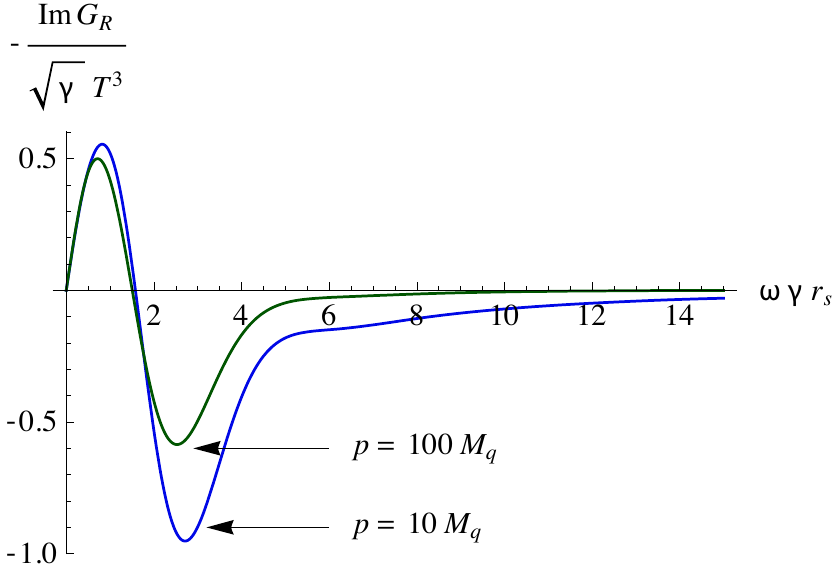}\caption{Velocity comparison (Charm at $T=T_c$)}\lab{vfix}\end{subfigure}~~~~
\begin{subfigure}{.5\textwidth}\includegraphics[height=6cm]{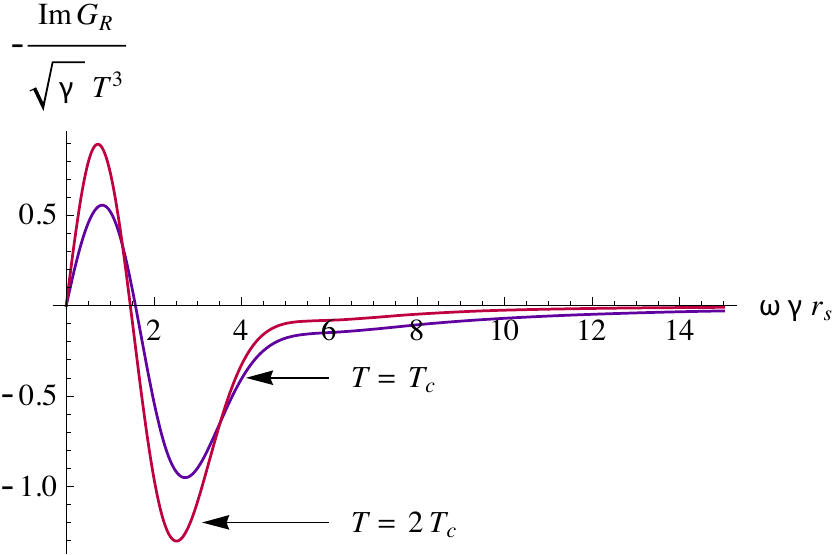}\caption{Temperature comparison (Charm with $p/M_q=10$)}\lab{Tfix}\end{subfigure}\\
\center{\begin{subfigure}{.5\textwidth}\includegraphics[height=6cm]{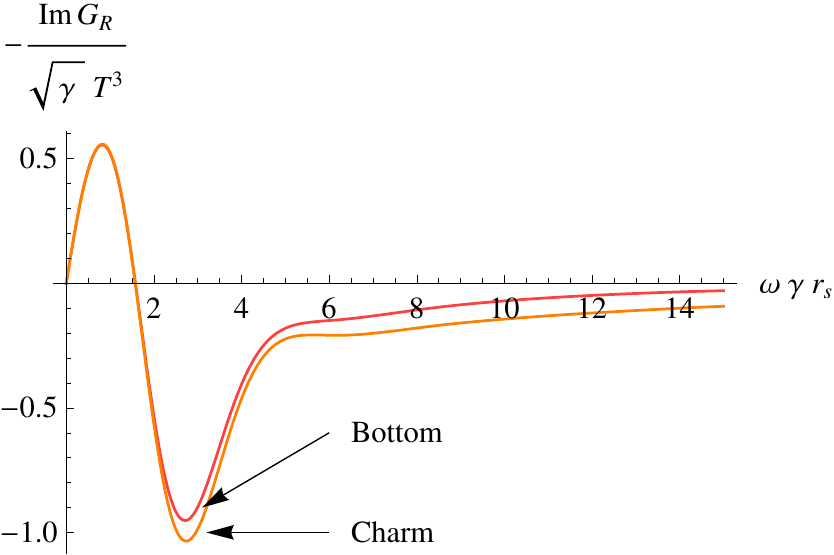}\caption{Mass comparison ($p/M_q=10$ and $T=T_c$)}\lab{Mfix}\end{subfigure}}
\caption{The imaginary part of the dressed retarded correlators, for different values of the temperature, momentum and mass, obtained by fixing the direction of the confining string to be (anti)parallel to the quark
velocity.}\lab{figfix}
\end{center} \end{figure}

These plots are obtained by numerically solving the equations for the imaginary part of the correlators, using the membrane paradigm (see equation (4.28) in \cite{langevin-1}). The equation is a first
order non linear differential equation with regularity condition in the infrared (at the worldsheet horizon, for finite temperature, and at the scale factor minimum $r_m$ for the confining phase). The
results for finite temperature were already discussed in \cite{langevin-1}, where the correlators have been computed for a non-conformal background, using the second--order
fluctuation equations. In \cite{langevin-2} the dressed correlator was computed for conformal and scaling backgrounds, where the zero--temperature classical world-sheet configuration is just a straight
string.

Here we compute the confining Langevin correlators $G^T$, as defined in equation \refeq{hatGperp}, and use equation \refeq{av15}, to obtain the dressed correlator by subtracting $G^T$ from the
finite--temperature result for the transverse modes:
\bea
G_{\pi}^\perp (\o;T) = G^\perp(\o;T) - G^T(\o). \lab{Gpi}
\eea

In figure \ref{figfix} we show the correlators normalized by the factor $\sqrt \g T^3$ as a function of the dimension--less frequency $\o_s \equiv \g \o r_s$. The normalization factor encodes the
dependence on the velocity and temperature of the finite temperature correlator at small frequencies (in the conformal limit). In fact, at small frequencies and for the velocities under consideration
(momenta such that $p/M_q\gtrsim 5$ correspond to ultra--relativistic velocities, $v\gtrsim .98$), the dressed correlator behaves as
\bea
\im G_{\pi} \simeq - \pi \,T_s\,\o_s\, \eta_{bh},
\eea
where $T_s$ is the world-sheet black--hole temperature
\bea
T_s= {1 \over 4\pi} \sqrt{ff'\le( { 4b' \over b} + {f' \over f} \ri)}\bigg|_{r_s} \simeq {T \over \sqrt \g}, \quad \mbox{for } v\to1.
\eea
%
Using \refeq{etafixbhperp} and $r_s \simeq 1/(\pi \sqrt\g T)$ in the ultra--relativistic approximation, we can evaluate
\bea
\im G_{\pi}^\perp \simeq - {\ell^2 \over \ell_s^2} \,{\pi^2 \over 2} \,h_s^2 \o_s \,\sqrt\g \,T^3,
\eea
where $h(r)$ is the subleading contribution as defined in \refeq{UVasy}.

\subsubsection{Dressed correlators averaged over the shadow quark velocity}

The discussion for the $w$--averaged correlators follows from the fixed--angle case, since, as we will argue below, the extra contributions that would arise from the longitudinal correlator
$G^L$ can be neglected for relativistic velocities of the quark, and the expression for the renormalized transverse correlator reduces to \refeq{Gpi}.

The general expression for the renormalized correlator in the $w$--average framework is simply derived using \refeq{av13b}:
\bea
G_{w}^\perp (\o;T) = G^\perp(\o;T) - \le( 1- { \td A(v) \over 2 \g^2} \ri) G^T(\o) - { \td A(v) \over 2 \g^2} { G^L(\o) \over \g^2} .
\eea
As we argued in subsection \ref{dressed eta}, the coefficient $\td A(v)/\g^2$ is very small for large velocities. More precisely, for the velocities corresponding to momenta $p/M\gtrsim 10$, i.e.
$v\gtrsim .995$, we can evaluate $\td A(v)/(2\g^2) \lesssim .02$ and starts becoming sensible for momenta smaller than $p\simeq3M$ (correspondingly, $v\simeq.95$ in fact yields $\td A(v)/(2\g^2)
\simeq .1$, as we can also guess from figure \ref{figA}).

So the plots in figure \ref{figfix} hold without relevant changes also for the $w$--averaged correlators $G^\perp_{R,w}$, given that for large momenta $\td A(v)/(2\g^2)\ll1$, $G^L\simeq\g^2 G^T$, and
\bea
G_{w}^\perp (\o;T) \simeq G^\perp(\o;T) - G^T(\o) = G_{\pi}^\perp(\o;T) .
\eea

\subsection{Effects of a finite quark mass, and limits of applicability of the Langevin approach}
\lab{finitemass}

Throughout this work we considered the confining trailing string as a semi-infinite string attached to a boundary point, and extending indefinitely in one of the spatial direction.
However, in physical situations where the quark has a finite mass $M_q$, the string cannot be considered truly infinite: when the string reaches a certain critical length,  it becomes unstable to breaking and producing another quark-antiquark pair. Roughly, for a heavy quark of mass $M_q$, the maximal length of the string is given by
\be\label{Lmax}
L_{max} \approx {2 \gamma M_q \over \sigma_c}
\ee
where $\sigma_c$ is the confining string tension and $2 M_q$ should be replaced by the chiral condensate $\Lambda_\chi$, for the case of light quarks, as
discussed in section \ref{physical}.
Therefore, for finite quark mass, the ``shadow'' quark is at a large but finite distance, and starts influencing the dynamics after a time of order $L_{max}$ from
the initial perturbation. The corresponding trailing string solution has a regular turn-around point at $r_0 < r_m$.

The finite length $L_{max}$ of the string acts as an infrared cutoff that effectively prevents dissipation from affecting modes of frequency smaller than $\sim 1/L_{max}$. Indeed, consider a situation in which we want to observe the response of the endpoint quark to a perturbation at $t=0$. Since we are interested in a retarded response, the influence of the shadow quark endpoint will not be felt until a time $t \sim L_{max}$, when the fluctuation has had time to propagate the full length of the string. At earlier times the solution will
be a travelling (infalling) wave going from the first to the second endpoint.
After a time of the order $L_{max}$ however, the fluctuation has traveled the full length of the string and we have to take into account the boundary conditions that have to be imposed on the second endpoint. Possible choices include for example keeping the second endpoint fixed (Dirichlet) or setting to unity the wave function at the second endpoint. But no matter which boundary condition we chose, the problem we are now solving is that of a wave equation along a string with one condition at each endpoint, and the wave-function for this problem is necessarily a real {\em standing} wave, which carries no flux. As a consequence, for $t \gtrsim L_{max}$, the imaginary part of the retarded correlator vanishes, and the length of the string $L_{max}$ acts
as a large-time, or small frequency cut-off.

Therefore, in practice, also for the transverse fluctuations the true retarded correlator in the vacuum background is strictly
zero at frequencies smaller than $1/L_{max}$, resulting in a vanishing diffusion constant.
In other words, equation \refeq{intro7} is valid, strictly speaking, only in the limit $M_q \to \infty$. For a finite $M_q$ on the other hand, we will have
$\im G^\perp(\omega) \equiv 0$ when $\omega < 1/L_{max}$. In particular, for any finite-mass quark the vacuum subtraction does not contribute to the diffusion constant, which is obtained in the zero-frequency limit of $\im G^\perp(\o)$. Moreover, the correlators we have computed from the confining trailing string solutions are reliable only above the infrared cutoff provided by $L_{max}$, i.e. for
\be\label{lower}
\omega \gtrsim \o_{min} \equiv {\sigma_c \over 2\g M_q},
\ee
or the improved bound with $2M_q$ replaced by the chiral condensate in the presence of light quarks.

On the other end of the frequency spectrum, a finite quark mass also implies a UV cutoff. Indeed, as shown in \cite{langevin-2}, for a finite quark mass the applicability of the Langevin description obtained holographically does not extend to infinitely high frequency, but rather it is limited by the condition
\be\label{upper1}
\omega \gamma r_q < 1.
\ee
The reason is that, when this bound is violated, the effective boundary action for the moving quark fails to be of the form appropriate for a relativistic particle, thus
one cannot write the simple relativistic Newton's equation with (generalized) friction and stochastic noise forces. Using the relation (\ref{rq}) between $r_q$ and the quark mass, the condition (\ref{upper1}) can be rewritten as the upper bound:
\be \label{upper2}
\omega < \o_{max} \equiv 2 \pi h_q^{-2} \le( {\ell_s \over \ell} \ri)^2 {M_q \over \g}.
\ee

Therefore, the range of $\omega$ we are describing must lie in the range between the lower bound (\ref{lower}) and the upper bound (\ref{upper2}).

Finally, a finite quark mass also imposes an upper bound on the quark momentum: recall that, in the thermal background, the trailing string solution has a world-sheet horizon
$r_s$, whose position moves towards the boundary as the quark velocity is increased. The trailing string solution becomes a bad approximation when $r_s$ becomes equal or smaller to the
endpoint of the quark $r_q$, which happens for large enough quark momentum $p$. In the ultra-relativistic limit, $r_s\simeq (\pi \sqrt\g T)^{-1}$,
with $\gamma \simeq p/M_q$, and
the constraint $r_s>r_q$ becomes an upper bound on the quark momentum in terms of the asymptotic background geometry, quark mass, and temperature:
\be \lab{pbound}
p< p_{max} \equiv 4 M_q \le({M_q \over T}\ri)^2 \le( {\ell_s \over \ell} \ri)^4 h_q^{-4},
\ee

Before we present numerical estimates of these bounds in the concrete example of IHQCD, for the sake of completeness we recall what is the regime
in which the local Langevin description (\ref{intro1}) is appropriate, and one does not need to use the full correlators. As discussed in
\cite{langevin-1}, this condition is that the thermal correlation time, (i.e. the width of the correlators, controlled by the world-sheet temperature
$T_s$) is much smaller than the relaxation time (roughly, the integrated correlators, controlled by the diffusion constant $\eta_{bh}$). Concretely,
this condition translates into $T_s \ll \eta_{bh}$, as showed in \cite{langevin-1}, and it depends on $p$ both through the dependence in $T_s$ and
the one in $\eta_{bh}$. In the ultra-relativistic regime, the local Langevin equation holds if we work under the condition
\be\lab{plocal}
p \ll \frac{p_{max}}{\pi^2}.
\ee
This is in fact almost the same bound as in \refeq{pbound}, except for the factor of $1/\pi^2$.
Notice that if the local Langevin equation \refeq{intro1} is valid, the consistency condition for the holographic description, equation
\refeq{pbound}, is automatically satisfied. On the other hand, in the non--local regime where we have to use the generalized Langevin equation
\refeq{intro2}, equation \refeq{plocal} is not satisfied, and condition \refeq{pbound} is non trivial.

\begin{figure}
\begin{center}
\begin{subfigure}{.5\textwidth}\includegraphics[height=6.3cm]{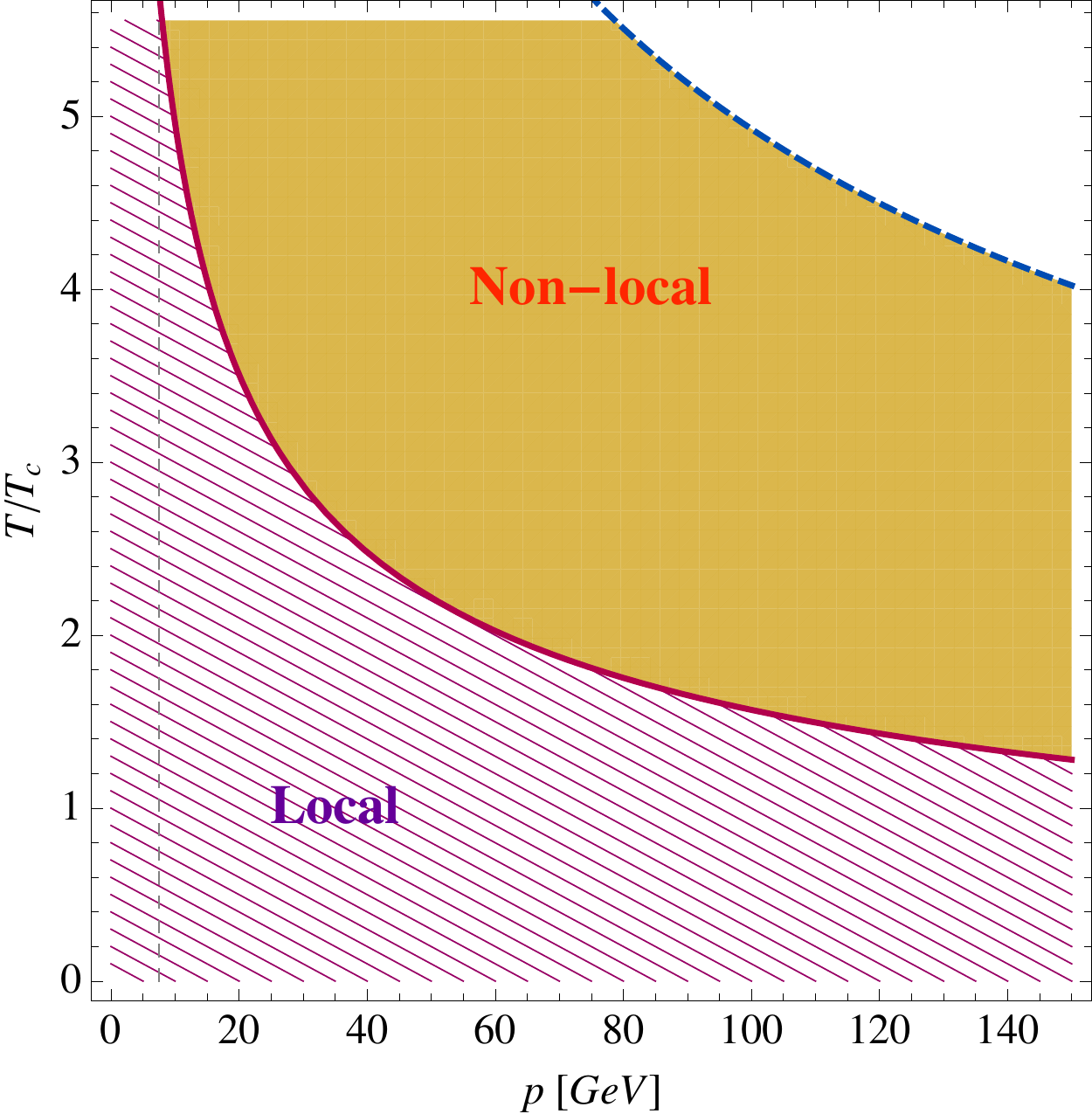}\caption{Charm}\lab{validityTpc}\end{subfigure}~~
\begin{subfigure}{.5\textwidth}\includegraphics[height=6.3cm]{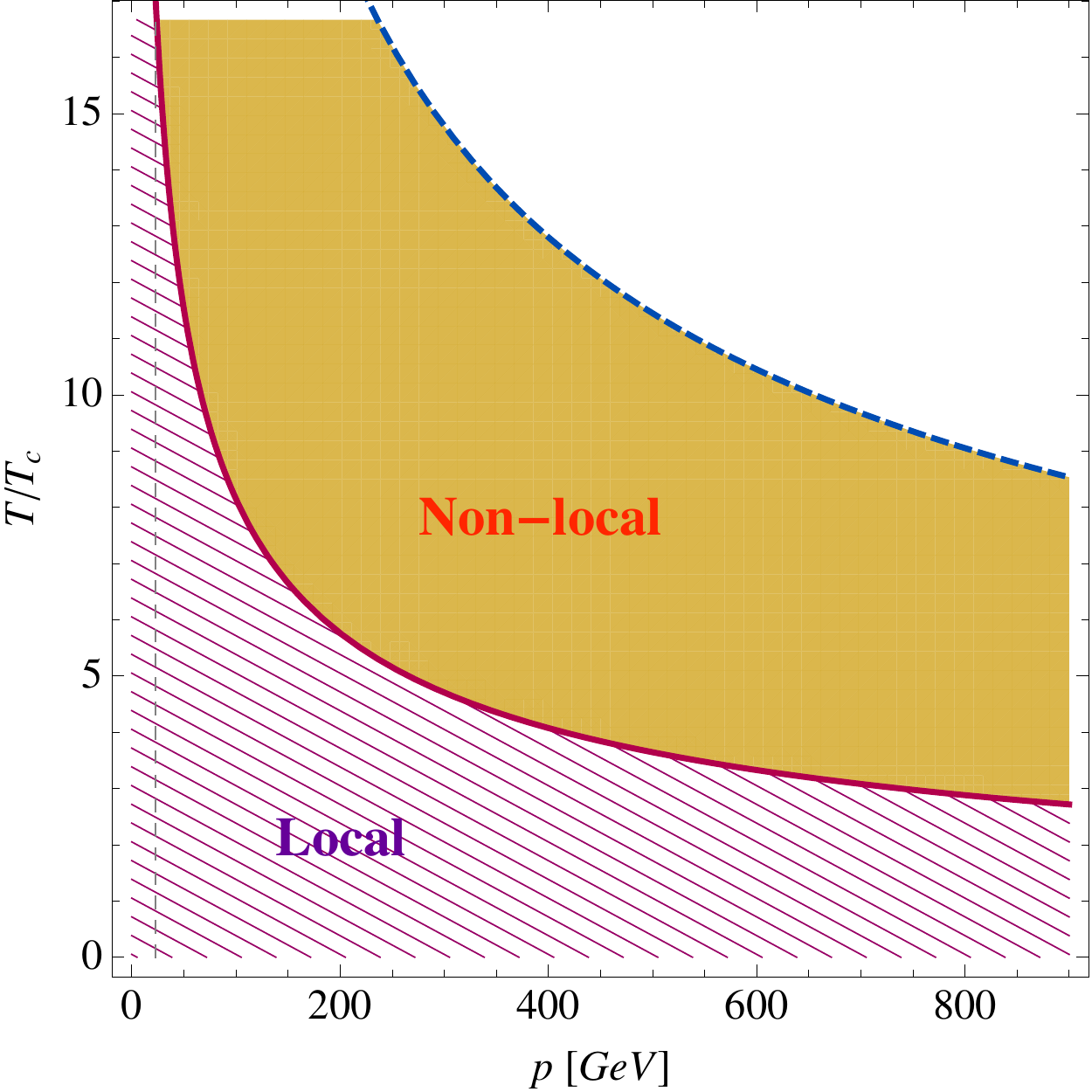}\caption{Bottom}\lab{validityTpb}\end{subfigure}
\caption{\small Validity of the local and non--local regime for Langevin diffusion. The red plain curve represents the right hand side of equation
\protect\refeq{plocal}, and the blue dashed curve is the bound (\protect\ref{pbound}).}
\lab{validityTp}
\end{center}
\end{figure}
The bounds discussed in this section are displayed graphically in figures \ref{validityTp} and \ref{validityw} for charm and bottom quarks, as
computed in the specific IHQCD model we are considering here. In figures \ref{validityTpc} and \ref{validityTpb}, the empty region in the $p-T$
plane is the one in which the holographic description via classical trailing string plus fluctuations breaks down, for charm and bottom quark
respectively. In those figures we also show the regions in which either the local or the generalized Langevin equation should be
used \cite{langevin-1}. While the local regime stays valid even for high momenta in the case of a bottom quark, it is clear from figure
\ref{validityTpc} that the diffusion of charm quarks through the quark--gluon plasma produced at LHC has to be described by the generalized Langevin
equation, according to our analysis.

\begin{figure}
\begin{center}
\begin{subfigure}{.5\textwidth}\includegraphics[height=6.3cm]{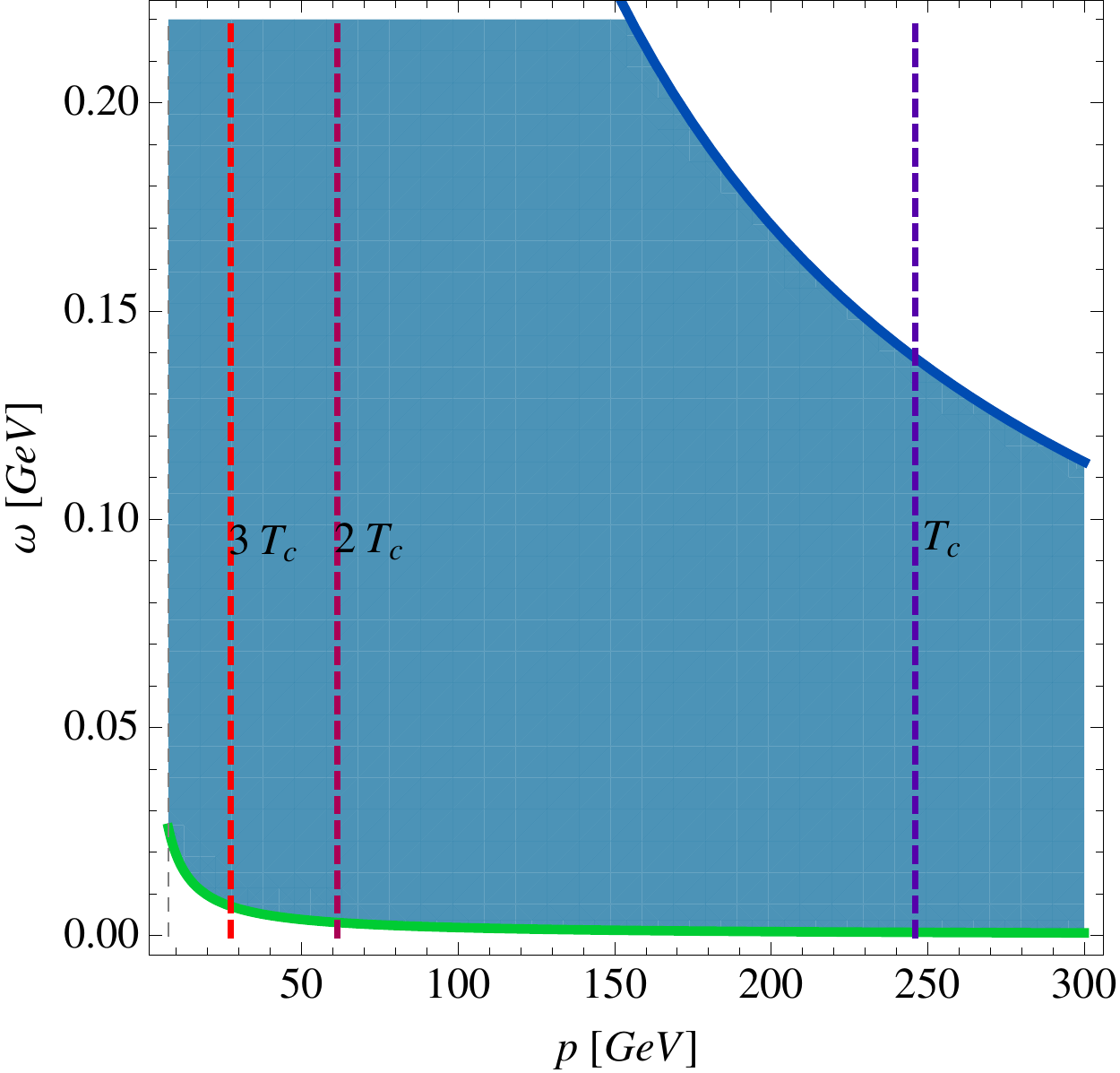}\caption{Charm}\lab{validitywpc}\end{subfigure}~~
\begin{subfigure}{.5\textwidth}\includegraphics[height=6.3cm]{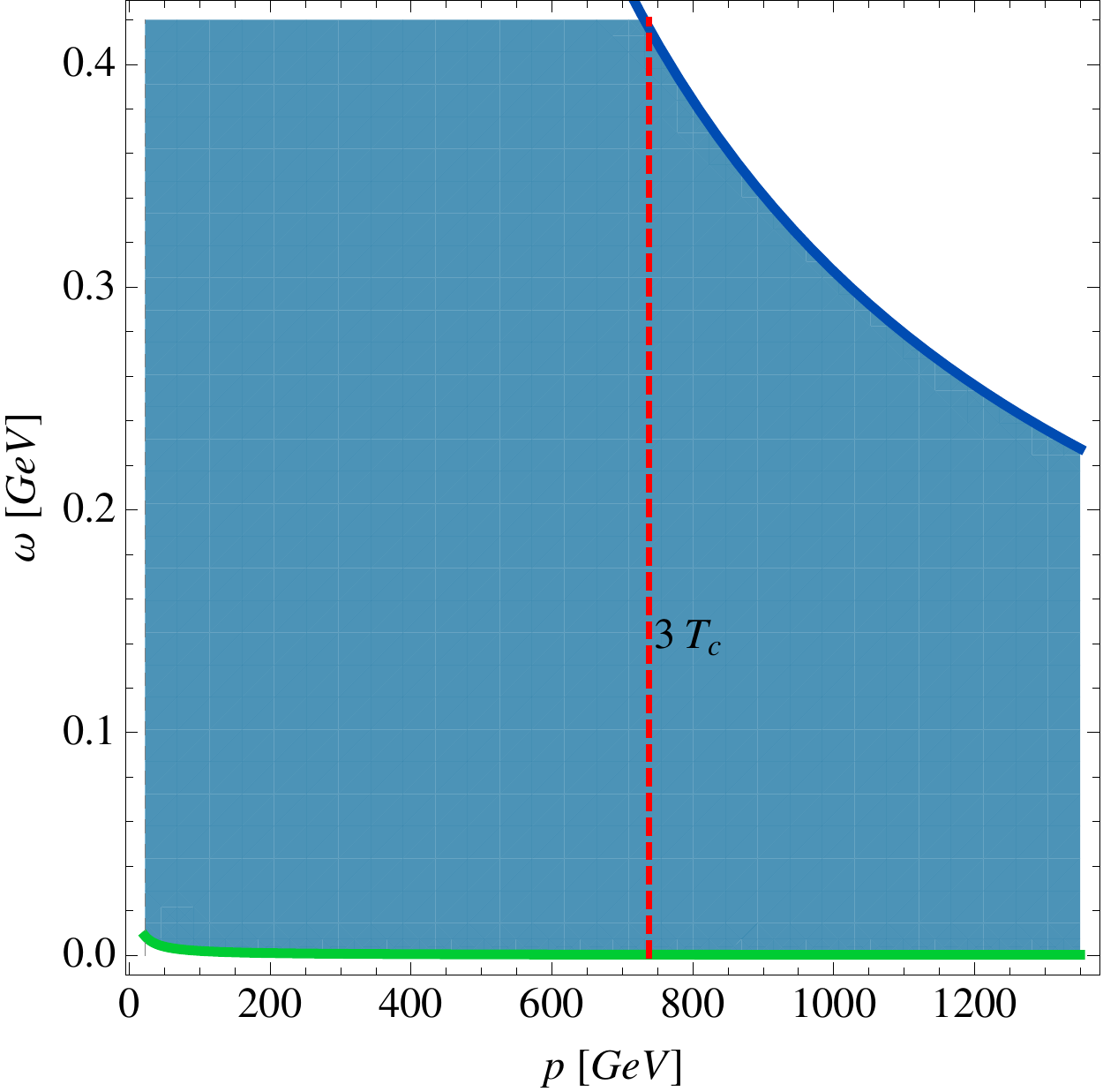}\caption{Bottom}\lab{validitywpb}\end{subfigure}
\caption{\small Allowed frequency
range for our trailing string picture. The blue curve represents the upper bound on the frequency (\protect\ref{upper2}), while
the green curve is the lower bound (\protect\ref{lower}).
The vertical (dashed) lines represent the bound (\protect\ref{plocal}). For momenta higher than these lines, only the generalized Langevin equation should be used, at each indicated temperature.}
\lab{validityw}
\end{center}
\end{figure}

On the other hand, the frequency range allowed by the bounds (\ref{lower}) and (\ref{upper2}) is graphically represented in figure \ref{validityw}, as a function of the momentum.
The vertical lines represent the momenta above which the local Langevin regime breaks down and the full correlator should be used, for three temperatures $T=T_c,2T_c,3T_c$.

\addcontentsline{toc}{section}{Acknowledgements}
\section*{Acknowledgements}
We would like to thank Jan de Boer, Alfonso Ramallo and Jacob Sonnenschein for discussion and comments. 

This work was supported in part by ESF
Holograv Network Short Visit Grant n.~5007, European Union's Seventh Framework Programme under grant agreements (FP7-REGPOT-2012-2013-1) n.~316165,
PIF-GA-2011-300984, the EU program ``Tha\-les'' MIS 375734, the European Commission under the ERC Advanced Grant BSMOXFORD 228169, and the program
``AdS - CMT - Holography and Condensed Matter Physics'' (ERC - 05), MIS 374071, which was co-financed by the European Union (European Social Fund,
ESF) and Greek national funds through the Operational Program ``Education and Lifelong Learning'' of the National Strategic Reference Framework (NSRF)
under ``Funding of proposals that have received a positive evaluation in the 3rd and 4th Call of ERC Grant Schemes''. 

This work is part of the D-ITP consortium, a program of the Netherlands Organisation for Scientific Research (NWO) that is
funded by the Dutch Ministry of Education, Culture and Science (OCW).

\newpage \addcontentsline{toc}{section}{Appendices} \appendix

\section*{Appendices}
\vskip 0.5cm 
\section{Black-hole trailing string revisited}
\lab{appBH}

The scope of this short appendix is to clarify why in the black hole there is a preferred direction in which the trailing string bends. We remind that the action in that case is
\be S = - {1\over 2\pi\ell_s^2} \int dtdr\, {b^2} \sqrt{(f - |\vec v|^2)(f^{-1} + |\vec \xi'|^2) + (\vec v \cd \vec \xi')^2}.\ee
If the vector $\xi$ is arbitrary it is convenient to introduce two variables, $\chi$ and $\chi_\perp$, defined by $\chi' = \sqrt{|\vec \xi'|^2}$ and $\chi_\perp'\equiv\sqrt{\chi'^2-(\vec v \cd \vec
\xi')^2}$. The action then takes the form
\be S = - {1\over 2\pi\ell_s^2} \int dtdr\, {b^2} \sqrt{1 - {f \over v^2} + f \chi'^2 + v^2 \chi_\perp'^2}. \ee
Note that when $\chi_\perp=0$ this reduces to the usual computation where the ansatz for the trailing string describes a string only laying along the direction of the velocity. We can find $\chi$ and
$\chi_\perp$ by solving the equations of motion, $\pi_\chi=C_\chi$ and $\pi_{\chi_\perp}=C_{\chi\perp}$, yielding
\bea \chi' = C_{\chi} {\sqrt{f-v^2} \over {f\sqrt{b^4 f - \le( C_{\chi}^2 - {f\over v^2} C_{\chi\perp}^2 \ri) }}}, \quad \chi_\perp' = C_{\chi\perp} {\sqrt{f-v^2} \over {v^2\sqrt{b^4 f - \le(
C_{\chi}^2 - {f\over v^2} C_{\chi\perp}^2 \ri) }}} . \non\\ \eea
Moreover, the onshell action reads
\be S = - {1\over 2\pi\ell_s^2} \int dtdr\, {b^2} \sqrt{f-v^2 \over b^4 f - \le( C_{\chi}^2 - {f\over v^2} C_{\chi\perp}^2 \ri) }. \ee
For the action to be regular at the world-sheet horizon $r_s$, where $f(r_s)=v^2$, we have to impose that
\be C_{\chi}^2 - { C_{\chi_\perp}^2 \over v^2} = b^4_s v^2 \equiv C^2, \ee
with $b_s = b(r_s)$. This relation still leaves $C_{\chi\perp}$ arbitrary within the range $0 \leq C_{\chi\perp} < v^2 C_{\chi}$ (this range excludes
the case where the string is orthogonal to the velocity direction, which is not consistent since it would imply $C=0$ for finite $v$).

However, the difference with the zero-temperature case is that here the on-shell action does depend on the two parameters $C$ and $C_{\chi\perp}$. In other words, the dependence on $C_{\chi}$ and
$C_{\chi\perp}$ does not appear in the combination that defines $C$, where $C$ is fixed by the regularity condition. Indeed, we can rewrite the on-shell action as
\be S = - {1\over 2\pi\ell_s^2} \int dtdr\, {b^2} \sqrt{f-v^2 \over b^4 f - C^2 - {1-f\over v^2} C_{\chi\perp}^2 }. \ee
It is a regular action that converges in the IR and it is minimized by the lowest possible value for $C_{\chi\perp}$, i.e. $C_{\chi\perp}=0$, just as in the zero-temperature AdS case $C=0$ minimizes
the world-sheet action. At zero temperature $f\equiv1$ and the term containing $C_{\chi\perp}$ vanishes. Finally, $C_{\chi\perp}=0$ means that the string trails bending in the direction of the
velocity, as expected.

\section{Background and fluctuations for the trailing string in the general case}
\label{kinkydetails}

We now consider a classical string attached to the boundary and bending in an arbitrary direction, with the end-point moving at velocity $\vec v$ with
components $v^i$:

\be\lab{classic kink app}
X^i = v^i t + \xi^i(r), \qquad i=1,2,3.
\ee

\subsection{Classical solution} \lab{Appgen}

The induced world-sheet metric and the classical world-sheet action then become

\be
g_{ab} = b^2 \le( \begin{array}{cc} -(1 - |\vec v|^2) & \vec v \cd \vec \xi' \\ \vec v \cd \vec \xi' & 1 + |\vec \xi'|^2 \end{array} \ri), \ee
\be S = - {1\over 2\pi\ell_s^2} \int dtdr\, {b^2} \sqrt{(1 - |\vec v|^2)(1 + |\vec \xi'|^2) + (\vec v \cd \vec \xi')^2},
\ee

The classical solution for $\vec \xi$ can be derived using $\chi$ and $\n$ defined by \be \chi' = |\vec \xi'|,\quad \vs = \vec v
\cd \vec \xi.\ee The action only depends on the $r$-derivative of $\chi$ and $\vs$, yielding the solutions
\be
\chi' = {C_\chi \over \sqrt{b(r)^4 - C^2}} , \quad
\vs' = {C_\vs \over \sqrt{b(r)^4 - C^2}}. \lab{chizeta}
\ee
with $C^2_\vs = (1- |\vec v|^2) (C^2 - C_\chi^2)$. In the Cartesian frame $\xi^i$ this translates into the solution
\be\lab{class sol-app}
\xi'^i = {c^i \over \sqrt{b(r)^4 - C^2}} , \quad |\vec c|^2 + {(\vec v \cd \vec c)^2 \over 1 - |\vec v|^2} = C^2.
\ee

The on-shell action then reads

\be S = - {1\over 2\pi\ell_s^2} \frac1\g \int dtdr\, {b^4 \over \sqrt{b(r)^4 - C^2}},
\ee
with $\g = (1 - |\vec v|^2)^{-1/4}$.
It only depends on $C$, leaving two free coefficients out of the three $c^i$, which are related to the angles that the confining trailing string forms with the
velocity. $C$ has to be set to $C=b(r_m)^2\equiv b_m^2$ in order to minimize the action.

\subsection{Fluctuations} \label{App-fluctua}

We can now fluctuate the ansatz \refeq{classic kink app}, by writing

\be
X^i = v^i t + \xi^i(r) + \delta X^i(r,t).
\ee

To second order in the fluctuations, the action reads:

\bea\lab{S2app}
S = - {1\over 2\pi\ell_s^2} \int dtdr\, \frac12 \, {\cal G}^{ab}_{ij} \dt_a \d X^i \dt_b \d X^j,
\eea
where the ${\cal G}^{ab}_{ij}$ are obtained by expanding the world-sheet action:
\bea
{\cal G}^{ab}_{ij} = {b^4 \over \sqrt{-\det g_{ab}}} \le[ b^2 g^{ab} \Xi_{ij} - 2 \e^{ab} \a_{ij} \ri].
\eea
Here $\Xi_{ij}$ and $\a_{ij}$ are respectively symmetric and antisymmetric:
\bea
\Xi_{ij} &=& \le[(1 - |\vec v|^2) (1 + |\vec \xi'|^2) + (\vec v \cd \vec \xi')^2\ri] \d_{ij}+ \non\\ &&+ (1 + |\vec \xi'|^2) v_i v_j - (\vec v \cd \vec \xi') (v_i
\xi'_j + v_j \xi'_i) - (1 - |\vec v|^2) \xi'_i \xi'_j , \\
\a_{ij} &=& v_i \xi'_j - v_j \xi'_i .
\eea

In the absence of $\a_{ij}$ ---e.g. at zero velocity or when the classic string is assumed to bend in the direction of the velocity--- ${\cal
G}^{ab}_{ij}$ is proportional to $g^{ab}$; hence diagonalizing the induced metric, implies that also ${\cal G}^{ab}_{ij}$ is diagonal in the $a,b$
indices. In the general case where the string can bend in an almost arbitrary direction (only constrained by the fixed value of $C$), $\a_{ij}$ is
non-zero. However, as we will later show more explicitly, if the antisymmetric part of ${\cal G}_{ij}^{ab}$ is constant in $r$, it becomes a
topological term in the action and hence does not contribute to the equations of motion. We can already anticipate that, indeed, this is the case,
since $b^4 \a_{ij}/\sqrt{-\det g_{ab}}$ can be easily shown to be $r$-independent using the worldsheet background solution \refeq{class sol-app}.

The equations for the fluctuations read:
\be \lab{generalfluct-b}
\dt_a \le[ {\cal G}^{ab}_{ij} \dt_b \d X^j \ri] = 0.
\ee
These are coupled equations for the fluctuations $\d X^i$. We note that it is not possible to diagonalize ${\cal G}_{ij}$, unless $\vec v$ and $\vec \xi'$
are parallel (as in the black hole case) or $\vec v = 0$.

Indeed, this can be seen by decomposing $\vec{\d X}$ on a suitable basis
\be \vec{\d X} = \d X_v \vec v + \d X_\xi \vec c + \d X_n \vec n,\ee
where $\vec n$ is a vector that lays in direction transverse to the plane formed by $\vec v$ and $\vec \xi$ and $\vec c$ is the vector pointing in the
direction of $\vec \xi'$, as shown in equation \refeq{class sol-app}. Were $\vec v$ and $\vec \xi'$  parallel, the fluctuations would be decomposed
into a longitudinal and transverse part with respect to the string, $\d X^\parl=\d X_\xi$, and $\d X^\perp=\d X_n$. The equations would decouple, as
in the black hole background.

In general one can show that $\Xi_{ij}$ and $\a_{ij}$ cannot be diagonalized simultaneously. In fact, $\a_{ij}$ has purely imaginary eigenvalues, if
$\vec v$ and $\vec \xi'$ are not parallel
\be \m_\pm = \pm \sqrt{(\vec v \cd \vec \xi')^2 - |\vec v|^2 |\vec \xi'|^2}, \quad \m_n = 0,\ee
where $\mu_n$ is the eigenvalue corresponding tho the vector transverse to the $\vec v,\vec\xi'$ plane. On the other hand $\Xi_{ij}$ has real
eigenvectors
\be \l_\pm = \frac12 \le[ 2 - |\vec v|^2 + |\vec \xi|^2 \pm \sqrt{(-|\vec v|^2 + |\xi'|^2)^2 - 4 \le( (\vec v \cd \vec \xi')^2 - |\vec v|^2 |\vec
\xi'|^2 \ri)} \ri],\ee
\be \l_n = \le[(1 - |\vec v|^2) (1 + |\vec \xi'|^2) + (\vec v \cd \vec \xi')^2\ri] = - {\det g_{ab} \over b^4}. \ee
Here $\l_n$ is the eigenvector corresponding to $\vec n$. While $\a_{ij}$ has eigenvectors with one imaginary component and a real one, $\Xi_{ij}$ has
completely real eigenvectors, as for the $\vec v, \vec c$ system. The transverse vector $\vec n$ is eigenvector for both $\Xi_{ij}$ and $\a_{ij}$ and
can be decoupled, as expected.

We will then have two coupled equations for $\d X_v$ and $\d X_\xi$ and one decoupled equation for $\d X_n$
\bea
\dt_a\le\{ g^{ab} {b^6 \over R} \dt_b \d X_n \ri\} &=& 0, \lab{eqn1}\\
\dt_a \le\{ g^{ab} {b^2 R} \le[ \le( 1 + {c_\perp^2 \over R^2} \ri) \dt_b \d X_v
+ {c_\parl \over v} \dt_b \d X_\xi \ri] - \e^{ab} { c_\perp^2 } \dt_b \d X_\xi \ri\} &=& 0, \lab{eqv1}\\
\dt_a \le\{ g^{ab} {b^2 R} \le[ {(1-v^2) C^2 \over v^2} \dt_b \d X_\xi + {c_\parl \over v} \dt_b \d X_v \ri]
+ \e^{ab} c_\perp^2 \dt_b \d X_v \ri\} &=& 0, \lab{eqxi1}
\eea
with
\be
c_\perp^2 = \frac{1}{| \vec v|^2} \le[ | \vec v|^2 | \vec c|^2 - (\vec v \cd \vec c)^2 \ri], \quad c_\parl = \frac{1}{| \vec v|} \vec v \cd \vec c,
\ee
and $R\equiv\sqrt{b^4-C^2}$.
To simplify notation we write $|\vec v|$ as $v$ and similarly for $|\vec c|$, $c^2=c_\perp^2+c_\parl^2$. The relation between $\vec c$ and $C$ in
\refeq{class sol-app} translates into
\be
C^2 = c_\perp^2 + {c_\parl^2 \over 1-v^2}. \lab{cperpcparl}
\ee

Given that the term proportional to $\e^{ab}$ in \refeq{eqv1}-\refeq{eqxi1} is constant in $r$ (and $t$), we note that the off-diagonal components in
the indices $a,b$ coming from this term have to vanish due to symmetry. In other words, as we noted earlier, this term is topological and contribute
only boundary terms to the worldsheet action.

Diagonalizing the induced metric by a suitable change in the time coordinate, $t \to \tau = t - \z(r)$, with $\z'=-g_{tr}/g_{tt}$, leads, after some
manipulations, to the following set of decoupled equations:
\bea
\dt_r \le[ R \, \dt_r \d X^T \ri] + \g^2 \o^2 {b^4 \over R} \, \d X^T &=& 0, \lab{eqn-b}\\
\dt_r \le[ {R^{3} \over b^4} \, \dt_r \d X^L \ri] + \g^2 \o^2 R \, \d X^L &=& 0, \lab{eqxi-b}
\eea
with
\bea
\d X^T = \d X_n, \d X_v, \quad
\d X^L &=& \d X_c \equiv \d X_\xi + {v c_\parl \over (1-v^2) C^2 } \, \d X_v,
\eea
%

\section{Correlators from the confining trailing string} \lab{correlators}

The aim of this section is to obtain the zero-temperature correlators that have to be subtracted from the finite-temperature results. For sake of
comparison to the finite-temperature setup, we have to derive the correlators $G_{ij}$ in the orthonormal basis with $i,j=1,2,3$, where
\bea X^1=v t + \xi^1 + \d X^1,\quad X^2=\xi^2 + \d X^2,\quad X^3 = \xi^3 + \d X^3, \eea
and the $\xi^i$'s are given by \refeq{class sol}, while the $\d X^i$'s will be linear combinations of the transverse and longitudinal wave functions
solving \refeq{eqn-b} and \refeq{eqxi-b} respectively.

The classical solution for the confining trailing string \refeq{class sol} is characterized by two free parameters that identify the string collective coordinates.
We will discuss below how to obtain consistent correlators that do not depend on them. These ``integrated'' correlators will be the correct candidates
for the subtraction yielding the finite-temperature dressed correlators.

The string collective coordinates can be chosen among the $c^i$'s that have to obey to the relation in \refeq{class sol}. Alternatively, we can
redefine them in terms of two angles, $\th$ and $\vf$, being $\th$ the angle between the string and the velocity (that can be chosen to lie entirely
on the $x_1$ axe) and $\vf$ the angle between the string projection on the plane orthogonal to the velocity and one axe in this plane (that, for
instance, can be chosen to be $x_3$). This redefinition turns out to be useful to analyze the confining trailing string correlators. The relation between $\{c^i\}$
and $\{\th,\vf\}$ is simply
\bea c^1 &=& |c| \cos \th, \quad c^2 = |c| \sin\th \cos\vf, \quad c^3 = |c| \sin\th \sin\vf. \eea
We also obtain
\bea |c| = \sqrt{ 1-v^2 \over 1-v^2 \sin^2 \th } \, C, \lab{modc}\eea
where we made use of the relation in \refeq{class sol} between $c$ and $C$. If $\th=0$ the classical string formally reduces to the finite-temperature
trailing string solution (provided that we substitute $r_m$ with $r_s$, $T_m$ with $T_s$), where the string only lies in the direction of the
velocity. Since for the confining trailing string $\th$ and $\vf$ are arbitrary, we shall integrate over them, as we will shortly describe in detail.

With this setup, the string fluctuations, $\d X^i$'s, will generally depend on $\th$ and $\vf$. However, from the previous section, we know that we
can choose the $\d X_v, \d X_c, \d X_n$ (non orthogonal) modes. These solve decoupled $\th$- and $\vf$-independent equations that reduce to the
transverse and longitudinal finite-temperature fluctuation equations, provided that now $C=b_m^2$, rather than $C=v b_s^2$. For future convenience, we
normalize them in a suitable way and define:
\bea & \d \hat X^v = {c \over C} \sin\th \,v \d X_v = \sqrt{1-v^2 \over 1-v^2 \sin^2 \th} \sin \th \, v \, \d X_v,& \non\\
& \d \hat X^c = C \, \d X_c, \quad \d \hat X^n = \d X_n .& \lab{normal}\eea
These modes diagonalize the fluctuation action, as we will show below (we will denote all the quantities that are related to this basis with a hat).
Furthermore, we need to write the partition function in terms of $\d \hat X^I$ (the index $I$ runs over $\{v,c,n\}$), since the associated correlators
can be derived by solving the decoupled set of fluctuations equations. The final step will be the integration over $\th$ and $\vf$, once we have
expressed the action in terms of the orthonormal basis $\d X^i$, for the sake of comparison to the finite-temperature case.

More specifically, the procedure we will adopt is the following.

\begin{enumerate}
\item We first express the second-order fluctuation action \refeq{S2app} in terms of the $\d X^I$'s. In the path integral this translates into
\bea Z &=& \int {d\O \over 4\pi} \int \cD\le[ \d \hat X^I \ri]^3 |J(\th,\vf)| e^{i S\le[ \d \hat X \ri]} \\
&=& \int {d\O \over 4\pi} \int \cD\le[ \d \hat X^I \ri]^3 \exp \le\{ - \frac{i}{2} \int dr dt \, \hat\cG_{IJ}^{ab}(r) \, \dt_a \d \hat X^I(r,t) \,
\dt_b \d \hat X^J(r,t) \ri\}, \non\eea
where $|J|=1$ is the Jacobian corresponding to the change of variables from $\d X^i$ to $\d \hat X^I$, $d\O = d\vf \, d\cos\th$ and $\hat
\cG_{IJ}^{ab}$ is a diagonal matrix that reads
\bea \hat \cG_{IJ}^{rr} = - {1 \over \g^2} {R^2 \over b^4}  \hat \cG_{IJ}^{\t\t} = {1 \over 2\pi\ell_s^2} {R \over \g} \, \diag \le( \g^2, {R^2 \over
b^4}, 1 \ri), \lab{calhatG} \eea
with $R=\sqrt{b^4 - C^2}$.
The matrix $J$ associated to the change of variables is given explicitly by
\bea (J^{-1})^I_j = O^I_j = \le( \begin{array}{ccc} {|c| \over C} \sin\th & - {|c| \over C} \cos\th \cos \vf & - {|c| \over C} \cos\th \sin \vf \\
\g^2 {|c| \over C} \cos\th & {|c| \over C} \sin\th \cos\vf & {|c| \over C} \sin\th \sin\vf \\
0 & - \sin\vf & \cos\vf \end{array} \ri), \lab{O} \eea
where $|c|$ is the function of $\th$ given by equation \refeq{modc}.

We may now observe that the normalization \refeq{normal} has been chosen so that $|J|=1$. Here we only include the symmetric part of the action \refeq{S2app}, $\propto g^{ab}\Xi_{ij}$. The vanishing of the
antisymmetric part, $\propto \e^{ab}\a_{ij}$, upon integration, can be checked easily by following the same steps that we describe below for the symmetric contribution (see appendix~\ref{antisym}).
\item We then make use of the equations of motion $\d \hat X^I(r,\o) = F_J^I(r,\o) \d \hat X_0^J(\o)$ in the Fourier transformed fields to evaluate the functional integral at the saddle point,
integrate by parts and evaluate the $dr$ integral. These steps lead to the following expression:
\bea Z &=& \int {d\O \over 4\pi} \lab{Zpath} \\ &&\times \exp \le\{- \frac{i}{2} \int d\o \, F_I^K(r_q,-\o) \hat\cG_{KM}^{rr}(r_q) \dt_r F_J^M(r_q,\o)\, \d \hat X_0^I(-\o) \d \hat X_0^J(\o) \ri\}.
\non \eea
Here $F_J^I$ is a diagonal kernel,
\bea F_J^I(r,\o) = \diag \le( F^T(r,\o), F^L(r,\o), F^T(r,\o) \ri), \lab{F} \eea
where $F^T$ and $F^L$ are the infalling wave-function solutions to the equations \refeq{eqn-b} and \refeq{eqxi-b} respectively, with unitary boundary conditions. The wave functions $\d \hat X^I$ take
boundary values $\d \hat X_0^I$. The boundary is at $r=r_q$, where $r_q=0$ for infinitely massive probes, while it is finite, $r_q \sim 1/M_q$, for probes with finite mass.
\item In order to obtain the correlators in the orthonormal basis $\{x_i\}$, we have to rotate the action back to this basis:
\bea Z
&=& \int {d\O \over 4\pi} \exp \le\{i \int d\o \, G_{ij}(\o;\th,\vf) \, \d X^i_0(-\o) \d X^j_0(\o) \ri\}.\eea
We have defined the $\th$- and $\vf$-dependent correlator as
\bea G_{ij}(\o;\th,\vf) = - \frac12 O^I_i(\th,\vf) F_I^K(r_q,-\o) \hat\cG_{KM}^{rr}(r_q) \dt_r F_J^M(r_q,\o) O^J_j(\th,\vf), \lab{tdGdef} \non\\ \eea
where $O$ is defined in \refeq{O}. Putting the equations \refeq{calhatG}, \refeq{F}, \refeq{tdGdef} and \refeq{O} together, we are led to the following expression for $G_{ij}$:
\bea G_{ij} &=& {1 \over 1-v^2 \sin^2\th} \left(H_{Tij} G^T + H_{Lij} G^L\right) , \lab{tdGapp} \eea
where $H_T$ and $H_L$ are $\o$-independent symmetric matrices:
\bea \begin{array}{rclcrcl}

H_{T}^{11} &=& \sin^2\th, &\quad& H_{L}^{11} &=& \cos^2\th, \\
H_{T}^{12} &=& - \cos\th \sin\th \cos\vf, &\quad& H_{L}^{12}&=& (1-v^2) \cos\th \sin\th \cos\vf, \\
H_{T}^{13} &=& - \cos\th \sin\th \sin\vf, &\quad& H_{L}^{13}&=& (1-v^2) \cos\th \sin\th \sin\vf, \\
H_{T}^{22} &=& \cos^2\th + (1-v^2) \sin^2\th \sin^2\vf , &\quad& H_{L}^{22} &=& (1-v^2)^2 \sin^2\th \cos^2\vf, \\
H_{T}^{23} &=& - (1 - v^2) \sin^2 \th \cos\vf \sin\vf ,&\quad& H_{L}^{23}&=& (1 - v^2)^2 \sin^2 \th \cos\vf \sin\vf, \\
H_{T}^{33} &=& \cos^2\th + (1-v^2) \sin^2\th \cos^2\vf , &\quad& H_{L}^{33}&=& (1-v^2)^2 \sin^2\th \sin^2\vf.
\end{array} ~\non\\\label{Hij} \eea
Here we have defined $G^T$ and $G^L$ as the correlators obtained for the diagonal system\footnote{These correlators formally reduce to the finite-temperature correlators if $r_m \to r_s$, $C \to b_s^2 v$, and the blackening function $f$
is introduced}: 
\bea G^T (\o) &=& - \frac12 {1\over2\pi\ell_s^2} {R \over \g} F^T(r,-\o) \dt_r F^T(r,\o) \big|_{r=r_q},
\lab{hatGperp-b}\\
G^L (\o) &=& - \frac12 {1\over2\pi\ell_s^2} {\g R^3 \over b^4} F^L(r,-\o) \dt_r F^L(r,\o) \big|_{r=r_q}.
\lab{hatGparl-b} \eea
We note that the $\th$- and $\vf$-dependent correlator \refeq{tdGapp} reduces to $G_{11} = G^L $, $G_{22} = G_{33} = G^T $, (all other components vanish) in the $\th\to0,\pi$
limit, as it is expected.

\end{enumerate}

\section{Averaging over the shadow quark velocity} \label{vaverage}

One way to perform the angular average is to integrate over all string direction with the standard solid angle integration measure. 
An alternative way to do obtain angle-independent boundary correlators is to average over the velocity of the quark at infinity. Suppose that, once a stationary regime is reached, the observed quark has constant velocity $\vec{v} = v \hat{z}$ (directed along $x$ without loss of generality) and the second (unobserved) quark has constant velocity $\vec{w}$.
 At long times, and regardless of initial conditions, we have:
\be\label{a1}
\vec{x}_1(t) = \vec{v}\, t + \ldots, \qquad \vec{x}_2(t) = \vec{w}\,t + \ldots
\ee
where $x_1$ and $x_2$ are the (classical) positions of, respectively, the observed and unobserved quark and $\ldots$ means subleading contributions in the $t\to \infty$ limit. In this limit, the direction of the string is the unit vector
\be\label{a2}
\hat{\xi} = {\vec{x}_2(t) - \vec{x}_1 (t)\over |\vec{x}_2(t) - \vec{x}_1(t)|} \rightarrow {\vec{w} -\vec{v} \over |\vec{w} -\vec{v}| }
\ee
and it is time-independent. In the stationary regime, the angle the string makes with the observed quark velocity (i.e. the angle $\theta$ of the previous sections) is fixed and is given by the relation:
\be\label{a3}
\cos \theta = {\vec{v} \over v} \cdot {\vec{w} -\vec{v} \over |\vec{w} -\vec{v}| } = {w \cos\theta_w - v \over (w^2 + v^2 -2 vw\cos\theta_w)^{1/2}}
\ee
where $w= |\vec{w}|$ and $\theta_w$ is the angle between $\vec{w}$ and $\vec{v}$, i.e. in spherical coordinates:
\be\label{a4}
\vec{w} = w \left( \begin{array}{c} \cos\theta_w \\ \sin\theta_w \, \cos\vf_w \\ \sin\theta_w \, \sin\vf_w \end{array}\right).
\ee

Assuming complete lack of knowledge of $\vec{w}$ amounts to averaging over all possible values of $\vec{w}$, i.e. on the 4-velocity vector $u_\alpha \equiv (u_0, \vec{u}) = (\g (w), \g (w) \vec{w})$, using the Lorentz-invariant measure:
\be\label{a5}
d\mu[u_\alpha] = d^4 u \, \delta(u^\alpha u_\alpha -1) = {d^3 \vec{u} \over 2 u_0} = {d^3 \vec{u} \over 2\sqrt{1+|\vec{u}|^2}}
\ee
Changing variables to $\vec{w}$ by the relation $\vec{u} = \vec{w}/\sqrt{1-w^2}$, the integration measure becomes:
\be\label{a6}
d\mu [\vec{w}] = {d^3{w} \over 2(1-w^2)^2} = d\Omega_w \, { w^2 dw \over 2(1-w^2)^2}
\ee
where the range of $w$ is now the interval $[0,1)$.

Given any $\vec{w}$-dependent observable $A(w,\theta_w,\vf_w)$ we define its $w$-average by:
\bea \label{a7}
\< A\>_w =&& \lim_{\epsilon \to 0}\left\{ \left[\int_0^{1-\epsilon} {w^2 dw \over 2(1-w^2)^2} \int_{-1}^1 d\cos\th_w \int_0^{2\pi} d\vf_w \; A(w,\theta_w,\vf_w) \right]\right. \nonumber \\ && \quad \quad \left. \Big/ \; \left[4\pi \int_0^{1-\epsilon} {w^2 dw \over 2(1-w^2)^2} \right] \right\}.
\eea
The regularization is required since $\int d\mu [\vec{w}] 1$ diverges. If the limit of (\ref{a7}) is finite, it should not depend on how exactly we regulate the integrals).
The normalization factor can be approximated as:
\be\label{a8}
4\pi \int_0^{1-\epsilon} {w^2 dw \over 2 (1-w^2)^2} = {\pi\over 2\epsilon} + O(1),
\ee
and as long as the numerator in (\ref{a7})  is divergent as $O(1 /\epsilon)$ or faster, the subleading terms are irrelevant.

For comparison, we will denote by $\< \>_{\Omega}$ the simple average over the direction of the string,
\be\label{a9}
\< A(v)\>_\Omega = \int {d\Omega(\theta,\vf) \over 4\pi} A(v,\theta,\vf)
\ee
Notice that, if $v=0$, equation (\ref{a3}) tells us that $\cos \theta_w = \cos \th$ , and
\be
\<A (v=0)\>_w = \< A(v=0)\>_\Omega .
\ee

\section{Antisymmetric contribution to fluctuation action} \label{antisym}

We here follow the steps 1-4 of section \ref{correlators}, giving the explicit expressions for the quantities associated to the antisymmetric
contribution to the second order fluctuation action, $\propto \e^{ab} \a_{ij}$. We will show that this contribution vanishes, once the integration over the angles
is performed.

\begin{enumerate}
\item If we express the antisymmetric contribution in terms of the eigenvalue basis, equation \refeq{normal}, the analogue of formula \refeq{calhatG} for the antisymmetric term is
\bea \hat \cG_{IJ}^{(A)r\t} = - \hat \cG_{IJ}^{(A)\t r} =
{1 \over 2\pi\ell_s^2} {2 v C \sin\th \over \sqrt{1-v^2 \sin^2\th}} \le( \begin{array}{ccc}0 & 1 & 0 \\ -1 & 0 & 0 \\ 0 & 0 & 0 \end{array} \ri). \lab{calhatGanti} \eea
\item We then integrate by parts and are left with a non trivial boundary term in $r$ and a trivial boundary term in $\t$. We obtain the result \refeq{Zpath} for the partition function, now with $\hat
\cG^{(A)r\t}$ replacing $\hat \cG^{rr}$ (the $\hat \cG^{(A)\t r}$ term gives the vanishing total derivative term in $\t$ and a bulk term that cancels the bulk contribution from $\hat \cG^{(A)r\t}$).
\item Rotating back to the Cartesian basis, making use of the matrix \refeq{O}, we obtain the analogue of formula \refeq{tdGapp}:
\bea G_{ij}^{(A)} = G^{TL} { 2 v C \sin\th \over \sqrt{1 - v^2 \sin^2 \th} } \le( \begin{array}{ccc}0 & \cos \vf & \sin\vf \\ -\cos\vf & 0 & 0 \\ -\sin\vf & 0 & 0 \end{array} \ri) . \eea
The off--diagonal factor $G^{TL}$ is defined by $$G^{TL}(\o) = - \frac12 {1\over 2 \pi \ell_s^2} (i\o) F^T(r_q,-\o)F^L(r_q,\o)= - \frac12 {1\over 2 \pi \ell_s^2} (i\o),$$
%
where we used the boundary condition $F(r_q,\o)=1$.
\item Averaging this expression yields a vanishing results in all cases, either due to the antisymmetry of the sinusoidal functions over the interval of integration (for the $\O$-- and $w$--average),
or because of the proportionality to $\sin \th$ in the fixed--angle case ($\th=\pi$).

\end{enumerate}

\section{Langevin equation and consistency relations} \label{Langevin-app}

Here we illustrate the derivation of the generalize Langevin equation, and
in particular show that
\be\label{consistency}
v{d\over dv} \< \cF^i \> = -i \left[dG^{ij} \over d\o\right]_{\o=0} v_j.
\ee
where $\cF^i$ is the plasma operator that couples to the quark position, $G^{ij}$ is the retarded correlator of the same operator, $v^i$ is the particle velocity (assumed to be constant) and $v = \sqrt{v^iv^i}$.

Assuming the correlator on the right hand side has an analytic expansion near $\o=0$, the right hand side equals the friction coefficient, $\eta^{ij}$, i.e,
\be
G^{ij} = - i \omega \eta^{ij} (v) + O(\omega^2)
\ee
On the other hand, the expectation value of the force operator $\< \cF^i \>$ equals, holographically, the classical drag force $\pi^i$ exerted by the trailing string on its endpoint. Therefore (\ref{consistency}) can be rewritten as:
\be\label{consistency1}
v {d\over dv} \pi^i = -\eta^{ij}v_j .
\ee

\subsection{Langevin equation for the fluctuations around a classical trajectory}

We start with the path-integral derivation of Langevin equation, using the double-field formalism. In this framework, the partition function for the position of a single quark $X^i(t)$ and coupled to an ensemble can be written as:
\be\label{cons1}
Z = \int {\cal D}[X^i_+(t)] {\cal D}[X^i_-(t)] e^{i (S_0[X_+] - S_0[X_-])} \Big\< \exp \left[ i\int dt\, {\cal F}_+\cdot X_+ - i \int dt\, {\cal F}_-\cdot X_- \right]\Big\>_{ens} \nonumber \\
\ee
where $S_0[X]$ is the classical action for a free particle, and $\cF^i$ accounts for the coupling between the ensemble and the particle.

The goal is to write the integral over the bath in the form of an effective action for the fields $X_{\pm}(t)$, which would change the dynamics from that
encoded in $S[X]$ to include the effect of the bath. This is standard, and
is achieved by expanding the exponentials, substituting linear and quadratic terms of $F$ with their ensemble averages (i.e. introducing one- and two-point functions) and re-exponentiating to obtain an effective action quadratic in $X$. For details the reader can is referred for example to \cite{kleinert} for a clean derivation. Here, since we are expanding around a constant-velocity background, and we want to keep
the result to all orders in $v$, we will repeat that derivation separating background and fluctuations.

First, we expand $X_+$ and $X_-$ around the constant $\vec{v}$ background. For this purpose, we write:
\be\label{cons2}
X^i_{\pm} = \bar{X}^i(t) + \delta X^i_{\pm}(t), \qquad \bar{X}^i(t) = v^i t
\ee
Expanding the exponential in (\ref{cons1}) to second order in $\delta X^i$ we obtain:
\bea\label{cons3}
Z \simeq && \int {\cal D}[X^i_+(t)] {\cal D}[X^i_-(t)] e^{i (S_0[X_+] - S_0[X_-])} \nonumber
\\ && \Big\< e^{i\int v^i t (\cF_+^i - \cF_-^i)} \left[1 + i\int dt \left(\delta X^i_+ \cF^i_+ - i \delta X^i_- \cF^i_- \right) + \right. \nonumber \\
&&
\left. -{1\over 2} \int dt dt' \left(\delta X^i_+\delta X^j_+ \cF_{+}^i\cF^j_+ + \delta X^i_-\delta X^j_- \cF_-^i\cF_-^j \right. \right. \nonumber \\ && \left. \left. - \delta X^i_+\delta X^j_- \cF_{+}^i\cF^j_- - \delta X^i_-\delta X^j_+ \cF_{-}^i\cF^j_+ \right) \right] \Big\>
\eea
This can be re-exponentiated by introducing the appropriate one- and two-point functions:
\be\label{cons4}
Z \simeq \int {\cal D}[X^i_+(t)] {\cal D}[X^i_-(t)] e^{i (S_0[X_+] - S_0[X_-]) + i S_{eff}[\d X_+,\d X_-]}
\ee
where
\be\label{cons5}
S_{eff} = \int dt\, \delta X_+^i \< \cF^i_{+} \>_v - \delta X_-^i \< \cF^i_{-} \>_v + {1\over 2}\int dt dt' G_{ab}^{ij}(t,t'; v) \delta X^i_a \delta X^j_b
\ee
where $,\< \>_v$ denotes correlators computed in the presence of the source $\bar{X}_{\m} = vt$ and
\be\label{cons6}
G_{ab}^{ij}(t,t'; v) = i \left(\begin{array}{cc}
\< \cF^i_+(t)\cF^j_+(t') \>_v & -\< \cF^i_+(t)\cF^j_-(t') \>_v \\
-\< \cF^i_-(t)\cF^j_+(t') \>_v & \< \cF^i_-(t)\cF^j_-(t') \>_v \end{array}\right)
\ee
are the {\em connected , path ordered} correlators of the operators $\cF_{\pm}$, and are related to the connected retarded and symmetric correlators by\footnote{These relations may be obtained by from the path ordering of the correlators, as well as the operator definitions of the correlators:
\be
G_{sym}(t,t') = {1\over 2}\<\{\cF(t), \cF(t')\}\>, \quad G_R(t,t') = -i \theta(t-t')\<[\cF(t), \cF(t')]\>.
\ee
}:
\bea
&& G_{sym} = {1\over 4} \left(\< \cF_+\cF_+\>_v + \<\cF_-\cF_- \>_v + \< \cF_-\cF_+\>_v + \< \cF_+\cF_-\>_v\right) \label{cons9i}\\
&& G_R = - {i\over 2}\left(\< \cF_+\cF_+\>_v - \<\cF_-\cF_- \>_v + \< \cF_-\cF_+\>_v - \< \cF_+\cF_-\>_v\right) \label{cons9ii}
\eea
(note that, with the exception of this and the following subsection in the appendix, we use $G$ for $G_R$, to simplify notation).
We now change to the $\d X_{cl}, \d X_r$ variables defined by
\be\label{cons7}
\d X_{cl} = (\d X_+ + \d X_- )/2, \qquad \d X_r = \d X_+ - \d X_-
\ee
and obtain
\bea\label{cons8}
&&S_{eff} = \int dt \,\delta X_r^i \< \cF^i_{+} + \cF^i_{-} \>_v + \delta X_{cl}^i \< \cF^i_{+} - \cF^i_{-} \>_v + \nonumber \\ &&\!\!\!\!\!\!\!\!\!\!\!\!\!\!\! - \int dt dt' G_R^{ij}(t,t'; v) \delta
X^i_{cl}(t) \delta X^j_r(t') + {i\over 2}G_{sym}^{ij}(t,t';v)\delta X^i_{r}(t) \delta X^j_r(t')
\eea

On the other hand, the classical part of the action can be expanded in the same fashion, giving:
\be\label{cons9}
S_0[X_+] - S_0[X_-] = \left[{\delta S_0 \over \delta X^i} \right]_{\bar{X}}\delta X^i_r + \left[{\delta^2 S_0 \over \delta X^i\d X^j}\right]_{\bar{X}} \delta X_{cl}^i \delta X_r^j
\ee
(the terms like e.g. $\delta X_{cl}^2$ cancel because of our choice of the background, i.e. $\bar{X}_-=\bar{X}_+$).

The total action appearing in the exponent of path integral (\ref{cons4}) contains a linear and a quadratic piece:
\bea
&&S[\delta X_a] = S^{(1)} + S^{(2)}, \nonumber \\
&&S^{(1)} = \int dt \,\left[\d X_r^i \left( {\delta S_0 \over \delta X^i} + \frac12 \< \cF^i_{+} + \cF^i_{-} \>_v \right) + \d X_{cl}^i \< \cF^i_{+} - \cF^i_{-} \>_v \right] \label{cons10}\\
&&
S^{(2)} = \int dt dt' \left[\, {i\over 2}\d X_r^i(t) G_{sym}^{ij}(t,t';v) \d X_r(t') \right.\nonumber \\
&&\left. + \d X_r^i(t) \left[\left.{\delta^2 S_0 \over \delta X^i(t)\d X^j(t')}\right|_{\bar{X}} - G_{R}^{ij}(t,t';v) \right] \d X_{cl}(t')\right]. \label{cons11}
\eea
At linear order, i.e. tadpole cancellation gives on the one hand the condition
\be\label{cons12}
\< \cF^i_{+}\>_v = \< \cF^i_{-} \>_v,
\ee
as expected since we have chosen a symmetric ansatz for the background; on the other hand it gives the classical equations of motions for the background,
\be\label{cons13}
{\delta S_0 \over \delta X^i} = - \< \cF^i\>_v.
\ee
The particle is subject to a (velocity dependent) external force $ \< \cF^i\>_v$, arising as the effect of the coupling with fluid.

At quadratic order, the action depends quadratically on $\d X_r$ and only linearly in $\d X_{cl}$. We resort to the usual trick of trading the variable $\d X_r$ for a random Gaussian noise $\xi(t)$ with variance $G_{sym}$, by the identity
\be\label{cons14}
\exp \left[-{1\over 2}\int \d X_r^i G_{sym}^{ij} \d X_r^j\right] = {\cal N}\int {\cal D}[\xi^i(t)] \exp \left[-{1\over 2}\int \xi^i ( G_{sym}^{-1})^{ij} \xi^j - i \int \d X_r^i \xi^i. \right]
\ee
where ${\cal N}$ is a normalization factor.
Now the action is linear in $\d X_r$ and the path integral over this variable can be performed resulting in a functional Dirac's $\delta$. We arrive at:
\bea\label{cons15}
Z = &&\int {\cal D}[\xi^i] {\cal D}[X^i_{cl}] \,\,\delta\left(\left.{\delta^2 S_0 \over \delta X^i\d X^j}\right|_{\bar{X}}\d X_{cl}^j(t) \,- \, \int dt'\, G_{R}^{ij}(t,t';v) \d X_{cl}^j(t') \,- \,\xi^i(t) \right) \nonumber\\
&&\times \exp\left[-{1\over 2}\int \xi^i ( G_{sym}^{-1})^{ij} \xi^j \right].
\eea
From this expression, one concludes that the fluctuation $\d X_{cl}(t)$ around the classical particle trajectory $\bar{X}^i(t)$ obeys the Generalized Langevin equation
\be\label{cons16}
\left.{\delta^2 S_0 \over \delta X^i\d X^j}\right|_{\bar{X}}\d X_{cl}^j(t) \,- \, \int dt'\, G_{R}^{ij}(t,t';v) \d X_{cl}^j(t') = \xi^i(t)
\ee
where the memory kernel $G_R(t,t')$ is the retarded correlator of the force operator $\cF$ and with Gaussian noise variance is the symmetric correlator of the same operator.

\subsection{A double-field Ward identity} \label{Appconst}

In the formalism of the previous subsection, it is straightforward to obtain the consistency relation (\ref{consistency}) between the ``classical'' and ``quantum'' terms arising from the interaction with the fluid. By definition,
and with the relation (\ref{cons12}), we have
\be\label{cons17}
\< \cF^i \>_{\bar{X}} = {1\over 2}\left\< \left(\cF_+^i + \cF_-^i\right) \exp \left\{i\int dt\, \bar{X}^i(t) (\cF_+^i - \cF_-^i) \right\} \right\>_{ens}
\ee
Specifying a uniform velocity background trajectory,
\be
\bar{X}^i(t) = v^i \, t
\ee
and taking a derivative of (\ref{cons17}) with respect to $v^j$ we obtain:
\be
{\de \< \cF^i \>_{\vec{v}} \over \de v^j} = {i\over 2}\left\< \int dt'\, t'\,\left(\cF_+^j(t') -\cF_-^j(t')\right) \left(\cF_+^i(t) + \cF_-^i(t)\right) \right\>_{\vec{v}} = \int dt' \,t'\, G_R^{ij}(t,t'; \vec{v})
\ee
where in the last equality we have used the relation (\ref{cons9ii}) between the retarded correlator and the double-field path-ordered correlators. Going to Fourier space, $G_R(t,t')=\int (d\omega/2\pi) e^{-i\omega(t-t')}G_R(\omega)$, this becomes:
\be
{\de \< \cF^i \>_{\vec{v}} \over \de v^j} = -i {dG^{ij}_R(\omega;v) \over d\omega}\Big|_{\omega=0}
\ee
From this general relation, projecting over $v^j$ and noting that $d/dv = (v^j/v) {\de/\de v^j}$ we obtain (\ref{consistency})

\end{document}